\documentclass[12pt, letterpaper, tikz, border=10pt]{article}

\usepackage{amsmath, amsfonts, amscd, amssymb}
\usepackage{fancybox, eufrak, euscript, oldgerm}
\usepackage{amsthm}
\usepackage{mathrsfs}
\usepackage[all,cmtip]{xy}
\usepackage{xcolor}
\usepackage{blindtext}

\usepackage{color,soul}

\usepackage[pdftex]{graphicx}
\usepackage{chronology}

\usepackage{tikz-cd}

\usepackage{smartdiagram}

\usepackage{graphicx}

\usetikzlibrary{decorations.markings,intersections} 

\usepackage{smartdiagram}
\smartdiagramset{font=\sffamily}

\smartdiagramset{font=\sffamily,
   text width = 3cm}
\tikzset{module/.append style=
  { top color=\col, bottom color=\col},
  every shadow/.style = {fill=none, shadow scale=0}}

\newcommand{\aconn}{\mathcal{A}}

\newcommand{\aut}{{\mathcal{A}}ut}

\newcommand{\conn}{\mathcal{D}}

\newcommand{\cons}{\mathbf{C}}

\newcommand{\curv}{R}

\newcommand{\eh}{\mathcal{E}\mathcal{H}}

\newcommand{\gauge}{\mathcal{U}}

\newcommand{\Hom}{{\mathcal{H}}om}
\newcommand{\kd}{\text{\texttt{d}}}

\newcommand{\modl}{\mathbf{\mathcal{E}}}

\newcommand{\omg}{\Omega}
\newcommand{\Omg}{\mathbf{\Omega}}

\newcommand{\cont}{\mathcal{C}^{0}}
\newcommand{\smooth}{\mathcal{C}^{\infty}}
\newcommand{\sconn}{\textsf{A}}

\newcommand{\struc}{\mathbf{A}}

\newcommand{\triad}{{\mathfrak{T}}}
\newcommand{\ctriad}{{\mathfrak{D}}\triad}

\newcommand{\wee}{\,{\scriptstyle\wedge}\,}

\newcommand{\ym}{\mathcal{Y}\mathcal{M}}
\newcommand{\com}{\mathbb{C}}
\newcommand{\mapto}{\longrightarrow}

\newcommand{\N}{\mathbb{N}}
\newcommand{\Z}{\mathbb{Z}}
\newcommand{\R}{\mathbb{R}}

\newcommand{\Q}{\mathbb{Q}}

\newcommand{\cas}{\mathcal{A}\mathcal{S}}
\newcommand{\geom}{\mathcal{G}\mathcal{M}_{\sts}}
 
\newcommand{\sts}{\mathscr{A}}
\newcommand{\mule}{\mathscr{M}}

\newcommand{\AR}{\sts\text{-}R}
\newcommand{\AI}{\sts\text{-}I}

\newcommand{\catcas}{\mathcal{C}\mathcal{A}\mathcal{S}}
\newcommand{\catdt}{\mathcal{C}\mathcal{D}\mathcal{T}}
\newcommand{\catf}{\mathcal{C}\mathcal{F}}
\newcommand{\catcs}{\mathcal{C}\mathcal{S}}
\newcommand{\catab}{\mathcal{S}h_{Ab}}

\newcommand{\Einst}{{\mathcal{E}\mathrm{inst}}}

\title{\bf $\sts$-Relativity: \\ A General Theory of Gravity\\ from Structure Sheaf\thanks{A proofread version of this paper has been recently published in the peer reviewed journal {\it Time Space} 2026, 2, 6. {\bf https://doi.org/10.3390/timespace2030006}}\thanks{Apart from its prefix `{\em $\sts$-Relativity}', the title of this paper closely emulates the title of Alexander Grothendieck's celebrated paper {\em A General Theory of Fiber Spaces with Structure Sheaf} \cite{groth1}. This is intentional and not just a superficial similarity done for effect: in this paper, we dissect, analyse, argue and explain in detail the title of our paper, but in a forthcoming paper \cite{rapX} we explore the remarkably similar, both conceptually (semantically-philosophically) and technically (structurally-mathematically), works of Alexander Grothendieck in {\em Algebraic} Geometry and Anastasios Mallios in {\em Differential} Geometry, respectively.}}

\author{Ioannis Raptis\thanks{Supply \& Substitute Secondary School Teacher of Mathematics, Physics and Chemistry, Milk Education, Chester, Cheshire, United Kingdom; email: {\it irapti11@gmail.com}}}

\date{Thursday, 16th of June 2026}        

\newpage

\begin{document}   

\maketitle

\pagestyle{myheadings}\markboth{\centerline {\small {\sc
{Ioannis Raptis}}}}{\centerline
{\footnotesize {\sc {{\bf I. Raptis:} $\sts$-Relativity: ADG-Gravity from Structure Sheaf}}}}
 
\pagenumbering{arabic}

\begin{abstract}

\noindent{\small The aim of this paper is to introduce to a wider readership of mathematicians, mathematical physicists and philosophers of mathematics and physics alike the basic theoretical, both conceptual and technical, tenets of Mallios's Abstract Differential Geometry (ADG), as well as to present and to summarise the main results from its applications in the last quarter of a century towards formulating an entirely homological algebraic, purely gauge-theoretic, finitistic, quantal and manifestly background geometrical smooth spacetime manifold independent vacuum Einstein gravity. This is an abstract and generalised version of the usual pseudo-Riemannian spacetime manifold based vacuum Einstein gravity of General Relativity (GR), here called {\em ADG-gravity}. ADG-gravity, like the mathematical theory of ADG on which it is based, is here shown to crucially rely on and, in a deeper sense that we discuss in detail, to derive from, a sheaf $\sts$ of commutative algebras representing the {\em structure sheaf} of generalised arithmetics or the module of sheaf cohomological coefficients in ADG, which is in turn physically interpreted as a sheaf of abelian algebras of generalised local coordinates in ADG-gravity. As a result, ADG-gravity is seen to support and be supported by abstract and generalised ADG-theoretic versions of both the Principle of General Covariance (PGC) of GR, here coined the {\em Principle of Algebraic Relativity of Differentiability} (PARD), and Einstein's General Principle of Relativity, here coined the {\em Principle of $\sts$-Relativity} ($\AR$). The PARD and the $\AR$ are seen to be effectively equivalent principles, both being categorical variants of what Mallios originally coined the {\em Principle of $\sts$-Invariance} ($\AI$), as both can be formulated {\em functorially} in terms of natural transformation type of morphisms between the relevant ADG-theoretic sheaf categories involved. In the last section, we discuss the physical import and significance of ADG-gravity, as well as of the PARD, the $\AR$ and the $\AI$ that it supports and is supported by, in various important technical (structural-mathematical) and conceptual (physico-philosophical) issues in current, and in potentially future, Classical (GR) and Quantum Gravity (QG) research. The paper closes with a recurring theme of ours, already touched and elaborated upon throughout the last two and a half decades of applying ADG to QG \cite{rap11,rap14,rap15,rap19}, namely: {\em the importance of developing new mathematics and associated theoretical concepts in QG research}. This paper is an extended, more mathematically slanted, sequel to our latest physico-philosophical review of applications of ADG in QG \cite{rap19} in the light of further recent categorical developments and results on the functorial character and nature of ADG-gravity and $\sts$-Relativity presented in \cite{rap15}.}

\vskip 0.1in               

\noindent{\footnotesize {\bf PACS numbers}: 04.60.-m, 04.20.Gz,
04.20.-q}

\noindent{\footnotesize {\bf Key words}: {\small Abstract Differential Geometry; Sheaf Theory; Sheaf Cohomology; Category Theory; Functoriality of Vacuum Einstein Dynamics; Dynamical Natural Transformations; Geometric Morphism Invariant Dynamics; Categorically Generalised Principle of General Covariance, The Principle of Algebraic Relativity of Differentiability; Structure Sheaf $\sts$-Relativity and $\sts$-Invariance; Background Geometrical Spacetime Manifold Independence; Vacuum Einstein Gravity as a Pure Gauge Theory; Half-Order Gravitational Gauge Field Formalism; Intrinsic Sheaf Cohomological Third Quantisation of Gravity and Gauge Field (Yang-Mills) Theories; Vacuum Einstein Gravitational Gauge Field as an External (Background) Spacetime Manifoldless, Fundamental Space-Time Scale-less, Autonomous and Closed Quantum Dynamical System; Pure Gauge Field Ontology and Realism}}  

\vskip 0.1in 
\centerline{$<.><.><.><.><.><.><.><.><.><.>$}      
\vskip 0.1in

\noindent $\bullet$\underline{\bf Author's Note 1:} {\em This paper is new, previously unpublished, original research.}

\end{abstract} 

\newpage

\tableofcontents

\newpage

\addcontentsline{toc}{section}{Opening Quote}

\section*{Opening Quote}

\bigskip\noindent (Q1)\hskip 0.9in
\begin{minipage}{11cm}
\noindent ``...Circa 1990, Anastasios Mallios used sheaf–theoretic methods to extend the mechanism of the
classical differential geometry (CDG) of smooth manifolds to spaces, which do not
admit the usual smooth structure (:smooth atlas). In this new setting of
abstract differential geometry (ADG) a large number of notions and results of CDG
have already been extended, becoming at the same time applicable to spaces
with singularities and to quantum physics.

{\em In ADG, the ordinary structure sheaf of smooth functions} [on a smooth manifold] {\em is replaced by a sheaf
of abstract algebras $\sts$, admitting a differential $\partial$ (in the algebraic sense), which takes
values in an $\sts$-module $\Omega$}. A triplet $(\sts , \partial , \Omega)$ like that is called a differential triad.
Suitably defined morphisms organize the differential triads into a category denoted by $\ctriad$. 
Every smooth manifold defines a differential triad and every smooth map
between manifolds defines a morphism of the respective differential triads, so that the
category $Man$ of smooth manifolds is embedded in $\ctriad$ (ibid.)...'' \cite{fragpap}    
\end{minipage}     

\vskip 0.1in
\centerline{$<.><.><.><.><.><.><.><.><.><.>$}      
\vskip 0.1in

\noindent $\bullet$ \underline{\bf Author's Note 2:} In the paper below, among various numbered {\em Quotes} and {\em Important Notes}, the reader will also encounter numerically ordered {\em Aphorisms} and {\em Apophthegms}. Aphorisms, first encountered in our recent paper \cite{rap19}, are {\em theoretical axioms} that are, in a sense, basic, conceptually irreducible and fundamental statements from an ADG-theoretic perspective.\footnote{In our latest mathematical endeavours and philosophical musings on applications of ADG to QG \cite{rap15,rap19}, in continuation and extension of the didactics (:lessons learned) in \cite{rap11,rap14}, we distilled certain {\em aphorisms} which encapsulate certain key ADG-theoretic concepts and results from applying ADG to QG research.} Apophthegms,\footnote{An {\em apophthegm}, {\em apophthegma}, or even {\em apothegm} (plural in Greek, {\em apophthegmata}, {\em apophthegms} or even {\em apothegms}; words which we will use interchangeably in the sequel).} on the other hand, are reduced down, concise and often terse statements, encapsulating in a nutshell a saying or proverb of wisdom, or even a pithy maxim, which aims to distill in a short space a lot of meaning or significance.

\section{A Brief Introduction to Abstract Differential Geometry}      

{\em Abstract Differential Geometry} (ADG), {\it alias}, the differential geometry of vector sheaves, has been with us for more than two and a half decades now, ever since its original inception and formulation by Anastasios Mallios in the two-volume research monograph \cite{mall1} and its, also two-volume, sequel containing further developments of the mathematical theory of ADG and its physical applications to abelian and non-abelian gauge theories of matter, including gravity \cite{mall4}.

In a nutshell, ADG abstracts and generalises in a simple and straightforward {\em axiomatic} fashion \cite{mall2} the Classical Differential Geometry (CDG) on smooth manifolds---the usual differential calculus on differential ({\it alias}, $\smooth$-smooth) manifolds---by showing that the essential differential mechanism of CDG, which revolves about the fundamental notion of a generalised differential operator $\mathbf{\partial}$, commonly known as the {\em connection} $\conn$, is not inextricably tied to or vitally dependent on a background (:base) geometrical locally Euclidean space---{\it i.e.}, a $\smooth$-smooth manifold $M$.

Instead, Mallios originally intuited and recognised early in the development of ADG that, since a manifold $M$, as a point set endowed with a differential structure, is effectively nothing but the algebra $\smooth(M)$ of smooth coordinate functions on its points (relative to an atlas of coordinate charts covering $M$), one can readily abstract and generalise CDG to ADG simply by assuming an abstract {\em structure sheaf} $\sts$ of (possibly non-functional) abelian algebras, on an in principle arbitrary (possibly non-point set) base topological space $X$,\footnote{See quote (Q1) opening this paper. The reader should note the two points made in the two brackets in this sentence: (i) $\sts$ may be a structure sheaf of non-functional algebras; and, (ii) the base topological space $X$, on which $\sts$ is localised, may be `pointless', like for instance a Grothendieck site \cite{macmo} or the finitary poset discretisations of continuous (:topological) $\cont$-manifolds {\it \`a la} Sorkin \cite{sork0,rap1,rap2,rapzap1,rapzap2}. We shall return to these two important points in the sequel, when we discuss various applications of ADG-gravity to QG research.} as long as $\sts$ also admits a linear and Leibnizian differential operator $\partial$ to act on its local sections, which is the archetype of a {\em flat $\sts$-connection}.

Mallios then defines {\em vector sheaves} $\modl$ to be locally free $\sts$-modules of finite rank and shows that, like in the case of the structure sheaf $\sts$ above, they too are carrier  (:representation) spaces of a generalised linear and Leibnizian differential operator $\conn$, commonly known as a {\em curved $\sts$-connection}.\footnote{Thus, the said vector sheaves $\modl$ are abstract and generalised {\em differential} $\sts$-modules of finite rank.}

Then, essentially based on the pair $(\sts ,\partial)$ and, {\it in extenso}, on its generalised curved counterpart $(\modl ,\conn)$---with the latter pair, as we shall see in the sequel, defining a so-called {\em $\sts$-connection field} in ADG---Mallios carefully defines and constructs all the fundamental notions and structures of (pseudo-)Riemannian CDG, such as the curvature $\curv(\conn)$ of an $\sts$-connection $\conn$ and a (pseudo-Riemannian) $\sts$-metric on $\modl$ compatible with the connection $\conn$. Then, based on these fundamental structures, he derives the vacuum Einstein equations for gravity holding on $\modl$ from a variational principle applied on an ADG-theoretic analogue of the usual Einstein-Hilbert action functional.\footnote{Mallios also explores the ADG-theoretic structure and dynamics of abelian and non-abelian gauge theories of matter, by deriving the vacuum Maxwell's equations and the free Yang-Mills equations \cite{mall4}, but in the present paper we will focus exclusively on {\em vacuum Einstein gravity}. The reader, however, should note that almost all our constructions and results presented in this paper, as well as their applications to QG research, carry {\it mutatis mutandis} to the case of free Yang-Mills theories, including the abelian Maxwellian vacuum electrodynamics \cite{mall4}.} 

Here, three fundamental characteristic features of ADG-gravity must be emphasised and highlighted upfront, and they will be further analysed and discussed in the sequel, namely that: 

\begin{enumerate}

\item The vacuum Einstein ADG-gravitational equations mentioned above, unlike those of the usual CDG-Riemannian geometry based GR, are formulated {\em purely gauge theoretically}---{\it i.e.}, solely in terms of the $\sts$-connection field $(\modl ,\conn)$;  

\item The vacuum Einstein ADG-gravitational equations are formulated {\em purely homological algebraically} (:categorically) as equations between sheaf morphisms such as the connection $\conn$ and its curvature $\curv(\conn)$, which are {\em functorial, natural transformation} type of morphisms between the relevant sheaf categories involved; but perhaps more importantly, 

\item The vacuum Einstein ADG-gravitational equations, unlike those of the usual CDG-Riemannian geometry based GR, are formulated in the {\em manifest absence of a base differential} (:$\smooth$-smooth) {\em manifold}, which is a fixed, locally Euclidean background geometrical point set, traditionally representing the curved spacetime continuum in GR.

\end{enumerate}

\noindent All three basic features of ADG-gravity outlined above have proven to be invaluable in applying ADG to various currently important, both conceptual (:qualitative physico-philosophical) and technical (:quantitative mathematical, calculational and structural ), issues in QG research over the last quarter of a century \cite{malrap1,malrap2,malrap3,rap1,rap2,rap5,rap7,rap11,rap13,rap14,rap15,rap19}, as we will selectively recall and argue for in the sequel.

In the following subsection, we will briefly introduce, define and discuss the five fundamental mathematical concepts and their associated structures on which the entire ADG-theoretic edifice rests and is built upon, and then, based on them, we will recount how these are systematically organised and synergistically come together in deriving the dynamical vacuum Einstein equations of ADG-gravity from a variational Lagrangian action principle. In the process, we will witness and highlight that:

\vskip 0.1in

\begin{quotation}
\noindent {{\em At the basis of it all lies the structure sheaf $\sts$}, which is the cornerstone structure {\em on which} the whole ADG is founded as a theory of {\em differential} geometry proper and, as a consequence, {\em from which} ADG-gravity derives as a physical application of the underlying mathematical theory; hence the title of this paper.} 
\end{quotation}

\vskip 0.1in

\noindent But before we delve head-on into the paper, and for the reader's convenience, we give a 3-paragraph, concise, yet detailed, summary-{\it cum}-account of the route of argument that we follow in this paper, together with a few key references in the published literature to back its claims.

\subsubsection{A Detailed Summary of the Route of the Argument Taken to Support the Title of the Paper}

The main route of argument that we follow in the present paper is that, from the perspective of Abtract Differential Geometry (ADG), {\em all Differential Geometry essentially boils down to, or originates from, the structure sheaf $\sts$ of abelian algebras of generalised arithmetics/coefficients/coordinates assumed and employed}. In particular, the fundamental structure and concept on which every `proper' theory of {\em differential} geometry rests---namely, that of a {\em connection} $\conn$ ({\it viz.}, generalised differential operator $\kd$)---essentially derives from $\sts$ \cite{mall1,mall4}. In ADG-gravity that is based on ADG, the gravitational field is identified with an algebraic $\sts$-connection field $\conn$ acting categorically as a sheaf morphism on (the local sections of) a vector sheaf $\modl$,\footnote{With the vacuum Einstein gravitational field being formally defined in ADG-gravity as the pair $\mathcal{F}_{Einst}:=(\modl ,\conn)$, as we will see shortly.} without any  dependence whatsoever on a background spacetime manifold for the geometrical interpretation of the theory; hence ADG-gravity is regarded as a purely algebraically\footnote{Homological algebraically ({\it i.e.}, sheaf and category-theoretically).} formulated gauge field theory, with no {\it a priori} background smooth geometrical spacetime continuum dependences or commitments, and with certain key algebraic quantum traits inherent in the structure and dynamics of the ADG-gravitational $\sts$-connection field $\conn$.\footnote{For instance, from a geometric prequantisation \cite{mall5} and second quantisation \cite{mall6} perspective, the local sections of $\modl$, which is locally isomorphic to $\sts^{n}$, represent local quantum particle (:`graviton') states of the ADG-gravitational field, while this author has proposed an extended purely algebraic, canonical type of sheaf cohomological {\em third quantisation} scenario for both vacuum Einstein gravity and free Yang-Mills theories along ADG-theoretic lines, which is also manifestly background geometrical spacetime manifold independent and inherently finitistic \cite{rap13,rap15}.} In the sequel, we explicate and explain all this in more detail.

From the point above, it follows that different choices and uses of structure sheaf $\sts$ of generalised coordinates in the theory result in different aspects, issues and problems of classical and quantum gravity that can be addressed, tackled and potentially get resolved by ADG-theoretic means.\footnote{The formulation of an inherently finitary (:locally finite), causal and quantal version of vacuum Einstein gravity and free Yang-Mills theories \cite{malrap1,malrap2,malrap3} which can resolve and evade altogether such stubborn and classically unmanageable from the smooth spacetime manifold based perspective of the CDG-based GR problems like the problem of the inner Schwartzschild singularity of a point-mass and its associated pathological field infinities \cite{mall3,mall7,mall9,malrap1,malrap2,malrap3,rap5,rap7,rap11}, as well as the formulation of vacuum Einstein gravity and free Yang-Mills theories on non-linear generalised functions' (:disctributions) spaces that are everywhere dense with singularities that the manifold and CDG-based GR simply cannot deal with \cite{malros1,malros2,malros3}.} It also means that the {\em categorical transformation theory} of $\sts$---{\it i.e.}, changing $\sts$ functorially---is tantamount to a homological algebraic generalisation of the PGC of GR, here coined as the {\em Principle of Algebraic Relativity of Differentiability} (PARD), which is formulated in terms of natural transformation type of mappings between the relevant sheaf categories involved \cite{rap14,rap15,rap19}. This is the mathematical essence of {\em a generalised version of Einstein's Principle of Relativity} \cite{einst3,einst4} as well, here coined the {\em Principle of $\sts$-Relativity} ($\AR$). In the ADG-gravitational vacuum Einstein dynamics, $\AR$ manifests itself as the $\aut_{\sts}\modl$-invariance of the ADG-theoretic version of the Einstein-Hilbert action functional and its associated $\aut_{\sts}\modl$-covariance of the vacuum Einstein equations that are derived from it by a variational action principle (on the affine space of $\sts$-connections). Our principle of  $\AR$ is equivalent to what Mallios, the original founder of ADG, called {\em $\sts$-Invariance} ($\AI$), which was subsequently seen to be equivalent to the existence of a categorical $\sts$-adjunction between the pair of (adjoint) functors $(\mathrm{Hom}_{\sts},\otimes_{\sts})$  \cite{mall14,mall15,mall13,malzaf1,zaf2,zaf3,zaf4}. In turn, the said pair of adjoint functors has recently been seen to define a {\em geometric morphism} $\geom:=(\mathrm{Hom}_{\sts},\otimes_{\sts})$  between the sheaf categories of ADG-gravitational connection fields and the category of ADG-gravitational curvature spaces, a morphism that ultimately leaves invariant the ADG-gravitational vacuum Einstein equations holding in the said curvature spaces \cite{rap14,rap15,rap19}. Again, in the paper that follows, we explicate and explain all of the above in great detail.

\vskip 0.1in

All in all, in this paper:

\begin{itemize}

\item We will explicitly show and argue structurally-mathematically starting from the very foundations of ADG which are rooted in $\sts$;

\item We will depict categorically-diagrammatically in various different, but equivalent, ways; and,

\item We will interpret physically the relevant mathematical structures involved, so as to justify the very title of this paper; that is to say:

\item We explain the sense in which $\sts$-Relativity---the natural transformation theory of $\sts$ supporting the PARD and the Principles of $\AR$ and $\AI$ of the ADG-gravitational vacuum Einstein dynamics---is a generalised, purely (homological) algebraic, genuinely background geometrical spacetime point set manifold independent, local relativistic gauge field theory of gravity (:ADG-gravity) that has its roots in and fundamentally derives, like the mathematical theory of ADG on which it is based, from a structure sheaf $\sts$ of commutative algebras of generalised arithmetics/coefficients/coordinates, together with some inherently built-in quantum traits, that can potentially address, tackle, resolve, or ultimately evade, several important issues/problems in current and future classical and quantum gravity research.

\end{itemize}

\subsection{Kinematical Structures I: A Synopsis of the Structural Aufbau of ADG and ADG-Gravity Starting from $\sts$}

As essential preparatory ground for showing that ADG-gravity is structurally based on, originates and essentially derives from the structure sheaf $\sts$ as the title of the present paper contends, below we give a sequence of five fundamental structures on which the entire ADG-theoretic edifice is founded and progressively built from the bottom up. As the title of this subsection indicates, one can regard these fundamental ADG-theoretic concepts and associated structures as the {\em basic kinematical scaffolding} on which the vacuum Einstein ADG-gravitational dynamics is based and derives from, as it will be explicitly shown subsequently.

\subsection{4+1 Fundamental Structural Pillars in the Progressive Bottom-Up Aufbau of ADG} 

The quintet of fundamental conceptual and technical-structural pillars on which the entire ADG theory is founded, progressively built upon and sequentially erected from the ground up, unfolds as follows:\footnote{For the sake of economy of exposition, for more technical details about any term involved in the presentation of the five fundamental structures underlying ADG and ADG-gravity below, the reader is referred to the original sources \cite{mall1,mall2,mall4}, or to this author's joint papers with Mallios \cite{malrap1,malrap2,malrap3,malrap4}.}

\vskip 0.1in

\noindent \underline{\bf Fundamental Structure 1.} At the very basis of ADG, we have the notion of a {\em
$\cons$-algebraised space} $\cas$, which consists of the following pair:

\begin{equation}\label{eq0}
\cas :=(X,\sts_{X})
\end{equation}

\noindent where $X$ is an in principle 
arbitrary and general base topological space,\footnote{For sheaf cohomological reasons, when $X$ is usually assumed to be a point set, the two mild technical conditions or requirements that ADG imposes on $X$ is that it is {\em paracompact} and {\em Hausdorff} \cite{mall1,mall2,mall4}. When $X$ is a more general pointless topological space, like a Grothendieck site or a locale \cite{macmo}, there are categorical conditions equivalent to paracompactness and Hausdorff points' separation that can be imposed on $X$ to ensure local-to-global glueing constructions that are essential for certain differential geometric sheaf cohomology results (like for example, de Rham's Theorem and Poincar\'{e}'s Lemma) to hold on $X$ \cite{mall1,mall2,mall4}.} and $\sts_{X}$ is assumed to be a sheaf of (real or complex) unital, {\em abelian} and associative $\cons$-algebras\footnote{With $\cons$ being the constant sheaf of the real $\R$ or the complex $\com$ scalars, which is thus naturally injected as a subsheaf into $\sts$: $\cons\hookrightarrow\sts$.} localised over $X$, called {\em the structure sheaf of generalised arithmetics or (sheaf cohomological) coefficients in the mathematical theory}.\footnote{The only mild technical constraint that ADG imposes on $\sts$ is that it is {\em flabby} \cite{mall1,mall2,mall4}, but we will not go into that technicality here.} 

\vskip 0.1in

\noindent $\bullet$\underline{\bf Important Note 0:} In physical applications of ADG to gravity and gauge theories \cite{mall1,malrap1,malrap2,malrap3,mall4}, {\em $\sts$ is physically interpreted as the sheaf of abelian algebras of generalised local coordinate measurements or determinations of the ADG-gravitational field}. In other words, and as we will see in more detail in the sequel, {\em $\sts$ physically represents  the sheaf of gauge localisations of the ADG-gravitational field} relative to a system $\gauge_{i}$ of local open gauges covering $X$.\footnote{See below for more technical details.} This is much in the same way that in the $\smooth$-smooth spacetime manifold $M$ based GR, the abelian algebra $\smooth(M)$ of smooth functions on $M$, which is the smooth manifold's natural structure sheaf $\sts\equiv\smooth_{M}$, physically represents the smooth coordinates of $M$'s spacetime point events (relative to a given atlas of smooth local coordinate charts or patches covering $M$) and, {\it in extenso}, the ten smooth components of the locally Lorentzian spacetime metric $g_{\mu\nu}$ (which physically represent the smooth gravitational potentials in GR) relative to a given locally Euclidean-Lorentzian coordinate frame prescribed on $M$.

\vskip 0.1in

\noindent \underline{\bf Fundamental Structure 2.} Once we have defined $\cas$ as above, the second fundamental ADG-theoretic structure is that of a {\em differential triad} $\triad$:

\begin{equation}\label{eq-3}
\triad=(\sts_{X} ,\partial ,\Omg_{X})
\end{equation}

\noindent which consists of a $\cons$-algebraised space $\cas$ as in (\ref{eq0}) above, a sheaf $\Omg_{X}$ of (differential) $\sts$-modules $\omg$
over $X$, and a $\cons$-linear derivation map (:differential operator) $\partial$ defined as the following {\em
sheaf morphism}: 

\begin{equation}\label{eq1}
\partial :~\sts\mapto\Omg
\end{equation}

\noindent which is a $\cons$-linear map additionally satisfying the following {\em Leibniz rule}:

\begin{equation}\label{eq2}
\partial(s\cdot t)=s\cdot\partial(t)+t\cdot\partial(s)
\end{equation}

\noindent for any local sections $s$ and $t$ of $\sts$ ({\it
i.e.}, $s,t\in\Gamma(U,\sts)\equiv\sts(U)$, with $U\subset X$ an
open set in a {\em system of local open gauges} $\gauge_{i}$ of $X$).\footnote{In ADG,  an open covering $\gauge_{i}=\{ U:\, U\subset X, \, \mathrm{open}\}$ of $X$ is called a {\em system of local open gauges} \cite{mall1,mall2,mall4,malrap1,malrap2,malrap3}. It is a collection of local open sets covering $X$ relative to which the generalised coordinate algebras in the structure sheaf $\sts$ are {\em sheaf theoretically localised} or {\em `gauged'}. From an ADG-theoretic standpoint, {\em localising}, or (the acts of) {\em locally measuring/locally gauging} the generalised coordinates in $\sts$ over $X$ relative to an open cover $\gauge_{i}$ of $X$, are regarded as synonymous concepts, hence our denomination of the structure sheaf $\sts$ in Important Note 0 earlier as {\em the sheaf of generalised local coordinate measurements or gauge localisations} (of the vacuum Einstein ADG-gravitational field). The importance of this point will be further highlighted in the sequel when we emphasise {\em the purely algebraic and purely gauge theoretic character of ADG-gravity}.}

\vskip 0.1in

\noindent \underline{\bf Fundamental Structure 3.} The third fundamental notion in the sequential {\it aufbau}\footnote{German for {\em building up} or {\em progressive construction from the ground up}.} of ADG, one that derives naturally once one has defined a differential triad as in (\ref{eq-3}) above, is that of an {\em $\sts$-connection} $\conn$ ({\it viz.} the generalised or `curved' version of the `flat' differential operator $\partial$ above).\footnote{The epithets `flat' and `curved' for $\partial$ and $\conn$ respectively, will become transparent shortly.} 

Thus, given a differential triad $\triad=(\sts,\partial
,\Omg)$, we let $\modl$ be a sheaf of (abstract) differential $\sts$-modules on $X$ \cite{mall1,mall2,mall4,malrap1,malrap2,malrap3}.

$\bullet$ \underline{\bf Note:} Parenthetically here, the reader should note the general formal definition of a {\em vector sheaf} $\modl$ in ADG: {\em a vector sheaf is a locally free $\sts$-module of finite rank $n\in\N$}. That is, for every $U\subset X$, $\modl_{U}\equiv\modl(U)\equiv\Gamma(U,\modl)\simeq\sts^{n}(U)$, with $\sts^{n}$ standing for the
$n$-fold Whitney sum of $n$ copies of $\sts$: $\modl=\bigoplus_{i=1}^{n}\sts_{i}$ . Thus, $\modl$ is {\em a locally free (differential) $\sts$-module of finite rank (or dimension) $n$}, an appellation synonymous to {\em vector sheaf} in ADG
\cite{mall1,mall2,mall4,malrap1,malrap2,malrap3}.\footnote{From the definition of a vector sheaf $\modl$ above, it follows that every
continuous local section $s$ of $\modl$ ($s\in\modl(U)$) can be
expressed or decomposed as a linear combination $\sum_{i=1}^{n}s_{i}e_{i}$
with coefficients $s_{i}$ in $\sts(U)$ relative to the local open gauge $U\in\gauge_{i}$ in $X$. In turn, the $n$-tuple $E(U)=\{ e_{i}(U)\}_{i=1}^{n} $ ($e_{i}(U)\in\Gamma(U,\modl)\equiv\modl(U)$) represents a local coordinate frame of the vector sheaf $\modl$ relative to the local open gauge $U$ in a system of local open gauges $\gauge_{i}$ covering $X$, as defined above.}

Then, the definition of an $\sts$-connection $\conn$ goes as follows:

\begin{equation}\label{eq3}
\conn:~\modl\mapto\modl\otimes_{\sts}\Omg\cong
\Omg\otimes_{\sts}\modl\equiv\Omg(\modl)
\end{equation}

\noindent which means that {\em $\conn$ is, categorically speaking, a $\cons$-linear sheaf morphism of the (real or complex) vector
sheaves involved},\footnote{Note in (\ref{eq3}) above that $\Omg$ is the {\em dual vector sheaf of abstract $\sts$-linear differential forms} on the vector sheaf $\modl$. That is: $\Omg\ni\omega:\, \modl\ni f\rightarrow \alpha\ni\sts;\, (\omega(f)=\alpha)$, so that: $\Omg\simeq\sts\otimes_{\sts}\modl$. Moreover, this sheaf of abstract (exterior) differential forms $\Omg$ is assumed to be graded: $\Omg=\bigoplus_{i}\Omg^{i}$, with higher order/grade extensions of the differential $\partial\equiv d$ in (\ref{eq1}) above effectuating sheaf morphisms: $d^{i}:\,\Omg^{i}\rightarrow\Omg^{i+1}$ thus defining an abstract version of the usual K\"ahler-de Rham sheaf cohomology on the corresponding spaces \cite{mall1,mall2,mall4,malrap1,malrap2,malrap3}. In fact, and crucially for sheaf cohomological reasons as we will see in the sequel, the fundamental differential geometric result known as {\em Poincar\'e's Lemma} is seen to hold on the vector sheaves involved in ADG-gravity  \cite{mall1,mall2,mall4,malrap1,malrap2,malrap3,malros1,malros2,malros3}.} 

The second {\em Leibniz product rule} type of condition that $\conn$ is assumed to satisfy is the following:

\begin{equation}\label{eq4}
\conn(\alpha \cdot s)=\alpha\cdot\conn(s)+s\otimes\partial(\alpha)
\end{equation}

\noindent for $\alpha\in\sts(U)$,
$s\in\modl(U)\equiv\Gamma(U,\modl)$, and $U$ an open gauge (:subset) of $X$, as usual.

One can then proceed a step further to find {\em the local form of $\conn$} and realise that it splits into the usual {\em flat connection} ({\it i.e.}, the differential $d\equiv\partial$) and the local connection $1$-forms:

\begin{equation}\label{eq5}
\conn=\partial +\aconn
\end{equation}

\noindent with $\aconn$ transforming affinely (:inhomogeneously or non-tensorially) like the usual local gauge connection potentials (:1-forms) under local gauge transformations \cite{mall1,mall2,mall4,malrap1,malrap2,malrap3}.\footnote{Parenthetically here, the reader should note that in ADG a {\em a local change of gauge} (:local gauge transformation) corresponds to a map $g_{ij}:\,U_{i}\rightarrow U_{j}$ between open sets (:local gauges) $U_{i}$ and $U_{j}$ in a local open covering (:gauge system) $\gauge$ of $X$, which plainly lifts to a change in the corresponding algebra sheaf of generalised arithmetics or coordinates on $X$: $\tilde{g}_{ij}:\,\sts(U_{i})\rightarrow \sts(U_{j})$. Without loss of generality, we identify $g_{ij}$ with $\tilde{g}_{ij}$ when we speak of {\em a local gauge transformation in ADG}.}

In turn, having defined an $\sts$-connection $\conn$ as in (\ref{eq3}) and (\ref{eq4}) above, one can then proceed to define categorically its associated {\em curvature morphism} $\curv(\conn)$ as the following \emph{$\sts$-morphism of $\sts$-modules}.\footnote{With a vector sheaf $\modl$, as explained before, regarded as a \emph{sheaf of differential $\sts$-modules, with structure sheaf $\sts$, that is locally isomorphic to $\sts^{n}(U)$}.}

\noindent To that end, we first define the {\em 1st prolongation of $\conn$} to be the following
$\cons$-linear vector sheaf morphism:

\begin{equation}\label{eq15}
\conn^{1}:~\Omg^{1}(\modl)\mapto\Omg^{2}(\modl)
\end{equation}

\noindent satisfying section-wise relative to $\conn$: 

\begin{equation}\label{eq16}
\conn^{1}(s\otimes t):=s\otimes\kd t-t\wee\conn
s,~(s\in\modl(U),t\in\Omg^{1}(U),U~\mathrm{open~in}~X)
\end{equation}

We are now in a position to define the curvature $\curv$ of an
$\sts$-connection $\conn$ by the following triangular commutative diagram:

\begin{equation}\label{eq17}
\xymatrix{
\modl \ar[d]|{\curv(\conn)=\conn^{1}\circ\conn} \ar[r]^{\conn}
& \Omg^{1}(\modl)\equiv\modl\otimes_{\sts}\Omg^{1} \ar[dl]^{\conn^{1}} \\
\Omg^{2}(\modl)\equiv\modl\otimes_{\sts} \Omg^{2} }
\end{equation}

\noindent from which we read directly that:

\begin{equation}\label{eq18}
\curv\equiv \curv(\conn):=\conn^{1}\circ\conn
\end{equation}

\noindent Therefore, any time we have the $\cons$-linear morphism
$\conn$ and its prolongation $\conn^{1}$ at our disposal, we can
define the curvature $\curv(\conn)$ of the connection
$\conn$.\footnote{In connection with (\ref{eq18}), one can justify
our earlier remark that the standard Cartan-K\"{a}hler (exterior) differential operator $\partial\equiv d\equiv d^{0}$ is a {\em flat} type of connection, 
since: $\curv(d)=d\circ d\equiv d^{2}=0$, which corresponds to the well known \emph{nilpotency} of the
usual Cartan-K\"{a}hler (exterior) differential operator $d$ \cite{gosch, mall1, mall2, mall4, malrap1, malrap2, malrap3}). In a (co)homological-algebraic sense, 
\emph{the curvature $\curv(\conn)$ of an algebraic $\sts$-connection $\conn$ measures the `obstruction' to, or the `deviation' from, nilpotency of the connection $\conn$} \cite{malrap1, malrap2, malrap3}.}

As a matter of fact, it is rather straightforward to see that, for
$\modl$ a vector sheaf, $\curv(\conn)$ is defined as {\em an $\sts$-functor of
$\sts$-modules}, in the following sense:

\begin{equation}\label{eq19}
\begin{array}{c}
\curv\in{\mathrm{Hom}}_{\sts}(\modl
,\Omg^{2}(\modl))=\Hom_{\sts}(\modl,\Omg^{2}(\modl))(X)\cr
\Omg^{2}({\modl}nd\modl)(X)=Z^{0}(\gauge,\Omg^{2}({\modl}nd\modl))
\end{array}
\end{equation}

\noindent where, as usual, $\gauge_{i}=\{ U_{\alpha}\}_{\alpha\in I}$ is an open
cover of the base topological space $X$ and $Z^{0}(\gauge,\Omg^{2}(\mathcal{E}nd\modl))$ the
$\sts(U)$-module of $0$-{\em cocycles} of
$\Omg^{2}(\modl nd\modl)$ relative to the
$\gauge_{i}$-covering of $X$.\footnote{One may wish to recall again
that, for a vector sheaf $\modl$ like the one involved in
(\ref{eq19}) above: ${\modl}nd\modl\equiv{\mathcal{H}}om_{\sts}(\modl
,\modl)\cong\modl\otimes_{\sts}\modl^{*}=\modl^{*}\otimes_{\sts}\modl$.}

\vskip 0.1in

\noindent \underline{\bf Fundamental Structure 4.} Having defined a vector sheaf $\modl$ and a connection $\conn$ on it, we arrive at the last fundamental ADG-theoretic notion, that of an {\em ADG-field}---{\it alias}, {\em an ADG $\sts$-connection field}. This is defined to be a pair:

\begin{equation}\label{eq6}
\mathcal{F}:=(\modl ,\conn)
\end{equation}

\noindent consisting of a vector sheaf $\modl$ and a $\cons$-linear Leibnizian $\sts$-connection $\conn$ categorically acting on its local sections as a sheaf morphism as in (\ref{eq3}).

\vskip 0.1in

In turn, associated with every ADG-field $\mathcal{F}$ as above, there is a corresponding ADG-theoretic {\em curvature space}, which is defined as the following triplet: 

\begin{equation}\label{eq7}
\mathcal{S}:=(\sts_{X},\conn ,\curv(\conn))
\end{equation}

\noindent which is similar to the definition of a differential triad $\triad$ in (\ref{eq-3}) above, but instead of involving the flat $\sts$-connection $\partial$ and the differential $\sts$-module $\Omega$, it involves the curved $\sts$-connection $\conn$ and its curvature form $\curv(\conn)$.

\vskip 0.1in

\noindent  $\bullet$\underline{\bf Important Note 1:} The reader should note that in the ADG-theoretic $\sts$-connection field structure $\mathcal{F}:=(\modl ,\conn)$ in (\ref{eq6}) above, $\modl$ may be regarded as the {\em associated vector sheaf} (or representation sheaf) of the {\em principal group sheaf} $\aut_{\sts}\modl$ of its own local $\sts$-automorphisms \cite{vas1,vas2,vas3,vas4,mall4,malrap3}. Since, as we saw earlier, $\modl(U)\simeq_{\mathrm loc.}\sts^{n}(U)$, the principal group sheaf  $\aut_{\sts}\modl\equiv\modl nd\modl^{\bullet}$ is locally isomorphic to the group sheaf $M_{n}^{\bullet}(\sts(U))$ of invertible $(n\times n)$-matrices with local entries in $\sts(U)\equiv\Gamma(U,\sts)$, the space of local sections of $\sts$ relative to the local open gauge $U\subset X$ \cite{mall1,malrap1,malrap2,malrap3,mall4}.\footnote{The bullet $\bullet$-superscript notation in $M_{n}^{\bullet}$ means {\em invertible}, while $\modl nd\modl$ indicates the $Hom$-set ${\modl}nd\modl\equiv{\mathcal{H}}om_{\sts}(\modl ,\modl)$ of {\em $\sts$-endomorphisms of $\modl$}. Clearly, $\modl nd \modl(U)\simeq_{\mathrm loc.}M_{n}(\sts(U))$; hence, $\aut_{\sts}\modl\equiv\modl nd\modl^{\bullet}\simeq_{\mathrm loc.}M_{n}^{\bullet}(\sts(U))$, as noted above.} We will return to this important salient point when we discuss the {\em purely gauge theoretic character of vacuum Einstein ADG-gravity} below.

\vskip 0.1in

\noindent \underline{\bf `Bonus' Auxiliary Structure 5.} As the title of this paragraph entails, there is a secondary, `bonus' auxiliary fifth structure in the progressive {\it aufbau} of ADG, which augments and completes the tetrad of fundamental structures outlined above. 

To define it, we let $\modl$ be a vector sheaf as above. By an {\em $\sts$-valued
pseudo-Riemannian inner product (metric) $\rho$ on $\modl$} (over $X$), we
mean the following {\em sheaf morphism}:

\begin{equation}\label{eq101}
\rho :~\modl\oplus\modl\mapto\sts 
\end{equation}

\noindent which is:

\noindent $\bullet$ i) {\em $\sts$-bilinear} between the
$\sts$-modules involved; 

\noindent $\bullet$ ii) symmetric ({\it i.e.},
$\rho(s,t)=\rho(t,s),~s,t\in\modl(U)$) and of indefinite
signature; as well as, 

\noindent $\bullet$ iii) {\em strongly non-degenerate}. 

\vskip 0.1in

\noindent That is to say,
we assume that $\rho(s,t)$, for any two local sections $s$ and $t$
in $\modl(U)$, is given via the
canonical isomorphism:

\begin{equation}\label{eq102}
\modl\stackrel{\tilde{\rho}}{\cong}\modl^{*}
\end{equation}

\noindent between $\modl$ and its dual $\modl^{*}$, as follows:

\begin{equation}\label{eq103}
\tilde{\rho}(s)(t):=\rho(s,t)
\end{equation}

\noindent with (\ref{eq102}) being true up to an
$\sts$-isomorphism\footnote{The epithet `strongly' to
`non-degenerate' above indicates that $\tilde{\rho}$ in
(\ref{eq102}) is also an {\em onto} map.}.

We further assume that for a vector sheaf $\modl$ (of finite
rank $n\in\N$) endowed with an $\sts$-connection $\conn$, the
vector sheaf $\Omg$ in the given differential triad
$\triad=(\struc ,\partial, \Omg)$ is the dual of $\modl$ appearing
in (\ref{eq102}) ({\it i.e.},
$\Omg=\modl^{*}\equiv{\Hom}_{\sts}(\modl ,\sts)$). 

Thus, in line with the usual Christoffel theory \cite{mall1,mall2,mall4}, we can
define a {\em linear connection} $\nabla$, as follows:

\begin{equation}\label{eq104}
\nabla :~\modl\times\modl\mapto\modl
\end{equation}

\noindent acting section-wise on $\modl(U)$ in the following way:

\begin{equation}\label{eq15}
\nabla(s,t)\equiv\nabla_{s}(t):=\conn(t)(s)
\end{equation}

Now, one says that {\em $\conn$ is a pseudo-Riemannian
$\sts$-connection} or that it is compatible with the indefinite
metric $g$ of the inner product $\rho$ in (\ref{eq101}), whenever
it fulfills the following two conditions:

\begin{itemize}

\item {\em Riemannian symmetry}: $\nabla(s,t)-\nabla(t,s)=[s,t]$;
for $s,t\in\modl(U)$ and $[\, .\, ,\, .\,  ]$ the usual Lie
bracket (product).

\item {\em Ricci identity}:
$\partial(\rho(s,t))(u)=\rho(\nabla(u,s),t)+\rho(s,\nabla(u,t))$;
for $s,t,u\in\modl(U)\simeq\sts^{n}(U)$, as usual.

\end{itemize}

In particular, for a pseudo-Riemannian (:Lorentzian) $\rho$ and its associated
$g$,\footnote{With respect to a {\em local (coordinate) gauge}
$e^{U}\equiv\{ U;~(e_{\mu})_{0\leq \mu\leq n-1}\}$ of the vector sheaf
$\modl$ of rank $n$,\footnote{For gravity (GR), we take $n=4$.}
$\rho(e_{\mu},e_{\nu})=g_{\mu\nu}=\mathrm{diag}(-1,+1,\cdots)$
\cite{mall1,mall2,malrap1,malrap2,malrap3,mall4}.} an $\sts$-connection $\conn$ is said to be
compatible with the Lorentzian $\sts$-inner product $\rho$ on
$\modl$\footnote{Such a metric connection is commonly known as
{\em Levi-Civita connection}.} when its associated Christoffel
$\nabla$ in (\ref{eq104}) satisfies:

\begin{equation}\label{eq106}
\nabla\rho=0
\end{equation}

\noindent which, in turn, is equivalent to the following `{\em
horizontality}' condition for the canonical isomorphism
$\tilde{\rho}$ in (\ref{eq102}) relative to the {\em connection}
$\conn_{\modl\otimes_{\sts}\modl^{*}}$ in the tensor product
vector sheaf $\Hom_{\sts}(\modl
,\modl^{*})=(\modl\otimes_{\sts}\modl)^{*}=\modl^{*}\otimes_{\sts}\modl^{*}$
induced by the $\sts$-connection $\conn$ on $\modl$:

\begin{equation}\label{eq107}
\conn_{\Hom_{\sts}(\modl ,\modl^{*})}(\tilde{\rho})=0
\end{equation}

\noindent It is worth reminding the reader who is familiar with
the usual theory that (\ref{eq107}) above implies that the
Levi-Civita $\sts$-connection $\conn$ that is compatible with the Lorentz
$\sts$-metric $\rho$ is {\em torsion-free} \cite{mall1,mall2,mall3,mall4,malrap1,malrap2,malrap3}.

\vskip 0.1in

\noindent $\bullet$\underline{\bf Important Note 2:} The reader should note that we called the fifth fundamental structure above---{\it i.e.}, the $\sts$-metric $\rho$ (:$g_{\mu\nu}$)---{\em auxiliary}, {\em secondary}, or {\em `bonus'}, because as we will see in the sequel, the sole dynamical variable in ADG-gravity is the Einstein $\sts$-connection field $\mathcal{F}_{Einst}=(\modl ,\conn)$ unlike in the standard background geometric Riemannian spacetime manifold based GR in which the smooth spacetime metric $g_{\mu\nu}$ is the only dynamical variable (2nd order formalism), with the ten independent components of the symmetric tensor $g_{\mu\nu}$ physically representing the gravitational potentials. We will return to this subtle point shortly, when we discuss the ADG-gravitational vacuum Einstein equations and the fundamental character of vacuum Einstein ADG-gravity as being {\em purely gauge} and its symmetries {\em purely internal to the ADG-gravitational field itself}, without any external (:background) geometrical spacetime manifold dependence or commitment.

\vskip 0.1in

\noindent  $\bullet$\underline{\bf Important Note 3:} With the four fundamental ADG-theoretic notions above, as well as with the optional fifth one coresponding to the $\sts$-metric $\rho$ and its compatibility with the $\sts$-connection $\conn$, one is able to reproduce {\em all} the basic concepts, structures, calculations and results thereof of the usual CDG---the standard Newtonian Differential Calculus (or Analysis) on Smooth (pseudo-)Riemannian Manifolds; albeit, entirely relationally (:homological-algebraically) and, more importantly {\it vis\`a-vis} ADG's applications to QG that we will discuss in the sequel, one is able to formulate ADG-gravity in the manifest absence of a smooth background geometrical spacetime manifold \cite{mall1,mall2,mall4,malrap1,malrap2,malrap3,rap14,rap5,rap7,rap13,rap15,rap19}.

\subsubsection{Structural Interregnum I: Derivative Derives from Algebra}

We close this subsection by making the following {\em key observation} on which the whole paper hinges and from which the whole paper effectively follows:

\vskip 0.1in

\noindent \underline{\bf Key Observation:} In ADG, {\em `differentiation'} and `differentiability'---{\it i.e.}, the usual (flat) differential operator $\partial$ in (\ref{eq1}) and its generalisation to the general (curved) connection $\conn$ in (\ref{eq3})---derives from and has as its source or {\em domain of definition} the structure algebra sheaf $\sts$ of generalised arithmetics, coefficients and theoretical local coordinate measurements, hence its ADG-theoretic denomination as an {\em $\sts$-connection}. In short, as the title of this subsection maintains:

\vskip 0.1in

\begin{quotation}

\noindent {{\em The derivative (:connection) derives from algebra} (pun intended); hence, it is an {\em $\sts$-derivative} or {\em $\sts$-derived connection}.\footnote{A structure that in the pure mathematics research literature on further developing ADG has been coined {\em Mallios's algebraic $\sts$-connection} \cite{vas1,vas2,vas3,vas4}. Thus, we could equivalently coin it here {\em Mallios's $\sts$-derived connection.}}}

\end{quotation}

\vskip 0.1in

We distil the Key Observation displayed above to the following pair consisting of an Aphorism and an Apophthegm:

\vskip 0.1in

\noindent\fbox{%
    \parbox{\textwidth}{%
\noindent\underline{\bf Aphorism 1:} One cannot do \underline{\em Differential} Geometry without a differential operator $\mathbf{d}$ like the flat $\partial$ in (\ref{eq1}) or its curved  generalisation, the algebraic $\sts$-connection $\conn$ in (\ref{eq3}). No derivative operator, no differentiation, no differential equations, no differential calculus or differential geometry. What qualifies ADG as a theory of {\em Differential} Geometry proper is the $\sts$-connection $\conn$, which is {\em the} fundamental concept and structural pillar in the {\it aufbau} of the mathematical theory of ADG \cite{mall1,mall2,mall4,rap19}.
}%
}

\vskip 0.2in

\noindent\fbox{%
    \parbox{\textwidth}{%
\noindent\underline{\bf Apophthegm 1:} ADG's central concept and structural backbone---the connection $\conn$ (or its flat counterpart $\partial$)---has its source and domain of definition in, thus it derives from, the structure algebra sheaf $\sts$, as (\ref{eq3}) for $\conn$, or (\ref{eq1}) for $\partial$, show; hence its ADG-theoretic denomination as {\em an algebraic $\sts$-connection} (cf. quote (Q1) opening this paper). Hence, in effect, {\em at the very basis and heart of ADG as a mathematical theory of {\em Differential} Geometry, as well as its application in formulating the law of vacuum Einstein ADG-gravity as a {\em differential} equation proper, lies the structure sheaf $\sts$ of generalised arithmetics, coefficients and theoretical local coordinate-measurements. Everything in ADG (and in ADG-gravity) originates and follows from, and in one way or another revolves about, the structure sheaf $\sts$}; hence the title of the present paper.
}%
}

\vskip 0.1in

\subsection{Philosophical Interregnum II: the Essentially Algebraico-Topological Character of the Notion of Connection}

In the early developmental stages of ADG, Mallios recognised that the archetypal example of the differential operator $\mathbf{d}$---{\it i.e.}, the standard (real or complex analytic) definition of the {\em derivative} in the spatial, locally Euclidean continuum (:manifold) setting of Newton's Differential Calculus (CDG):\footnote{For example, for simple real valued functions on the real  (flat) $1$-dimensional manifold $\R$---the real number line: $f:\; \R\rightarrow\R$, this is the usual $\delta\text{-}\epsilon$ definition of the derivative in standard (Real) Analysis.} 

\begin{equation}\label{eq100}
\frac{df}{dx}:=\lim_{x\rightarrow x_{0}}\left(\frac{f(x)-f(x_{0})}{x-x_{0}}\right)=\lim_{\Delta x\rightarrow 0}\left(\frac{\Delta f}{\Delta x}\right)
\end{equation}

\noindent essentially involves in its definition the following two important mathematical-structural characteristics:

\begin{enumerate}

\item \underline{\bf Algebraic Character-Structure}: as $\frac{\Delta f}{\Delta x}$ in (\ref{eq100}) above involves the algebraic (structure) operations of subtraction (:the inverse of addition) and division (:the inverse of multiplication) afforded by the number field $\R$; and,

\item \underline{\bf Topological Character-Structure}: as the notion of limit $\lim_{\Delta x\rightarrow 0}$ in (\ref{eq100}) above presumes (or requires) the presence of an underlying topology (:topological structure) being defined (or assumed) on the (base) Euclidean domain space ($\R$) of $f$.

\end{enumerate}

\noindent We distil the two fundamental structural traits of the derivative operator above to the following Aphorism, our second one:

\vskip 0.1in

\noindent\fbox{%
    \parbox{\textwidth}{%
\noindent\underline{\bf Aphorism 2:} The usual notion of derivative or (flat) differential operator $\mathbf{d}\equiv\partial$, or its generalisation to a (curved) connection operator $\conn$, is essentially of an {\em algebraico-topological} (or equivalently, of a {\em topologico-algebraic}) character.
}%
}

\vskip 0.1in 

\noindent which leads us to the following Apophthegm, our second one:

\vskip 0.1in

\noindent\fbox{%
    \parbox{\textwidth}{%
\noindent\underline{\bf Apophthegm 2:} In ADG, Mallios's algebraic $\sts$-connections $\conn$ derive from the structure sheaf $\sts$ of (abelian) algebras over an underlying (in principle arbitrary) topological space, as the structure of such a {\em sheaf $\sts$ of algebras} over an arbitrary base topological space $X$ possesses all the essential algebraico-topological structure that is necessary to define an abstract and generalised differential (:derivative) operator such as $\conn$ in the first place, as in the defining expression (\ref{eq3}) earlier. As such, {\em ADG is a purely relational (:algebraic) way of doing differential geometry that does not involve at all the mediation of a background locally Euclidean space (:a point set manifold $M$ that locally looks like $\R^{n}$) for defining and setting up differential geometric structures and doing differential geometric calculations (:Calculus---{\it e.g.}, setting up and solving differential equations) based on them}.
}%
}

\vskip 0.1in

\noindent \underline{\bf Important Note 7:} By contrast to Apophthegm 2 above, it must be emphasised here that in the usual Newtonian Differential Calculus and Riemannian Differential Geometry (CDG), the source or origin of the standard differential operator $\mathbf{d}$---the (flat) derivative $\partial$ in (\ref{eq1}) and/or the $\frac{d}{dx}$ operator in (\ref{eq100})---is the base locally Euclidean differential point set manifold $M$; or equivalently, the structure sheaf $\smooth_{M}$ of smooth coordinate functions of its points. Such a smooth background geometrical manifold $M$, however, is manifestly absent from ADG. To emphasise it once again, {\em ADG is a genuinely background geometrical manifold independent way of doing differential geometry} \cite{mall1,mall2,mall4,malrap1,malrap2,malrap3,rap14,rap15,rap19}.

\vskip 0.1in

\noindent \underline{\bf Addendum to Important Note 7:} In addition to the Important Note above, it must be highlighted, as a key historical type of remark on the development of ADG, that Mallios's original\footnote{{\it Circa} 1990, as (Q1) in the beginning of the paper recounts.} intution and inception of an algebraic $\sts$-connection $\conn$ as the key structure for developing ADG was made in the context of {\em topological algebra theory}---{\it i.e.}, when {\em the original structure sheaves $\sts$ were assumed to be sheaves of general (possibly non-normed) topological algebras} \cite{mall-7,mall-6,mall-5,mall-4,mall-3,mall-2,mall-1}. As such, they were {\em the} archetypal structure sheaves $\sts$, ones that combine the essential both {\em topological} and {\em algebraic} characters/structures noted in Aphorism 2 and Apophthegm 2 above. As noted above, those topologico-algebraic characters are necessary for defining and deriving the $\sts$-connections $\partial$ and $\conn$ as in (\ref{eq1}) and (\ref{eq3}), respectively. {\it A fortiori}, shortly after the publication of the first 2-volume monograph on ADG \cite{mall1}, Mallios continued to apply ADG on mathematical physics topics, from GR \cite{mall3, mall7} to QFT \cite{mall5, mall6}, by using (possibly non-functional or even generalised functional) topological algebra structure sheaves $\sts$ that carry the requisite ADG-theoretic $\sts$-connections in the manifest absence of a background geometrical smooth spacetime manifold \cite{mallX1}. 

\subsection{Kinematical Structures II: Three Important ADG-Theoretic Functor (Sheaf) Categories ($\catdt$, $\catf$, $\catcs$)}

After this brief, albeit important, digression about the algebraico-topological and background manifold-free character of the ADG-theoretic $\sts$-connections $\conn$, we note that, together with the five fundamental kinematical ADG-theoretic structures outlined above, and anticipating our musings in the next section about the vacuum Einstein ADG-gravitational equations, their dynamical functoriality \cite{rap15} and the categorically (:functorially) generalised version of the PGC of GR that we have coined PARD, in this subsection we discuss PARD's theoretical precursor---the fundamental {\em kinematical functoriality} underlying ADG-field theory.

We thus start this subsection by the following important technical remark: since by its very definition, {\em a sheaf is essentially a functor},\footnote{We briefly recall from \cite{bredon,macmo} the two equivalent definitions of a sheaf: (i) a sheaf is a {\em local homeomorphism}---which effectively reveals the local topological character of the notion of sheaf (to which we are going to return shortly); and, (ii) a sheaf is a {\em sheafified presheaf}---that is, {\em a contravariant functor} ({\it e.g.}, a presheaf of sets is a contravariant functor from the category of open subsets of a topological space, to the poset category of structureless sets, ordered by $\subseteq$-inclusion) {\em together with suitable compatibility, gluing or stitching up, conditions for the overlapping regions of the underlying covering open sets}.} the {\em three fundamental structural-kinematical ADG-theoretic functor categories supporting ADG-gravity} \cite{pap1,mall4,pap2,malrap3,rap11,rap7,rap15,rap19} are the following: 

\begin{enumerate}

\item {\em The Category of Differential Triads} $\catdt$: This category has for objects differential triads $\triad$, like those defined by equation (\ref{eq-3}) earlier, and differential triad morphisms as arrows between them \cite{pap1,pap2,pap3,pap4,pap5,fragpap}. Since the elements of differential triads are functorial objects (:sheaves) as explained above, $\catdt$ is essentially a {\em functor category}, thus the differential triad morphisms $\mathcal{N}_{\mathcal{T}}$ are {\em morphisms of functors}, {\it alias} {\em natural transformation type of maps} \cite{macmo}.

\item {\em The Category of ADG-Fields} $\catf$: Likewise for the category of the dynamical ADG-fields $\mathcal{F}$, like the ones defined by equation (\ref{eq6}) earlier.\footnote{Originally in \cite{mall4}, and subsequently in \cite{rap14,rap15}, the ADG-theoretic (abelian gauge) Maxwell fields, (non-abelian gauge) Yang-Mills fields and (non-abelian gravitational gauge) Einstein fields, are all seen to be organised into respective categories called {\em Maxwell} (abelian gauge fields), {\em Yang-Mills} (non-abelian gauge fields) and {\em Einstein category}, respectively.} Like $\catdt$ above, $\catf_{Einst}$ too has vacuum Einstein ADG-gravitational fields $\mathcal{F}_{Einst}$ as objects and natural transformation type of arrows $\mathcal{N}_{\mathcal{F}}$ between them as ADG-field morphisms.\footnote{Plainly, the natural transformation type of map $\mathcal{N}_{\mathcal{F}}$ preserves the connection sheaf morphisms $\conn$ of the respective ADG-gravitational fields as the latter are defined by (\ref{eq3}) above. We are going to witness such natural transformation type of maps in the sequel, when we discuss the Priciple of Algebraic Relativity of Differentiability (PARD), which is the ADG-theoretic generalisation of the Principle of General Covariance (PGC) of GR.}

\item {\em The Category of Curvature Spaces} $\catcs$: {\it Mutatis mutandis} for the category $\catcs$ of curvature spaces: its objects are curvature spaces $\mathcal{S}$ like the ones defined in (\ref{eq7}) earlier, while its arrows are functorial, natural transformation type of maps $\mathcal{N}_{\mathcal{S}}$ between them.\footnote{Like it was posited in our last footnote, the natural transformation type of map $\mathcal{N}_{\mathcal{S}}$ preserves the curvature $\sts$-morphisms $\curv$ of the corresponding ADG-curvature spaces as defined by equation (\ref{eq7})  earlier. As it was mentioned in the previous footnote, we are going to meet such functorial, natural transformation type of maps when we discuss the PARD below.} 

\end{enumerate} 

\vskip 0.1in

\noindent \underline{\bf Important Note 8:} At this point it must be emphasised that, since we have already seen how the structure sheaf $\sts$ of abelian algebras of generalised arithmetics, coefficients or theoretical generalised local coordinate determinations (measurements), by constituting what we earlier defined in (\ref{eq0}) as a $\cons$-algebraised space $\cas :=(X,\sts_{X})$, fundamentally underlies both the notion of differential triad $\triad=(\sts_{X} ,\partial ,\Omg_{X})$ {\em and} the notion of an ADG-field $\mathcal{F}=(\modl ,\conn)$ and its associated curvature space $\mathcal{S}=(\sts_{X},\conn ,\curv(\conn))$, we deduce the following apophthegm, our third one:

\vskip 0.1in

\noindent\fbox{%
    \parbox{\textwidth}{%
\noindent\underline{\bf Apophthegm 3:} A possible {\em structure sheaf morphism} $\mathcal{N}_{\sts}:\, \sts_{1}\rightarrow\sts_{2}$---as it were, a natural transformation type of map between sheaves of algebras of generalised arithmetics, coefficients, or theoretical local coordinate measurements---{\em induces or lifts to natural transformation type of maps} (:morphisms) {\em between the corresponding three important functor (:sheaf) categories} $\catdt$, $\catf$ and $\catcs$ above.
}%
}

\vskip 0.1in

\noindent We coin the functorial transformation mappings (:changes) like $\mathcal{N}_{\sts}$ in Apophthegm 3 above {\em fundamental structure sheaf morphisms} or {\em fundamental changes of generalised arithmetics, coefficients or theoretical local coordinate measurements}. Alternatively, but equivalently, we may view them as {\em fundamental natural transformation type of morphisms between $\cons$-algebraised spaces}, themselves natural transformation type of functors in the category  $\catcas$ of $\cons$-algebraised spaces\footnote{By definition, the category $\catcas$ has the $\cons$-algebraised spaces $\cas$ defined by equation (\ref{eq0}) as objects, and $\cas$-structure preserving sheaf morphisms between them as arrows.} underlying the three kinematical-structural categories $\catdt$, $\catf$ and $\catcs$ above:

\begin{equation}\label{eq8}
\xymatrix{
\sts_{1} \ar[d] \ar[r]^{\mathcal{N}_{\sts}} & \sts_{2} \ar[d]\\
\cas_{1} \ar[r]_{\hat{\mathcal{N}}_{\mathcal{A}\mathcal{S}}} & \cas_{2} }
\end{equation}

\vskip 0.1in

\noindent In other words, the fundamental structure sheaf morphisms $\mathcal{N}_{\sts}:\, \sts_{1}\rightarrow\sts_{2}$ above, induce or lift to natural transformation type of morphisms between their corresponding $\cons$-algebraised spaces in the category of $\cons$-algebraised spaces underlying the three fundamental kinematical categories $\catdt$, $\catf$ and $\catcs$ above.

\vskip 0.1in

\noindent \underline{\bf Important Note 9:} Categorically speaking, the natural transformations (morphisms) $\mathcal{N}_{\sts}$ above are members (:categorical arrows) in the class $\mathrm{Hom}(\sts,\sts)$ of (abelian algebra) {\em structure sheaf morphisms}. As we will see and posit below, they constitute {\em the fundamental $\sts$-Relativity structure underlying our abstract and generalised theory of ADG-gravity}, which $\sts$-Relativity ($\AR$), in turn, and in a dynamical context, supports and has as a functorial consequence both the aforementioned PARD and Mallios's Principle of $\sts$-Invariance ($\AI$).

\vskip 0.1in

By the foregoing discussion and results, we thus arrive at the following {\em Fundamental Question}:

\vskip 0.1in

\noindent\fbox{%
    \parbox{\textwidth}{%
\noindent\underline{\bf Fundamental Question (FQ):} How can we understand and physically interpret in the context of ADG-gravity changes in structure sheaves of generalised arithmetics, coefficients or theoretical generalised local coordinate determinations like the morphisms $\mathcal{N}_{\sts}\in\mathrm{Hom}(\sts,\sts)$ depicted in (\ref{eq8}) above?
}%
}

\vskip 0.1in

\noindent The next section comes to shed more light on, to discuss, to answer, as well as to further generalise precisely the FQ above from a dynamical-functorial point of view that was originally anticipated in \cite{rap14} and was recently worked out in detail in \cite{rap15}, which will lead us to the aforementioned {\em Principle of Algebraic Relativity of Differentiability} (PARD) and Mallios's {\em Principle of $\sts$-Invariance} ($\AI$). In turn, the PARD is a novel ADG-theoretic (:sheaf and category-theoretic) generalisation of the PGC of GR, which is one of the main reasons for our calling ADG-gravity {\em a general theory of gravity} in the title of the present paper.

\section{Dynamical Functoriality: the Principles of $\sts$-Relativity ($\AR$) and Algebraic Relativity of Differentiability (PARD) of the Vacuum Einstein ADG-Gravitational Field Dynamics}

In this section, we move on from the aforedescribed kinematical structures supporting the {\it aufbau} of the entire edifice of ADG-gravity, to the build-up of the ADG-gravitational dynamics in the form of the vacuum Einstein field equations, leading in turn to the principles of $\sts$-Relativity ($\AR$) and PARD supporting them, which, as we shall see, are homological algebraic ({\it i.e.}, categorical/functorial) abstractions and generalisations of the PGC of the spacetime manifold based GR, which, in turn, {\em are the two main reasons for our regarding ADG-gravity as a generalisation of the classical relativistic field theory of gravity (GR) as the the title of the present paper contends}. 

\subsection{ADG-Theoretic Vacuum Einstein Equations, Functorial $\sts$-Invariance and $\aut_{\sts}\modl$-Covariance: The Forebearers of the $\sts$-Relativity and the PARD Principles}

In order to introduce the $\AR$ and the PARD principles, we first recall a series of {\em nine key facts and results}, as well as certain important conceptual issues and structural characteristics that have been established about the categorical (:functorial) ADG-gravitational field dynamics \cite{malrap3,rap15,rap19}---{\it i.e.}, the ADG-theoretic version of the vacuum Einstein and free Yang-Mills equations---in our research work thus far.\footnote{Below, we also give the ADG-theoretic version of the free Yang-Mills equations in order for the reader to appreciate its formal structural similarities with the vacuum Einstein equations, thus discern what we have maintained and highlighted throughout our research so far: namely that, {\em from the ADG-theoretic point of view, (vacuum Einstein) gravity is a pure gauge theory, in the sense that its dynamics iinvolves solely the gravitational $\sts$-connection field variable}. See further remarks below.} On purpose, for the sake of economy, we do not give any rigorous technical details, mathematical expressions, equations and derivations of, or quantitative calculations leading to, the results being quoted below, but we give instead complete references to the published literature where these results have been rigorously worked out and derived, as well as physically interpreted and discussed in great detail:\footnote{Occasionally, however, if we feel we need to recall a certain key technical/mathematical concept, expression or equation, we do indeed present it in sufficient detail for this exposition.}

\subsubsection{Nine key facts about the vacuum Einstein ADG-gravitational field dynamics}

\begin{enumerate}

\item \underline{\bf Fact 1:} In ADG, the dynamical laws of vacuum Einstein gravity and free Yang-Mills theory derive from a variational extremum principle applied to an action functional---the vacuum Einstein-Hilbert and free Yang-Mills Lagrangian action functionals, respectively \cite{mall1,mall4,malrap1,malrap2,malrap3,rap5,rap7,rap11,rap13,rap14,rap15,rap19,zaf3}; as depicted below:

\begin{equation}\label{eq9}
\begin{array}{c}
\eh_{\modl}(\conn)=\int_{X}tr(\curv_{Ric}(\conn))\;\stackrel{\delta\aconn}{\mapto}\;\curv_{Einst}(\conn)|_{\modl}=0 \cr
{}\cr
\ym_{\modl}(\conn)=\frac{1}{2}\int_{X}tr(\curv_{YM}\wedge\star\curv_{YM})\;\stackrel{\delta\aconn}{\mapto}\;\Delta_{\modl
nd\modl}^{2}(\curv_{YM})=0
\end{array}
\end{equation}

\noindent with $tr(\curv_{Ric}(\conn))$ the ADG-theoretic version of the Ricci curvature scalar\footnote{This is the usual trace of the Ricci curvature tensor $\curv_{Ric}(\conn)$ of the connection $\conn$ on $\modl$.} of the Einstein curvature $\otimes_{\sts}$-tensor $\curv_{Einst}(\conn)|_{\modl}$ of the $\sts$-connection field $\conn$ on the vector sheaf $\modl$.

{\it A propos}, the reader should notice that {\em both scalar action functionals in} (\ref{eq9}) {\em are $\sts$-valued}, so this is another instance of our general motto in this paper that the whole of ADG and ADG-gravity involves, and revolves about, the structure sheaf $\sts$ in one way or another.

\item \underline{\bf Fact 2:} The physical laws---{\it i.e.}, the dynamical equations of vacuum Einstein gravity and free Yang-Mills theory in (\ref{eq9}) above---are {\em differential} equations proper that are homological algebraically (:categorically) expressed ADG-theoretically as equations between the relevant $\sts$-connection sheaf morphisms $\conn$ and their curvatures $\curv(\conn)$, which, as we saw earlier, are natural transformation type of maps between the relevant sheaf categories involved---the category $\catdt$ of differential triads, the category $\catf$ of ADG-fields and its associated category $\catcs$ of curvature spaces  \cite{mall1,mall4,malrap1,malrap2,malrap3,rap5,rap7,rap11,rap13,rap14,rap15,rap19,zaf3,malzaf1};

\item \underline{\bf Fact 3:} As we see in (\ref{eq9}) above, {\em the dynamical laws are actually expressed via the curvatures} $\curv(\conn)$ {\em of the corresponding $\sts$-connection fields} (:gravitational and Yang-Mills), which are $\otimes_{\sts}$-tensors.\footnote{Whereby, $\otimes_{\sts}$ is the {\em homological tensor product functor} with respect to the generalised arithmetics or coefficients in $\sts$ \cite{mall1,mall4,malrap3,rap15,rap19}.} Thus, the curvatures involved in the integrands of the two Lagrangian action functionals in (\ref{eq9}) are $\aut_{\sts}\modl$-invariant, hence also the laws that they define (or obey) as differential equations proper, are  $\aut_{\sts}\modl$-covariant\footnote{By {\em covariant laws}, we mean {\em form invariant laws} under $\aut_{\sts}\modl$ self- or auto-symmetry transformations of the corresponding ADG-fields (:vacuum Einstein gravitational and free Yang-Mills fields).} \cite{mall1,mall4,malrap1,malrap2,malrap3,rap5,rap7,rap11,rap13,rap14,rap15,rap19,zaf3,malzaf1};

\item \underline{\bf Fact 4:} From the two facts above it transpires that the {\em internal local gauge symmetry group sheaf} in ADG-gravity is the {\em principal sheaf} $\aut_{\sts}\modl$ of local automorphisms of the {\em associated (representation) vector sheaf} $\modl$ \cite{vas1,vas2,vas3,vas4,mall4,malrap3,rap14,rap15,rap19}.\footnote{See Important Note 1 earlier.} Notice that $\aut_{\sts}\modl$, as a group sheaf of local automorphisms of $\modl$, {\em acts internally}---{\it i.e.}, within the vacuum Einstein ADG-gravitational field $\mathcal{F}_{Einst}=(\modl ,\conn)$. It is the group sheaf of local dynamical self-symmetries of the vacuum Einstein ADG-gravitational field $\mathcal{F}_{Einst}=(\modl ,\conn)$ and of the dynamical equations  that it defines via its $\sts$-invariant curvature form (:$\otimes_{\sts}$-tensor) $\curv(\conn)$ in (\ref{eq9}).

\item \underline{\bf Fact 5:} Due to the manifest absence of an external (:to the gravitational and Yang-Mills ADG-theoretic $\sts$-connection fields) background geometrical spacetime manifold, the said dynamics is {\em purely gauge}. That is to say, as noted in Fact 4 above, the ADG-gravitational field's symmetries are purely internal---{\it i.e.}, intrinsic to the ADG-gravitational field itself. Equivalently stated, in ADG-gravity, there is no external (or background) to the ADG-gravitational field itself geometrical spacetime manifold with its own, also external to the gravitational field itself, spacetime symmetries. It follows that {\em there is no external geometrical spacetime interpretation of the theory} either, thus from an ADG-theoretic viewpont, {\em gravity is a pure gauge theory in which the sole dynamical variable is the ADG-gravitational $\sts$-connection Einstein field} $\mathcal{F}_{Einst}=(\modl ,\conn)$, with $\aut_{\sts}\modl$ its structure group sheaf of self-symmetries of the vacuum Einstein ADG-gravitational field law (\ref{eq9}) that it defines via its curvature $\curv(\conn)$ \cite{malrap1,malrap2,malrap3,mall4,rap13,rap14,rap15,rap19};

\item \underline{\bf Fact 6:} The background spacetime manifoldless gauge field theory that ADG-gravity is has been coined {\em gauge theory of the third kind}, because it is still a local (:sheaf-theoretic) gauge field theory of gravity based solely on the gravitational $\sts$-connection field variable $\conn$, but without any background spacetime manifold geometry to support either its mathematical representation by differential geometric (:CDG) means, or its physical interpretation in terms of spacetime concepts, constructions and associated `geometrical pictures'. This variant of gauge theory should be distinguished from the usual external (background) smooth spacetime manifold based gauge theories of matter of the first (:global) and second (:base geometrical spacetime manifold localised) kind \cite{malrap1,malrap2,malrap3,rap13,rap14,rap15,rap19}; 

\item \underline{\bf Fact 7:} In addition to the six facts above, the ADG-theoretic local gauge connection field theory used to formulate ADG-gravity has been coined {\em half-order formalism}, to distinguish it from the usual CDG and external differential spacetime manifold based gravitational theories of Einstein (:second order formalism) and the smooth {\it vierbein-cum-connection} based Palatini-Ashtekar scheme (:first order formalism). In the ADG-theoretic half-order formalism for ADG-gravity, {\em the sole dynamical variable is the gravitational $\sts$-connection field $\conn$}; hence, {\em the kinematical space of the theory is the moduli space} $\sconn_{\sts} /\aut_{\sts}\modl$---the orbifold of the affine space $\sconn_{\sts}$ of $\sts$-connections quotiented by the local gauge group $\aut_{\sts}\modl$, which is physically interpreted as {\em the space of gauge equivalent $\sts$-connections}  \cite{malrap1,malrap2,malrap3,rap13,rap14,rap15,rap19};   

\item \underline{\bf Fact 8:} From the seven facts above, it follows that the principal (gauge group) sheaf $\aut_{\sts}\modl$ of local dynamical gauge (internal) self-symmetries of ADG-gravity comes to replace the (external) spacetime diffeomorphism group $\mathrm{Diff}(M)$, which is the dynamical symmetry group of the spacetime manifold $M$ based GR, implementing the Principle of General Covariance (PGC) in the classical, CDG-based field theory of gravity (GR). In turn, as noted in Fact 4 above, $\aut_{\sts}\modl$ can be thought of as the {\em principal}, internal to the ADG-gravitational vacuum Einstein field $\mathcal{F}_{Einst}=(\modl ,\conn)$ itself, {\em gauge group sheaf of auto-symmetries of the ADG-gravitational field auto-dynamics} in (\ref{eq9}). Moreover, as explained earlier in Important Note 1, since the vector sheaf $\modl$ is, by definition, locally isomorphic to $\sts^{n}$ ($\modl(U)\simeq\sts^{n}(U)$), $\aut_{\sts}\modl(U)\simeq\modl nd\modl^{\bullet}(U)\simeq M_{n}(\sts(U))^{\bullet}$, with $\modl nd\modl^{\bullet}(U)$ the local group of invertible endomorphisms of $\modl$, which is in turn locally isomorphic to the linear group $M_{n}(\sts(U))^{\bullet}$ of invertible $(n\times n)$-matrices with entries in $\Gamma(U,\sts)\equiv\sts(U)$---the local sections of the structure sheaf $\sts$ \cite{malrap1,malrap2,malrap3,rap13,rap14,rap15,rap19};\footnote{The reader should also note that if the rank $n$ of $\modl$ is equal to $4$, $\aut_{\sts}\modl(U)\equiv M_{n}(\sts(U))^{\bullet}=\mathcal{G}\mathcal{L}(4,\sts(U))$---the local linear gauge group sheaf of invertible $(4\times 4)$-matrices with entries in $\sts(U)$, which is the (sheaf-theoretic) ADG-gravitational analogue of the general linear group $GL(4,\R)$ of the usual $4$-dimensional smooth spacetime manifold based GR, implementing the PGC as {\em the group of general coordinate transformations}.}

\item \underline{\bf Fact 9:} From Fact 8 above, we note that, in ADG-gravity, the PGC of GR is expressed by the following three different, but mutually equivalent \cite{malrap3,rap15,rap19}, {\em smooth background geometrical spacetime manifoldless} ways:  

\vskip 0.1in

\noindent $\bullet$ {\bf (i)} By what Mallios coined $\sts$-invariance ($\AI$) or $\sts$-functoriality: this means that the ADG-gravitational field dynamics in (\ref{eq9}) is expressed via the curvature $\curv(\conn)$ of the $\sts$-connection $\conn$, which is an $\otimes_{\sts}$-tensor---{\it i.e.}, it is a quantity that transforms $\sts$-tensorially under the homological tensor product functor $\otimes_{\sts}$. This corresponds to the fact that $\curv(\conn)$ is an $\sts$-morphism which transforms homogeneously under the, intrinsic to the vacuum Einstein ADG-gravitational field $\mathcal{F}_{Einst}=(\modl ,\conn)$, principal gauge group sheaf $\aut_{\struc}\modl$, unlike the connection $\conn$ from which it derives, which is {\em not} an $\sts$-morphism hence it transforms inhomogeneously (affinely) under local gauge transformations in  $\aut_{\struc}\modl$ \cite{mall10,mall14,mall15,mall13,mall17,rap14,rap15,rap19,malzaf1,zaf3,zaf4}; 

\vskip 0.1in

\noindent $\bullet$ {\bf (ii)} By what we originally coined  $\aut_{\sts}\modl$-invariance of the Einstein-Hilbert Lagrangian action functional and $\aut_{\sts}\modl$-covariance of the vacuum Einstein dynamical equations of motion, which result from extremising the corresponding action functionals with respect to the $\sts$-connection field $\conn$ as depicted in (\ref{eq9}) earlier  \cite{malrap1,malrap2,malrap3,rap14,rap15,rap19}. These are the internal, purely gauge $\aut_{\sts}\modl$-effectuated self-symmetries (:auto-invariances) of the vacuum Einstein ADG-gravitational field $\mathcal{F}_{Einst}=(\modl ,\conn)$ and of the dynamical equations that it defines in (\ref{eq9}) above \cite{malrap3,rap14,rap15,rap19}; and,  

\vskip 0.1in

\noindent $\bullet$ {\bf (iii)} By the natural transformation type of functorial equivalence between the category of vacuum Einstein ADG-fields $\catf$ and the category of their corresponding curvature spaces $\catcs$ \cite{mall14,mall15,mall13,mall17,malzaf1,rap15,rap19,zaf3,zaf4}. Recently, in \cite{rap15}, it was shown that the said natural functorial equivalence is represented by the following {\em pair of adjoint functors}:
 
\begin{equation}\label{eqGM}
\mathcal{G}\mathcal{M}_{\sts}=(\otimes_{\sts},\mathrm{Hom}_{\sts})
\end{equation}

\noindent which is commonly known in category theory as a {\em geometric morphism} \cite{macmo}, between the sheaf categories of fields $\catf$ and curvature spaces $\catcs$ that we saw earlier. That is to say, since the vacuum Einstein ADG-gravitational dynamics in (\ref{eq9}) is expressed via the curvature $\curv(\conn)$ of the $\sts$-connection variable $\conn$, which is an $\otimes_{\sts}$-tensor, the dynamics is $\aut_{\sts}\modl$-invariant.\footnote{Notice, as indicated by their subscripts, that the two adjoint functors involved in the geometric morphism $\mathcal{G}\mathcal{M}_{\sts}$ above are {\em functors relative to} $\sts$: that is to say, {\em they are $\sts$-respecting} or {\em $\sts$-preserving functors}. Hence the dynamics is $\sts$-functorial or $\sts$-invariant, as maintained in point (i) above.} The geometric morphism $\mathcal{G}\mathcal{M}_{\sts}$ in (\ref{eqGM}) above is what Mallios refers to in \cite{mall10,mall14,mall15,mall13,mall17,rap15,rap19} as {\em the fundamental $\sts$-adjunction representing the $\sts$-functoriality and the $\sts$-invariance} ($\AI$) of the mathematical ADG-theoretic constructions and, as a result, of the physical vacuum Einstein ADG-gravitational field dynamics in (\ref{eq9}), which is fundamentally based on those mathematical structures.

\end{enumerate} 

\noindent In the following subsection, we go one step further to extend and generalise ADG-theoretically the PGC of the spacetime manifold based GR by diagrammatically representing and discussing in detail the aforementioned ADG-field theoretic {\em Principle of Algebraic Relativity of Differentiability} (PARD), which was first conceived primitively in the heptalogy \cite{malrap1,malrap2,malrap3,malrap4,rap5,rap11,rap7}, then explicitly stated in \cite{rap14} and recently functorially distilled and expressed in terms of the geometric morphism $\mathcal{G}\mathcal{M}_{\sts}$ in (\ref{eqGM}) \cite{rap15,rap19}.      

\subsection{A Functorial Diagrammatic Representation of the ADG-Gravitational Field Theoretic Principle of Algebraic Relativity of Differentiability (PARD)}

We now come to answer and to discuss the {\bf Fundamental Question} (FQ) that concluded our penultimate subsection above.

From the previous section (cf. Apophthegms 1-2), we recall that in ADG the `differential geometric mechanism' or `differentiability machninery'---{\it i.e.}, the main ADG-theoretic $\sts$-connection field $\mathcal{F}=(\modl ,\conn)$ of (\ref{eq6}), which is the main conceptual and structural tool of ADG regarded as a theory of {\em differential} geometry proper----has as its topologico-algebraic source and domain of definition, thus it effectively derives from, the structure algebra sheaf $\sts$ of generalised arithmetics and (sheaf cohomological) coefficients in ADG regarded as a mathematical theory, which, in turn, is physically interpreted as our theoretical generalised local coordinate measurements (:local gauge determinations) of the ADG-gravitational $\sts$-connection field in ADG-gravity.

Thus, in ADG-gravity, Einstein's Principle of Relativity \cite{einst3,einst4}, which is tantamount to the PGC of GR when generalised to the employment of any arbitrary system of abstract local coordinates represented by the structure sheaf $\sts$ relative to a system $\gauge_{i}$ of local open gauges covering the aribtrary base localisation topological space $X$, may be stated as the following theoretical imperative or {\it Ur}-principle of $\sts$-Relativity ($\AR$):

\vskip 0.1in

\noindent\fbox{%
    \parbox{\textwidth}{%
\noindent\underline{\bf Fundamental Ur-Principle of $\sts$-Relativity ($\AR$):} The ADG-gravitational field law of vacuum Einstein gravity, which is differential geometrically represented by homological-algebraic means ({\it i.e.}, sheaf and category-theoretically) by differential equations involving the relevant $\sts$-connection sheaf morphism $\conn$ and its corresponding $\otimes_{\sts}$-functorial curvature form $\curv(\conn)$ in (\ref{eq9}), is independent of any choice of, hence of any potential (functorial) changes in, structure sheaf $\sts$ of generalised arithmetics, coefficients or theoretical local coordinate measurements that {\em we choose in the first place} in order to coordinatise and measure (:locally gauge) the ADG-gravitational field (relative to a chosen system $\gauge_{i}$ of local open gauges covering $X$). That is to say, the vacuum Einstein ADG-gravitational field law in (\ref{eq9}) is {\em $\sts$-transformations' form-invariant}, which in turn, as explained earlier, for a particular choice of structure sheaf $\sts$, is equivalent to it being $\aut_{\sts}\modl$-gauge covariant. Then, as noted above, the corresponding $\sts$-valued Lagrangian Einstein-Hilbert action functional, from which the vacuum Einstein differential equations derive variationally by extremising it with respect to variations $\delta\conn$ of the $\sts$-connection field as depicted in (\ref{eq9}) earlier, should be $\aut_{\sts}\modl$-gauge invariant \cite{malrap1,malrap2,malrap3,rap14,rap15,rap19}. As also noted earlier, this is effectively the content and physical meaning of Mallios's {\em Principle of $\sts$-Invariance} ($\AI$) \cite{mall10,mall14,mall15,mall13,mall17,rap14,rap15,rap19,malzaf1,zaf3,zaf4} as well as of the {\em $\sts$-functoriality of vacuum Einstein ADG-gravity and free Yang-Mills theories} recently observed in \cite{rap15, rap19} principally in connection with the geometric morphism depicted in (\ref{eqGM}) earlier.
}%
}

\vskip 0.1in

\noindent The fundamental Ur-principle of $\AR$ above is an abstract and generalised ADG-theoretic version of Einstein's Principle of Relativity \cite{einst3,einst4} maintaining that {\em the laws of Physics should be independent of our observations, our local field determinations---in effect, of our local coordinate measurement gauges, thereof}, which in the context of GR is reflected by the PGC stating that {\em the gravitational field laws should be independent (invariant) under arbitrary (smooth) coordinate frame changes}, with local coordinate frames understood to be the `local measurement laboratory instruments' (or the `local gauges') employed to localise (:measure and determine locally) the $\smooth$-smooth spacetime coordinates of the spacetime manifold's point events, relative to which the smooth gravitational potentials---the ten components of the smooth metric tensor $g_{\mu\nu}$ in the original formulation of GR (:2nd order formalism)---take their values.

\vskip 0.1in

The $\sts$-Relativity principle ($\AR$) is an {\em abstract and generalised version of Einstein's Principle of Relativity} \cite{einst3,einst4} in the following two ways:

\begin{itemize}

\item The vacuum Einstein gravitational field law in (\ref{eq9}) is not only invariant with respect to, or independent of, arbitrary changes within a particularly chosen structure algebra sheaf $\sts$ of generalised coordinate measurements of the ADG-gravitational field $\mathcal{F}_{Einst}=(\modl ,\conn)$;\footnote{This is the ADG-theoretic notion of $\sts$-invariance and $\aut_{\sts}\modl$-covariance that generalise the PGC of GR mentioned earlier.}

\item But it is also {\em differential form invariant} under aribtrary changes in structure sheaves like those depicted in (\ref{eq8}): one may use different, or whatever may be suitable to a particular problem or situation in hand, structure sheaves $\sts$ to describe, measure or locally coordinatise the vacuum Einstein ADG-gravitational field $\mathcal{F}_{Einst}=(\modl ,\conn)$ and the dynamical law (\ref{eq9}) that it defines, and the latter's form of expression, as a {\em differential} equation proper, remains form-invariant ({\it alias}, covariant) under such changes.

\end{itemize}

Below, we are going to give an answer to the {\bf Fundamental Question} (FQ) in the light of the {\em Fundamental Ur-Principle of $\sts$-Relativity} ($\AR$) above by presenting diagrammatically a sequence---a top-down hierarchical tower of fundamental ADG-theoretic structures and closely entwined $\sts$-functorial natural transformation type of maps between them, and discuss in detail their structural-mathematical and physico-philosophical significance.

The top-down tower of functorial and natural transformation type of morphisms below represents the basic architectonic, homological algebraic skeletal backbone {\it aufbau} of ADG-field theory and ADG-gravity in particular: from its roots in the chosen, but in principle arbitrary, $\sts$ at the very top, to an envisaged, but yet to technically materialise, purely and fully gauge group $\aut_{\sts}\modl$-covariant and self-symmetric (:$\sts$-functorial) quantum path integral type of dynamics over the aforementioned kinematical moduli (orbifold) space $\sconn_{\sts}$ of $\aut_{\sts}\modl$-gauge equivalent $\sts$-connections at the very bottom. In the diagram that follows, this progressive building-up of ADG-gravity starts, in reverse order, at the very top of the tower and at the very basis of ADG regarded as a theory of {\em differential} geometry, where the category $\catcas$ of $\cons$-algebraised spaces is situated---from the structure sheaves of generalised arithmetics, coefficients and theoretical local coordinate measurements from which Mallios's algebraic $\sts$-connections $\conn$ derive. In turn, the latter are the very heart of ADG as they provide us with an $\sts$-functorial differential mechanism from which all the rest of the tower springs out and unfolds. 

\vskip 0.1in

\noindent \underline{\bf Important Note 6:} In the tower of ADG-theoretic structures below and their functorial interdependences, the reader should note that the structure sheaf $\sts$ is involved in one way or another, explicitly or implicitly, at every level. This is another instance `justifying' the title of this paper that ADG, and {\it in extenso} ADG-gravity, comes or originates from the structure sheaf $\sts$.

\vskip 0.1in

\subsubsection{Diagram I: The ADG-theoretic architectonic tower from top to bottom.}

We start the tower from the very top, focusing our attention on the functor category $\catcas$ of $\cons$-algebraised spaces, by considering functorial changes in structure sheaves of generalised arithmetics or generalised local coordinate measurements of the type we saw earlier in (\ref{eq8}),\footnote{We called structure sheaf changes such as those in (\ref{eq8}), {\em  fundamental changes of generalised arithmetics, coefficients or theoretical local coordinate measurements or determinations}.} which we recap here:

\begin{equation}\label{eq22}
\xymatrix{
\sts_{1} \ar[d] \ar[r]^{\mathcal{N}_{\sts}} & \sts_{2} \ar[d]\\
\cas_{1} \ar[r]_{\hat{\mathcal{N}}_{\mathcal{A}\mathcal{S}}} & \cas_{2} }
\end{equation}

\noindent The diagram above depicts a functorial sheaf morphism $\mathcal{N}_{\sts}$ effecting a {\em structure sheaf change}---{\em a transformation of generalised arithmetics, coefficients or local coordinate measurements}, which in turn induces (or lifts to) an {\em $\cons$-algebraised space morphism} $\hat{\mathcal{N}}_{\mathcal{A}\mathcal{S}}$ between the $\cons$-algebraised spaces to which the corresponding structure sheaves belong.\footnote{In (\ref{eq22}) above, we tacitly assume, without loss of generality, that the base topological space $X$, on which the structure sheaves are localised, remains the same, so that the said structure sheaf morphism $\mathcal{N}_{\sts}$ lifts effectively to a {\em functorial automorphism} $\hat{\mathcal{N}}_{\mathcal{A}\mathcal{S}}$ of the category $\catcas_{X}$ of $\cons$-algebraised spaces soldered on an in principle arbitrary, but fixed, base topological space $X$, as explained earlier. $\catcas_{X}$ may in turn be regarded as being isomorphic to the category $\catab(X)$ of sheaves of abelian associative algebras on $X$. In the same way, we shall henceforth assume that the background arbitrary topological space $X$ on which all sheaves are soldered remains constant throughout all our subsequent deliberations as, as we saw and argued earlier, it plays no role whatsoever in the ADG-theoretic {\em differential geometric mechanism} or {\em differential structure}, from which in turn all our physical constructions, dynamical equations and their associated calculations/solutions/results---our ADG-theoretic Calculus---derive from. Moreover, we have seen that the said ADG-theoretic {\em local differential mechanism} originates from the algebraic stalks of the structure sheaf $\sts$, not from the base topological space $X$, while; {\it a fortiori}, that it is $\sts$-functorial, hence $\sts$-invariant \cite{mall14,mall15,mall13,malzaf1,rap15,rap19,mall17}.}

Thus, starting from such {\em fundamental changes in structure sheaves of generalised arithmetics, coefficients and theoretical local coordinate-measurements}, we produce the following {\em Top-Down Tower of ADG-Theoretic Natural Transformations and Equivalences}:

\newpage

\begin{equation}\nonumber
\begin{tikzcd}[row sep=huge, column sep=huge] 
\sts_{1}(X)\arrow[r, "{\mathcal{N}_{\sts}}" description]\arrow[dd, dashed, "{\bigoplus_{i=1}^{n=\mathrm{rank}}}" description] & \sts_{2}(X)\arrow[dd, dashed, "{\bigoplus_{i=1}^{n=\mathrm{rank}}}" description] &[-1em] &[-2em]\\
&&& \\[-4ex]
\modl_{1}(X)\arrow[r, "{\hat{\mathcal{N}}_{\modl}}" description]\arrow[dd, dashed, "{\conn_{1}}" description] & \modl_{2}(X)\arrow[dd, dashed, "{\conn_{2}}" description] & \\[-3ex]
&&& \\
\mathcal{F}_{1}:=(\modl_{1},\conn_{1})_{X}\arrow[r, "{\hat{\mathcal{N}}_{\mathcal{F}}}" description]\arrow[dd, dashed, "{\mathcal{G}\mathcal{M}_{\sts}}" description] & \mathcal{F}_{2}:=(\modl_{2},\conn_{2})_{X}\arrow[dd, dashed, "{\mathcal{G}\mathcal{M}_{\sts}}" description] & \\[-3ex]
&&& \\
\mathcal{S}_{1}:=(\modl_{1},\curv_{1}(\conn_{1}))_{X}\arrow[r, "{\hat{\mathcal{N}}_{\mathcal{S}}}" description]\arrow[dd, dashed, "{\mu_{\sts_{1}}\equiv\int_{\sconn(\modl_{1}) /\aut_{\sts_{1}}\modl_{1}}}" description] & \mathcal{S}_{2}:=(\modl_{2},\curv_{2}(\conn_{2}))_{X}\arrow[dd, dashed, "{\mu_{\sts_{2}}\equiv\int_{\sconn(\modl_{2}) /\aut_{\sts_{2}}\modl_{2}}}" description] & \\[-3ex]
&&& \\
\int_{{\sconn(\modl_{1}) /\aut_{\sts_{1}}\modl_{1}}}tr(\curv_{1}(\conn_{1}))\arrow[r, "\mathcal{N}_{\sts}" description]\arrow[dd, dashed, "{\delta\aconn_{1}}" description] &\int_{{\sconn(\modl_{2}) /\aut_{\sts_{2}}\modl_{2}}}tr(\curv_{2}(\conn_{2}))\arrow[dd, dashed, "{\delta\aconn_{2}}" description] & \\[-3ex]
&&& \\
\curv_{1}(\modl_{1})=0\arrow[r, "\hat{\mathcal{N}}_{\modl}" description]\arrow[dd, dotted, "\mathcal{QPI}\equiv{e^{\imath S_{EH}/\hbar}}" description]  & \curv_{2}(\modl_{2})=0\arrow[dd, dotted, "\mathcal{QPI}\equiv{e^{\imath S_{EH}/\hbar}}" description] & \\[-3ex]
&&& \\
e^{\frac{\imath}{\hbar}\int_{\sconn(\modl_{1}) /\aut_{\sts_{1}}\modl_{1}}[tr(\curv_{1}(\conn_{1}))]d\mu[\aconn_{1}]}\arrow[r, "\hat{\mathcal{N}}_{QPI}" description] & e^{\frac{\imath}{\hbar}\int_{\sconn(\modl_{2}) /\aut_{\sts_{2}}\modl_{2}}[tr(\curv_{2}(\conn_{2}))]d\mu[\aconn_{2}]}
\end{tikzcd}
\end{equation}

\vskip 0.1in

\centerline{\underline{\small\bf Diagram I: Vertical Top-Down Architectonic Aufbau of ADG-Gravity}}

\newpage

Below, we itemise our explanation of each morphism storey-by-storey in the top-down skeletal architectonic tower above:

\begin{itemize}

\item\underline{\bf Horizontal Level 1:} In the top storey, the horizontal map:

\begin{equation}\label{d1}
\mathcal{N}_{\sts}:\, \sts_{1}(X)\longrightarrow\sts_{2}(X)
\end{equation}

\noindent may be thought of as a functorial, natural transformation type of morphism (between the corresponding sheaf categories of $\cons$-algebraised spaces to which the structure sheaves belong) corresponding to {\em changes of structure sheaves of generalised arithmetics and coefficients} in ADG, regarded as a mathematical theory, and to {\em changes of theoretical generalised local coordinate measurements} in ADG-gravity, regarded as a physical theory.

Since, as we have seen earlier, the entire differential geometric mechanism of ADG derives from the structure sheaf $\sts$ of generalised arithmetics employed, these are {\em structure sheaf changes that effect `functorial changes of differentiability'}---{\it i.e.}, changes of the entire `differential structure' or `differential mechanism' of ADG that leave the ensuing structures (: differential triads, fields, curvature spaces, $\sts$-metrics, {\it etc.}), that issue from $\sts$ and trickle all the way down the tower, `{\em differential form invariant}'. 

\item \underline{\bf The Physical Significance of the Structure Sheaf Morphisms $\mathcal{N}_{\sts}$:} The theoretical physics import and physical significance of 
such {\em `functorial changes of differential structure or differentiability'} effectuated by $\mathcal{N}_{\sts}$ as depicted in (\ref{d1}) can be appreciated in the light of two quite successful applications of ADG to quantum spacetime structure, quantum GR and QG:

\vskip 0.1in

\begin{enumerate}

\item \underline{\bf Finitary and Quantal ADG-Gravity:} If we change structure sheaf of generalised coordinates from the usual sheaf $\smooth_{M}$ of smooth coordinates on a differential spacetime manifold $M$, on which one can write the usual vacuum Einstein equations by CDG means, to {\em finitary spacetime sheaves of discrete differential incidence algebras associated with Sorkin's finitary poset substitutes of continuous manifolds} \cite{sork0,rap2,rapzap1,rapzap2}, we are able to formulate a locally finite, causal and quantal version of vacuum Einstein-Lorentzian gravity \cite{malrap1,malrap2,malrap3}. Moreover, we are able to `resolve' (better, {\em evade} altogether) both the exterior, but more importantly, the interior Schwarzschild singularities of GR \cite{rap5} without any `breakdown' of the physical law of gravity (which is represented by a differential equation proper) in their vicinity, unlike what the usual CDG (:the standard Newtonian Differential Calculus) and background geometrical $\smooth$-smooth manifold-based Analysis of those spacetime singularities purports to show \cite{clarke3,clarke4}. 

\item \underline{\bf Spacetime Foam Dense Singularities:} The second successful application of ADG involves again switching from the usual structure sheaf $\sts=\smooth_{M}$ of a differential spacetime manifold $M$ to the so-called {\em Rosinger algebra sheaf of spacetime foam dense singularities} $\sts_{nd}$ \cite{malros1,malros2,malros3,mall10}. This is a flabby structure sheaf of {\em differential algebras of generalised functions}\footnote{The epithet `{\em differential}' to the noun `{\em algebras}' here pertaining to the fact that these algebras provide us with the essential differential mechanism of ADG---Mallios's $\sts$-connections $\partial$ and $\conn$ \cite{malros1,malros2,malros3}.} (:non-linear distributions) that are teeming with singularities of the most unmanageable kind when viewed from the usual perspective of the featureless smooth manifold based CDG,\footnote{The subscript `{\em nd}' to Rosinger's algebra sheaf above means `{\em nowhere dense}', while from \cite{malros1,malros2,malros3,mall10} we read that Rosinger's non-linear generalised functions not only include $\smooth_{X}$ as a subset, but they also include the Dirac $\delta$-functions and the usual Schwartz linear distributions $\mathcal{D}_{X}$ as a proper subsheaf on $X$.} yet the whole differential mechanism and the technical-{\it cum}-conceptual panoply of ADG applies to them in full force,\footnote{For instance, Poincar\'{e}'s Lemma and a full-fledged de Rham Cohomology are seen to hold intact on the very pathological and problematic, when viewed from the perspective of a $\smooth$-smooth manifold $M$, structure sheaf  $\sts_{nd}$.} one can do to the extent that the vacuum Einstein equations are seen to hold intact, and in no differential geometric sense are seen to break down, at their presence \cite{malros1,malros2,malros3,mall10}.

\item In addition to the two applications above, we are able to regard and cast ADG-gravity (and ADG-Yang-Mills theory) as an inherently 3rd gauged (:background geometrical spacetime manifoldless and purely gauge) and 3rd quantised (:canonically sheaf cohomologically quantised) field theory \cite{rap13,rap14,rap15,rap19}.\footnote{See remarks on Sheaf Cohomology and Third Quantisation in the second half of the paper below.}

\end{enumerate}

\end{itemize}

\noindent Thus, from a physical applications of ADG-field theory point of view, we may distill the gist of the fundamental structure sheaf changes in (\ref{d1}) above to our next `utilitarian' Apophthegma:

\vskip 0.1in

\noindent\fbox{%
    \parbox{\textwidth}{%
\noindent\underline{\bf Apophthegm 4:} Whenever we encounter a conceptual issue or technical problem associated with, say, a particular choice of structure sheaf $\sts$ of generalised arithmetics ({\it e.g.}, the $\sts\equiv\smooth_{M}$ associated with a smooth base spacetime manifold $M$, with all its pestilential `geometrical pathologies' in the form of singularities and associated unphysical field infinities either at the classical or at the quantum level of description of the theory), we can switch to a more useful or appropriate structure sheaf $\sts^{'}$ for the physical situation/model or problem at issue, with respect to which the physical field laws---which are still modelled after differential equations involving $\sts^{'}$-invariant connection and curvature sheaf morphisms with respect to the new structure sheaf $\sts^{'}$---still hold intact. That is, the laws of vacuum Einstein gravity and free Yang-Mills theory will still be $\sts^{'}$-invariant and $\aut_{\sts '}\modl$-covariant in the form of (\ref{eq9}). Moreover, the $\sts$-functoriality of all the main ADG-theoretic concepts, constructions and associated structures ({\it e.g.}, $\cons$-algebraised spaces, differential triads $\triad$, differential $\sts$-modules/vector sheaves $\modl$ and $\sts$-connections $\conn$ on them, {\it etc}.) secures and guarantees that the latter will still be in force in the new `differential geometric setting' that is based on, and derives from, the new structure algebra sheaf $\sts^{'}$ employed.}%
}

\vskip 0.1in

\noindent In this light, we quote a short passage from \cite{mall10}\footnote{Quote (4.18) in \cite{mall10}.} that on the one hand totally corroborates our Apophthegm 4 above, and on the other, it captures perfectly our interpretation of the {\em Top-Down Tower of ADG-Theoretic Natural Equivalences} representing categorically what Mallios coined {\em $\sts$-invariance of the entire ADG-theoretic differential geometric mechanism} \cite{mall10,mall14,mall15,mall13,mall17,rap14,rap15,rap19,malzaf1,zaf3,zaf4}:\footnote{The morphisms presented below are called `{\em Natural}', because of an essential characteristic of theirs. Technically speaking, they are mappings between {\em sheaf functors} (:sheaf categories); hence they are {\em Natural Transformation} type of maps \cite{macmo}.}

\bigskip\noindent (Q2)\hskip 0.9in
\begin{minipage}{11cm}
\noindent ``...Starting \emph{from any} basic \emph{``differential triad''},
    in the sense of ADG (even a classical one, as e.g. a
    \emph{``locally euclidean} one, this is the case, herewith,
    \emph{we} can then \emph{perform any} (functorial)
    \emph{operation, provided within the category of differential
    triads}, to get thus at a new one [occasionally, more
    useful/flexible than the initially given one!]...'' 
\end{minipage}

\vskip 0.1in

\noindent Thus, in what follows, we will witness the import and usefulness of this `{\em functorial conservation of differential geometric structure and mechanism}' effect, which follows from the $\sts$-functoriality observed in the {\it aufbau} of the entire ADG-theoretic edifice, storey-by-storey, in the {\em Top-Down Tower of ADG-Theoretic Natural Transformations and Categorical Equivalences} above.

\begin{itemize}

\item\underline{\bf Horizontal Level 2:} The second storey (level) in the top-down tower involves seeing the original fundamental structure sheaf change $\mathcal{N}_{\sts}$ lift to (or induce) the following morphism between the respective vector sheaves (:differential $\sts$-modules of finite rank):

\begin{equation}\label{d2}
\hat{\mathcal{N}}_{\modl}:\, \modl_{1}(X)\longrightarrow\modl_{2}(X)
\end{equation}

\noindent which is a natural transformation type of functorial correspondence within the category of differential triads $\ctriad$ \cite{pap1,pap2,pap3} as Mallios points out in (Q2) above, or within its associated category of vector sheaves $\modl$.

\vskip 0.1in

On the other hand, the syncopated vertical arrow from Level 1 to Level 2:

\begin{equation}\label{d3}
\bigoplus_{i=1}^{n=\mathrm{rank}}:\, \sts_{1}(X)\dashrightarrow\modl_{1}(X)
\end{equation}

\noindent simply represents the local assembly of a vector sheaf $\modl$ as the $n$-th Whitney sum of $n$-copies of its corresponding underlying structure sheaf $\sts$, as it was defined in the first section of the paper.

\vskip 0.1in

\item\underline{\bf Horizontal Level 3:} The third storey (level) in the top-down tower involves the vector sheaf change $\mathcal{N}_{\modl}$ in the previous level lifting to (or inducing) the following ADG-field morphism:

\begin{equation}\label{d4}
\hat{\mathcal{N}}_{\mathcal{F}}:\, \mathcal{F}_{1}=(\modl_{1},\conn_{1})_{X}\longrightarrow \mathcal{F}_{2}=(\modl_{2},\conn_{2})_{X}
\end{equation}

\noindent which is again a natural transformation type ($\mathcal{N}_{\mathcal{F}}$) of functorial correspondence within the category of vacuum Einstein gravitational ADG-fields $\catf$---a natural ADG-field transformation (:functorial morphism) \cite{mall1,mall4,rap15}.

\vskip 0.1in

On the other hand, the syncopated vertical arrow from Level 2 to Level 3:

\begin{equation}\label{d5}
\conn_{1}:\, \modl_{1}(X)\dashrightarrow(\modl_{1},\conn_{1})_{X}
\end{equation}

\noindent simply denotes the endowment of a vector sheaf $\modl$ with an $\sts$-connection $\conn$ to form the vacuum Einstein ADG-gravitational field pair $\mathcal{F}_{\Einst}:=(\modl,\conn)_{X}$ on $X$ (relative to a chosen local coordinatising structure sheaf $\sts$), as defined earlier in (\ref{eq3}).

\vskip 0.1in

\item\underline{\bf Horizontal Level 4:} The fourth storey (level) in the top-down tower involves the vector sheaf change $\mathcal{N}_{\modl}$ of Level 2 lift to the following morphism:

\begin{equation}\label{d6}
\hat{\mathcal{N}}_{\mathcal{S}}:\, (\modl_{1},\curv_{1}(\conn_{1}))_{X}\longrightarrow(\modl_{2},\curv_{2}(\conn_{2}))_{X}
\end{equation}

\noindent which is yet again a natural transformation type ($\mathcal{N}_{\mathcal{S}}$) of correspondence within the category of vacuum Einstein gravitational ADG-curvature spaces $\catcs$---a functorial morphism type of change between ADG-curvature spaces \cite{mall1,mall4,rap15}.

\vskip 0.1in

On the other hand, the syncopated vertical map from Level 3 to Level 4:

\begin{equation}\label{d7}
\mathcal{G}\mathcal{M}_{\sts}:\, (\modl_{1},\conn_{1})_{X}\dashrightarrow(\modl_{1},\curv_{1}(\conn_{1}))_{X}
\end{equation}

\noindent corresponds to the aforementioned fundamental {\em geometric morphism} (:pair of adjoint functors) $\mathcal{G}\mathcal{M}_{\sts}=(\otimes_{\sts},\mathrm{Hom}_{\sts})$ that we saw earlier in (\ref{eqGM}), which maps the Einstein ADG-gravitational field pair $\mathcal{F}_{Einst}=(\modl_{1},\conn_{1})$ to its corresponding curvature space $\mathcal{S}_{Einst}=(\modl_{1},\curv(\conn_{1}))$ on $X$ in their respective categories $\catf$ and $\catcs$ \cite{rap15,rap19}.

\vskip 0.1in

\item\underline{\bf Horizontal Level 5:} The fifth storey (level) in our top-down tower involves the morphism $\hat{\mathcal{N}}_{\mathcal{S}}$ in the previous horizontal level inducing the following morphism:

\begin{equation}\label{d8}
\mathcal{N}_{\sts}:\, \int_{{\sconn(\modl_{1}) /\aut_{\sts_{1}}\modl_{1}}}tr(\curv_{1}(\conn_{1}))\longrightarrow\int_{{\sconn(\modl_{2}) /\aut_{\sts_{2}}\modl_{2}}}tr(\curv_{2}(\conn_{2}))
\end{equation}

\noindent which represents an $\sts$-functorial natural transformation type of map between the ADG-gravitational Einstein-Hilbert action functional on the moduli space $\sconn(\modl_{1})/\aut_{\sts_{1}}$ of $\aut_{\sts_{1}}$-gauge equivalent $\sts_{1}$-connection fields, to its corresponding counterpart dynamical action functional on the space $\sconn(\modl_{2})/\aut_{\sts_{2}}$ of gauge equivalent connections.\footnote{The reader should recall that, as it was noted earlier and originally emphasised in \cite{malrap3,rap13,rap14,rap15,rap19}, the relevant `kinematical configuration space' in ADG-gravity, regarded as a 3rd-gauged and 3rd-quantised pure gauge field theory, is the affine moduli space $\sconn(\modl)/\aut_{\sts}\modl$ of $\aut_{\sts}\modl$-equivalent $\sts$-connection vacuum Einstein fields $\mathcal{F}_{Einst}=(\modl ,\conn)$ in the relevant Einstein category $\catf_{Einst}$.}

\vskip 0.1in

On the other hand, the syncopated vertical map from Level 4 to Level 5 in our top-down tower diagram above:

\begin{equation}\label{d9}
\mu_{\sts_{1}}\equiv{\int_{\sconn(\modl_{1}) /\aut_{\sts_{1}}\modl_{1}}}:\, (\modl_{1},\curv_{1}(\conn_{1}))_{X})\dashrightarrow\int_{{\sconn(\modl_{1}) /\aut_{\sts_{1}}\modl_{1}}}tr(\curv_{1}(\conn_{1}))
\end{equation}

\noindent corresponds to a Radon-type of measure map on the affine moduli space $\sconn(\modl_{1})/\aut_{\sts_{1}}\modl_{1}$ of $\aut_{\sts_{1}}\modl_{1}$-gauge equivalent dynamical $\sts_{1}$-connection fields implementing the Einstein-Hilbert dynamical action functional on the said moduli space of ADG-gravitational $\sts_{1}$-connection fields  \cite{mall1,malrap3, mall4}.

\vskip 0.1in

\item\underline{\bf Horizontal Level 6:} The sixth storey (level) in our top-down tower involves the natural transformation type of morphism $\mathcal{N}_{\sts}$ between the dynamical action functionals relative to two different structure sheaves at the previous level lifting to the following morphism:

\begin{equation}\label{d10}
\hat{\mathcal{N}}_{\sts}:\, \int_{{\sconn(\modl_{1}) /\aut_{\sts_{1}}\modl_{1}}}tr(\curv_{1}(\conn_{1}))\longrightarrow\int_{{\sconn(\modl_{1}) /\aut_{\sts_{2}}\modl_{2}}}tr(\curv_{2}(\conn_{2}))
\end{equation}

\noindent which represents a natural transformation-type of functor linking the Einstein-Hilbert action functional holding on vector sheaf $\modl_{1}$ relative to (or derived from) $\sts_{1}$, to the Einstein-Hilbert action functional holding on $\modl_{2}$ relative to (or derived from) $\sts_{2}$.

\vskip 0.1in

On the other hand, the syncopated vertical map from Level 5 to Level 6 in our top-down tower diagram above:

\begin{equation}\label{d11}
\delta\aconn_{1}:\, \int_{{\sconn(\modl_{1}) /\aut_{\sts_{1}}\modl_{1}}}tr(\curv_{1}(\conn_{1}))\dashrightarrow\curv_{1}(\modl_{1})=0
\end{equation}

\noindent represents the variation of the Lagrangian Einstein-Hilbert action functional with respect to the local ADG-gravitational $\sts_{1}$-connection field variable that yields as an extremum expression the vacuum Einstein equations holding on the vector sheaf $\modl_{1}$.

\vskip 0.1in

\item\underline{\bf Horizontal Level 7:}  The penultimate seventh storey (level) in our top-down tower involves the natural transformation type of morphism:

\begin{equation}\label{d12}
\hat{\mathcal{N}}_{\modl}:\, \curv_{1}(\modl_{1})=0\longrightarrow\curv_{2}(\modl_{2})=0
\end{equation}

\noindent which, in turn, represents an $\sts$-functorial map between the vacuum Einstein equations holding on the vector sheaves $\modl_{1}$ (relative to $\sts_{1}$) and $\modl_{2}$ (relative to $\sts_{2}$) in their respective vacuum Einstein curvature spaces $\mathcal{S}_{1}$ and $\mathcal{S}_{2}$.

\vskip 0.1in

On the other hand, the {\em faintly syncopated} last vertical map from Level 6 to Level 7 in our top-down tower diagram above:

\begin{equation}\label{d13}
\mathcal{QPI}\equiv{e^{\imath S_{EH}/\hbar}}:\, \curv_{1}(\modl_{1})=0\dashrightarrow e^{\frac{\imath}{\hbar}\int_{\sconn(\modl_{1}) /\aut_{\sts_{1}}\modl_{1}}[tr(\curv_{1}(\conn_{1}))]d\mu[\aconn_{1}]}
\end{equation}

\noindent represents the application of a `fiducial' (:still not explicitly constructed) {\em Quantum Path Integral} ($\mathcal{QPI}$) involving a quantal Radon type of measure on the affine moduli space of $\aut_{\sts}\modl$-gauge equivalent ADG-gravitational connection fields that is expected to ultimately yield a {\em fully $\sts$-invariant} (:$\sts$-functorial, $\aut_{\sts}\modl$-covariant and $\aut_{\sts}\modl$-invariant) {\em quantum gravitational ADG-theoretic vacuum Einstein dynamical equations}. 

\vskip 0.1in

\item\underline{\bf Horizontal Level 8:} Finally, the last storey (level) in our top-down tower involves the natural transformation type of morphism $\hat{\mathcal{N}}_{QPI}$\footnote{The subscript `{em QPI}' standing for `{\em Quantum Path Integral}'.} between envisaged quantum dynamical path integral action functionals over the corresponding moduli spaces of gauge-equivalent $\sts$-connections relative to different structure sheaves (:$\sts_{1}$ and $\sts_{2}$, respectively), as follows:

\begin{equation}\label{d14}
\hat{\mathcal{N}}_{QPI}:\,e^{\frac{\imath}{\hbar}\int_{\sconn(\modl_{1}) /\aut_{\sts_{1}}\modl_{1}}[tr(\curv_{1}(\conn_{1}))]d\mu[\aconn_{1}]}\longrightarrow e^{\frac{\imath}{\hbar}\int_{\sconn(\modl_{2}) /\aut_{\sts_{2}}\modl_{2}}[tr(\curv_{2}(\conn_{2}))]d\mu[\aconn_{2}]}
\end{equation}

\end{itemize}

\subsubsection{The Principle of Algebraic Relativity of Differentiability Distilled}

In closing this subsection on the top-down tower of ADG's architectonic backbone skeleton of basic concepts and structures supporting the {\it aufbau} of ADG-gravity presented above, we give a concise statement of the {\em Principle of Algebraic Relativity of Differentiability} (PARD) in the form of a Fundamental Apophthegm, our fifth one:

\vskip 0.1in

\noindent\fbox{%
    \parbox{\textwidth}{%
\noindent\underline{\bf Fundamental Apophthegm (5):} The Principle of Algebraic Relativity of Differentiability (PARD) maintains that one can naturally transform (:change or switch) from one structure sheaf $\sts_{1}(X)$ (or $\cons$-algebraised space $\cas_{1}$ or even differential triad $\triad_{1}$ in their respective categories $\catcas$ and $\ctriad$) on a fixed and in principle arbitrary base topological space $X$, to another one $\sts_{2}$ (or $\cons$-algebraised space $\cas_{2}$, or even differential triad $\triad_{2}$) on the same base space $X$, and the entire inherently algebraic ({\it i.e.}, $\sts$-based and derived) ADG-theoretic differential geometric technical machinery and structural mechanism remains intact (:invariant). This PARD is secured and guaranteed to hold due to the aforementioned Mallios' $\sts$-functoriality and $\sts$-invariace ($\AI$) of the said differential geometric mechanism of ADG, which is in turn manifested by the $\sts$-functorial natural transformation type of morphisms at each horizontal level of the Top-Down Tower of the architectonic skeletal backbone of ADG-categorical structures supporting the {\it aufbau} of the entire ADG-theoretic edifice, as it was ostensibly depicted and explained above. 
}%
}

\vskip 0.1in

\noindent An immediate corollary apophthegma, which we explained earlier in the guise of the two successful applications of ADG to Rosinger's {\em Spacetime Foam Dense Singularities of Generalised Functions} \cite{malros1,malros2,malros3} and to this author's {\em Finitary, Causal and Quantal Vacuum Einstein-Lorentzian Gravity} \cite{rap1,rap2,malrap1,malrap2,malrap3,rap5,rap11,rap14,rap15}, its associated {\em Finitary-Algebraic Resolution of the Inner Schwarzschild Singularity} in \cite{rap7} and its {\em Sheaf Cohomological Third Quantisation} scenario in \cite{rap13}, is the following:

\vskip 0.1in

\noindent\fbox{%
    \parbox{\textwidth}{%
\noindent\underline{\bf Corollary Apophthegm (6): The Physical Utility of PARD} One can naturally transform (:change or switch functorially) from a given structure sheaf $\sts_{1}$ of generalised arithmetics or coordinates in which the vacuum Einstein ADG-gravitational field law, which is expressed as a differential equation proper via the curvature $\sts$-morphism $\curv(\conn)$ as in (\ref{eq9}), may appear to be `singular', `problematic', or `difficult' to handle ({\it e.g.}, by the manifold based and CDG-dependent means, when $\sts_{1}\equiv\smooth_{M}$ and the base space $X$ is a differential manifold $M$), to another one $\sts_{2}$, which is better suited to the problem or issue in hand, while the entire ADG-theoretic mechanism---as well as the physical connection field law that it supports---remains intact and still in force, {\em without `suffering' any loss, let alone breakdown, of differential geometric structure whatsoever}. That is, we are still able to do Differential Geometry in the very presence of singularities, infinities and other differential geometric anomalies and pathologies, which only appeared to be insuperable analytic obstacles and irrerarable breakdown effects of the physical field law of gravity \cite{clarke3,clarke4}, when the latter is formulated by the smooth geometrical spacetime manifold based means of CDG (:$\sts_{1}\equiv\smooth_{M}$) \cite{malrap3,rap5,rap7,rap13,rap14,rap15,rap19}. Equivalently stated, all the apparent differential geometric anomalies and pathologies inherent in $\sts_{1}\equiv\smooth_{M}$ can be `{\em naturally transformed away}'---{\it i.e.}, directly evaded simply by naturally transforming the whole inherently algebraic ADG-theoretic differential geometric mechanism to a different, more `suitable' (:less pathological and better suited to a physical situation in hand) structure sheaf of generalised coordinates $\sts$. 
}%
}

\vskip 0.1in

\noindent\underline{\bf Important Note 10:} In connection with the natural transformation type of `{\em change of structure sheaf}' morphism $\mathcal{N}_{\sts}$ above, we note that the category-theoretic epithet {\em natural} in front of transformation, apart from its usual mathematical (:category-theoretic) meaning as `{\em a functor preserving functors}' \cite{macmo}, in ADG-gravity it has the additional {\em physical} meaning and significance of {\em preserving the differential geometric form of the vacuum Einstein $\sts$-connection field equations that mathematically represent the {\em Natural} law of gravity,\footnote{More commonly referred to as the {\em Physical} law of gravity. {\em Physis} means {\em Nature} in both Ancient and Modern Greek.} regardless of what structure sheaf $\sts$ of generalised arithmetics, coefficients or theoretical generalised local coordinate measurements one employs in order to localise (:locally `measure' or `gauge') and thus `coordinatise' the ADG-theoretic vacuum Einstein $\sts$-connection field} $\mathcal{F}_{Einst}=(\modl ,\conn)$.\footnote{Always relative to a chosen system $\gauge_{i}$ of open local gauges covering $X$.} Thus, we arrive at our seventh apophthegm: 

\vskip 0.1in

\noindent\fbox{%
    \parbox{\textwidth}{%
\noindent\underline{\bf Apophthegm 7:} The {\em naturality} of $\mathcal{N}_{\sts}$ reflects its {\em physicality}: on the one hand, as a categorical, natural transformation type of morphism, it preserves the homological-algebraic (:sheaf and category-theoretic) differential geometric structure and mechanism of ADG (in the respective categories), and on the other, it preserves the {\em `differential geometric form'} of the ADG-vacuum Einstein field equations that represent the physical vacuum Einstein gravitational field law. In line with the $\sts$-functoriality and Mallios's principle of $\sts$-invariance ($\AI$) that we saw earlier, the form of the physical law remains invariant no matter what structure sheaf $\sts$ one chooses to employ in order to theoretically locally coordinatise and measure ({\it i.e.}, locally gauge) the vacuum Einstein ADG-gravitational $\sts$-connection field $\mathcal{F}_{Einst}=(\modl ,\conn)$. {\em Natural transformations} (in a categorical sense) are {\em physical changes}, in the physical sense that they preserve the form of  the dynamical physical laws as the aforementioned Fundamental Ur-Principle of $\sts$-Relativity ($\AR$) requires.              
}%
}

\vskip 0.1in

\noindent\underline{\bf A Weak Analogy from GR about Preferred Coordinate Frames.} A loose analogy with the local Principle of Relativity in GR \cite{einst3,einst4,mtw} is that there one can switch (transform) to a particular coordinate system---a so-called `locally inertial' (or free-falling) frame---in which the gravitational field is `gauged' or `transformed away' (thus manifesting the local Principle of Equivalence).\footnote{This corresponds to the fact that {\em locally, the curved spacetime of GR can be reduced to the flat Minkowski space of SR}---or geometrically pictured: {\em at any point of the curved smooth spacetime manifold of GR, the tangent space is isomorphic to flat Minkowski space and the gravitational field, which is represented by the smooth spacetime metric $g_{\mu\nu}$, can be `diagonalised' to the Minkowski one $\eta_{\mu\nu}=diag(-1,1,1,1)$ relative to a locally inertial frame.} \cite{mtw}.} Among all the possible general coordinate frames, the local inertial ones are preferred for locally `factoring out' or `gauging away' the gravitational field.

\vskip 0.1in

\noindent\underline{\bf A Stronger Analogy from GR about Preferred Coordinate Frames.}  An even stronger analogy, again coming from GR, is the case of the outer, so-called {\em coordinate}, Schwartzschild singularity of the spherically symmetric gravitational field of a point-mass \cite{df,mtw}. There, an ingenious switch from the usual cartesian or spherical coordinates to the so-called Eddington-Finkelstein frame, shows that the exterior Schwartzschild singularity is not a real or genuine gravitational singularity---a site where the gravitational field breaks down or blows up without bound, but rather that acts as a unidirectional membrane (:black hole horizon) across which physical signals can travel only one way, towards the black hole interior \cite{df,mtw,rap5}. Among all the possible general coordinate frames, the Eddington-Finkelstein ones are `suitable' or `preferred' for revealing the black hole horizon nature of the exterior Schwartzschild singularity, which is thus merely a so-called {\em coordinate singularity}.

\vskip 0.1in

\noindent\underline{\bf The Strongest Analogy from ADG-gravity about Preferred Coordinate Frames.} {\it A fortiori}, and as briefly alluded to earlier, in \cite{rap5} we apply the ADG-machinery to a finitary (:locally finite), causal and quantal setting already developed in \cite{malrap1,malrap2,malrap3} to show that a finitary version of vacuum Einstein-Lorentzian gravity---when one switches from the structure sheaf $\sts\equiv\smooth_{M}$ of smooth functions on a base differential spacetime manifold, to finitary spacetime sheaves \cite{rap2} of differential incidence algebras \cite{rapzap1,rapzap2} associated with Sorkin's finitary poset substitutes of continuous (:$\cont$-topological) manifolds \cite{sork0}---still holds intact, and in no way ({\it i.e.}, in any differential geometric sense) the field law of gravity breaks down or blows up to infinity (in the usual analytic sense of \cite{clarke3,clarke4}), even in the immediate topological vicinity (neighbourhood) of the inner (:black hole) singularity of the point mass source of the gravitational field. Moreover, even in generalised function spaces teeming with singularities of the most unmanageable kind from the CDG-theoretic perspective, like in {\em Rosinger's algebra sheaf of spacetime foam dense singularities} $\sts_{nd}$ mentioned earlier \cite{malros1,malros2,malros3,mall10}, the ADG-gravitational field equations (\ref{eq9}) are still seen to be in full force and they do not break down in any differential geometric sense. They only seem to do so from the analytic perspective of the spacetime manifold based CDG supporting GR \cite{clarke3,clarke4}.

In the next subsection, we forget about the horizontal natural transformation type of mappings in the top-down diagram of ADG-architectonics above and we present an inverted, bottom-up diagram depicting again the architectonic {\it aufbau} of ADG-gravity which emphasises that it all starts from, and progressively builds on, the structure sheaf $\sts$ of generalised coordinates, as the present paper contends.

\newpage

\subsection{Diagram II: Linear 10-Storeys Bottom-Up Tower of the Basic Skeletal Backbone Structures Supporting the Aufbau of ADG-Gravity}


\begin{equation}\label{eqad7}
{\fontsize{0.13in}{0.13in}
\begin{CD}
\boxed{\mathbf{Level~9:}~\mathrm{Fully~Covariant~Quantum~Path~Integral~Quantisation~(\sts_{X},\conn ,\curv(\conn),\curv_{Einst}(\modl)=0, e^{\imath S_{EH}})}}\\
@AAA\\
\boxed{\mathbf{Level~8:}~\mathrm{Canonical~Sheaf~Cohomological~Third~Quantisation~of~Vacuum~Einstein~Gravity}}\\
@AAA\\
\boxed{\mathbf{Level~7:}~\mathrm{Solution~Curvature~Space~~(\sts_{X},\conn ,\curv(\conn),\curv_{Einst}(\modl)=0)}}\\
@AAA\\
\boxed{\mathbf{Level~6:}~\mathrm{Vacuum~Einstein~gravitational~equations:~\curv_{Einst}(\modl)=0}}\\
@AAA\\
\boxed{\mathbf{Level~5:}~\mathrm{Einstein-Hilbert~action~functional:~S_{EH}=\int_{X}tr(\curv_{Ric}(\conn))}}\\
@AAA\\
\boxed{\mathbf{Level~4:}~\mathrm{ADG~curvature~space:~\mathcal{S}=(\sts_{X},\conn ,\curv(\conn))}}\\
@AAA\\
\boxed{\mathbf{Level~3:}~\mathrm{ADG-gravitational~field:~\mathcal{F}_{Einst}=(\modl ,\conn)}}\\
@AAA\\
\boxed{\mathbf{Level~2:}~\mathrm{Vector~sheaf~\modl~and~curved~connection~\conn~on~it}}\\
@AAA\\
\boxed{\mathbf{Level~1:}~\mathrm{Flat~Kaehler-Cartan~differential~d\equiv\partial~and~differential~triad~\triad=(\sts_{X} ,\partial ,\Omg_{X})}}\\
@AAA\\
\boxed{\mathbf{Level~0:}~\mathrm{Structure~sheaf~\sts~and~\cons-algebraised~space~\cas=(X,\sts_{X})}}\\
@AAA\\
\boxed{\mathbf{Level~-1:}~\mathrm{The~arbitrary/general~base~topological~space:~X}}
\end{CD}}
\end{equation}

\vskip 0.05in

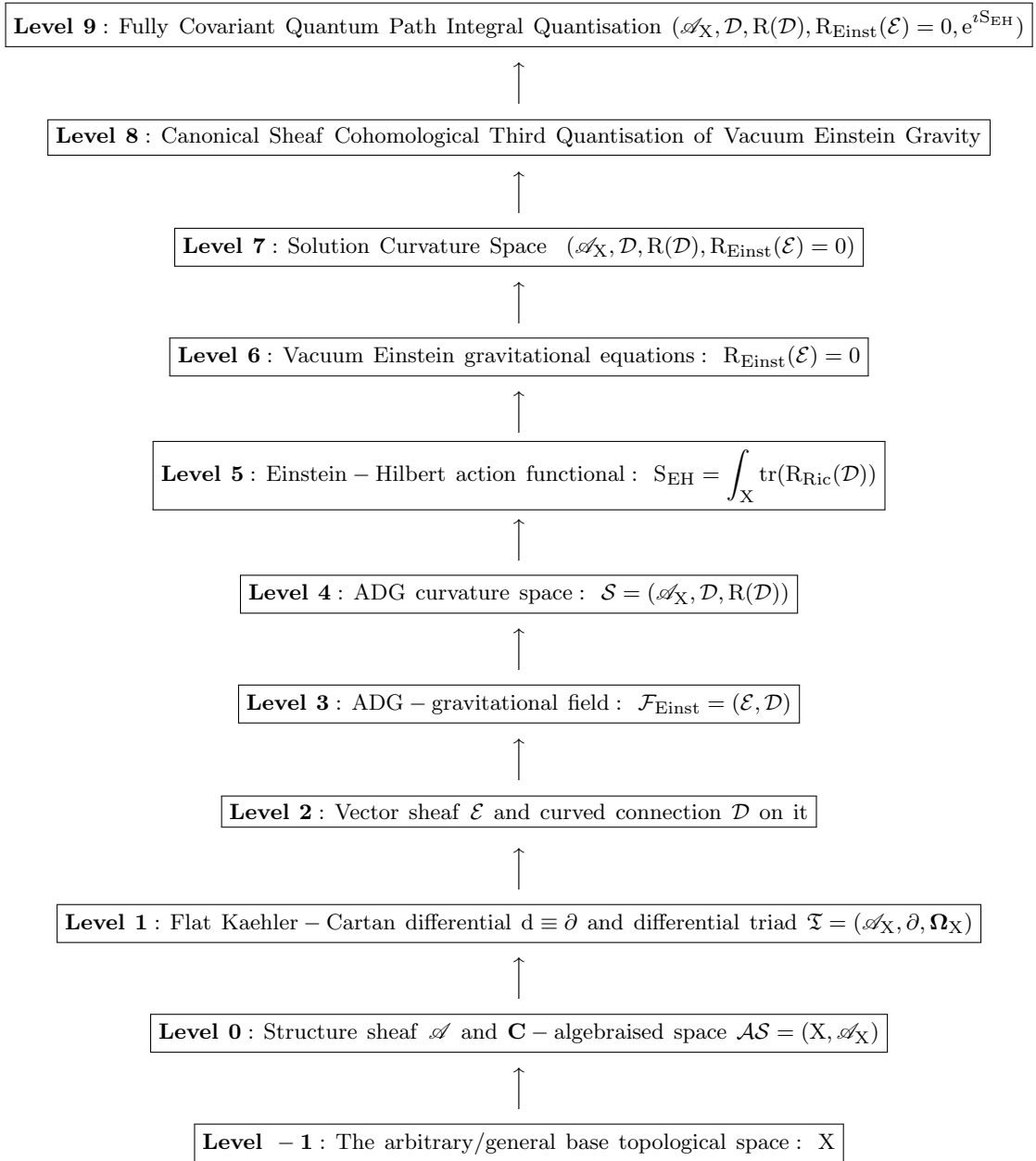
\centerline{\underline{\small\bf Diagram II: Vertical Bottom-Up Aufbau of ADG-Gravity}}

\newpage

A quartet of remarks are due here about the diagram depicting the bottom-up {\it aufbau} of vacuum Einstein ADG-gravity above:

\begin{itemize}

\item The general and in principle arbitrary base topological space $X$ is relegated to `subterranean'  Level $-1$ as its sole functional role is as a kind of surrogate space for the sheaf theoretic localisation of our variable generalised local coordinate measurements in $\sts$ of the dynamically variable vacuum Einstein ADG-gravitational gauge $\sts$-connection field $\mathcal{F}_{Einst}=(\modl ,\conn)$. As such a surrogate base space, $X$ plays no actual role in the vacuum Einstein ADG-gravitational dynamics in (\ref{eq9}), but it secures that the ADG-gravitational dynamics is continuous with respect to its (:$X$'s) topology, no matter what topology {\em we} choose that $X$ has.\footnote{As noted at the start of the paper, one could in principle choose any base topological space for the sheaf-theoretic localisation and gauging of the generalised coordinate measurement algebras in $\sts$ and, {\it in extenso}, of the $\sts$-modules in $\modl$.}

\item At the fundamental, ground Level $0$ lies the structure sheaf $\sts$, which in turn defines a $\cons$-algebraised space, {\em on which the entire aufbau of ADG-gravity rests and from which the entire aufbau of ADG-gravity derives}, as the present paper contends and shows.\footnote{The tower effectively starts from $\sts$ and progressively builds the whole mathematical (ADG-theoretic) structure supporting ADG-gravity.}

\item The penultimate Level 8 at the top pertains to the autonomous (:self-closed, self-sustaining and external spacetimeless) sheaf cohomological 3rd quantisation theoretical scenario for vacuum Einstein gravity and free Yang-Mills theories originally proposed and developed in \cite{rap13} and recently refined in \cite{rap15}. We will revisit it briefly in the next section when we discuss the intrinsically and inherently quantal character of the vacuum Einstein ADG-gravitational field $\mathcal{F}_{Einst}=(\modl ,\conn)$.

\item At the very top (Level $9$) we find a still highly speculative fully covariant (path integral) quantisation scheme for vacuum Einstein gravity briefly mentioned earlier, which has already been anticipated to involve a Radon type of measure on the kinematical affine moduli space $\sconn(\modl) /\aut_{\sts}\modl$ of $\aut_{\sts}\modl$-gauge equivalent vacuum Einstein ADG-gravitational $\sts$-connections \cite{malrap3,rap13,rap14,rap15,rap19}. Alas, we still lack the full mathematical development of ADG-theoretic integration on such orbifold spaces of Mallios's gauge equivalent $\sts$-connections. One of our guiding principles in developing an abstract integration theory along ADG-theoretic lines is that the relevant integrals, effectuated by Radon functional type of measures on the relevant algebra and vector sheaves, should be, like the inherently algebraic mechanism of ADG, $\sts$-functorial---what Mallios coined $\sts$-invariant \cite{mall14,mall13,mall17,rap14,rap15,rap19}. 

\end{itemize} 

\vskip 0.1in

\section{Five `Innate' and `Intrinsic' Properties of the Vacuum Einstein ADG-Gravitational Field}

In this section we wish to highlight five fundamental characteristics of the vacuum Einstein ADG-gravitational field that, in the last section, we will contend that can go a long way in addressing and tackling a few central issues in current and future QG research. So far in this paper, we have established the following three `innate' and `intrinsic' properties of the vacuum Einstein ADG-gravitational gauge $\sts$-connection field $\mathcal{F}_{Einst}=(\modl ,\conn)$:

\begin{enumerate}

\item $\mathcal{F}_{Einst}$ is purely algebraic (relational), as it derives from the structure algebra sheaf $\sts$, or its associated differential $\sts$-module $\modl$, and it is defined homological algebraically (:categorically) as a sheaf morphism acting on the relevant algebra and vector sheaves' (local) sections;\footnote{The flat connection $\partial$ derives from and acts on $\sts$'s local sections as in (\ref{eq1}), while the curved connection $\conn$ derives from and acts on $\modl$'s local sections as in (\ref{eq3}), respectively.}

\item $\mathcal{F}_{Einst}$ is external (background geometrical) spacetime manifoldless, as it needs no base locally Euclidean space to support it, unlike the usual GR which is fundamentally based on a background pseudo-Riemannian spacetime manifold $M$;

\item $\mathcal{F}_{Einst}$ is purely gauge, as there is no background geometrical spacetime manifold external to it, carrying its own, external to  $\mathcal{F}_{Einst}$, $\mathrm{Diff}(M)$ symmetries. The local dynamical symmetries of the vacuum Einstein law (\ref{eq9}) that the field defines as a differential equation proper are purely internal or intrinsic to the field itself, being represented by the principal group sheaf $\aut_{\sts}\modl$ of  $\mathcal{F}_{Einst}$-automorphisms (relative to a chosen structure sheaf $\sts$ of generalised local coordinate measurements or gauge localisations of $\mathcal{F}_{Einst}$). These are the {\em dynamical auto-symmetries} or {\em self-invariances} of the vacuum Einstein ADG-gravitational field $\mathcal{F}_{Einst}$ and of the field law (\ref{eq9}) that it defines. 

\end{enumerate}    

\vskip 0.1in

\noindent Characteristics 1-3 above can be subsumed under the following apothegmatic characterisation of $\mathcal{F}_{Einst}$:   

\vskip 0.1in         

\noindent\fbox{%
    \parbox{\textwidth}{%
\noindent\underline{\bf Apophthegm 8:} {\em  The vacuum Einstein ADG-gravitational field $\mathcal{F}_{Einst}$ is a purely algebraic, purely gauge, external (:background geometrical) spacetime manifoldless and dynamically autonomous field system}. 
}%
}

\vskip 0.1in

Having established the three characteristics of the ADG-gravitational field above, in the following subsection we present two additional key {\em quantum characteristics} that, as we shall argue, are, like the dynamically autonomous, purely algebraic and purely gauge-theoretic character of $\mathcal{F}_{Einst}$ described above, {\em `inherent'}, {\em `innate'}, or perhaps better, {\em `intrinsic'}, built into the vacuum Einstein ADG-gravitational field $\mathcal{F}_{Einst}=(\modl ,\conn)$.\footnote{These three equivalent epithets---{\em `inherent'}, {\em `innate'} and {\em `intrinsic'}---that we give to the five fundamental characteristics of $\mathcal{F}_{Einst}$, can be justified as follows: (i) they are {\em `inherent'}, because $\mathcal{F}_{Einst}$ inherits them from the structure sheaf $\sts$ from which it derives; (ii) they are {\em `innate'}, because $\mathcal{F}_{Einst}$ is born with them from $\sts$; and (iii) they are {\em `intrinsic'}, because they are internal, algebraic or relational, structural characteristics within $\mathcal{F}_{Einst}$ itself, without dependence on or recourse (:reference) to any external or background to the dynamical field itself geometrical spacetime manifold structure.}  These two additional intrinsic quantum characteristics have been explored and developed in a long series of papers spanning more than two decades now \cite{mall5,mall6,malrap1,malrap2,malrap3,rap7,rap11,rap13,rap14,rap15,rap19}, from which we recall and borrow certain key ideas, structures and arguments in our presentation below.  

\subsection{Long Shots in the Quantum Deep: Two `Inherent' and `Intrinsic' Quantum Traits of Vacuum Einstein ADG-Gravity}

The two `innate' quantum traits of $\mathcal{F}_{Einst}=(\modl ,\conn)$ are:

\begin{enumerate}

\item The intrinsic, autonomous, self-contained and self-consistent {\em sheaf cohomological 3rd quantisation scenario} for the vacuum Einstein gravitational (and free Yang-Mills) ADG-fields that was originally explored in \cite{rap13} and further developed recently in \cite{rap15}. We will recall and outline its main tenets and characteristics in the next subsection.

\item In a subtle sense, which we are going to explicate and explain below, $\mathcal{F}_{Einst}$ can be regarded as an {\em $\sts$-closed quantum system} relative to our generalised local coordinate measurements in $\sts$ that, furthermore, because of its inherent spacetime manifoldlessness mentioned above, it draws no fundamental spacetime Planck scale `cut-off' nor it distinguishes or posits a theoretically {\it ad hoc} Heisenberg {\it schnitt} between a classical exosystem and a quantum endosystem like the (quantum state $\Psi$ in the) usual Copenhagen Interpretation of Quantum Theory (CIQT) does.

\end{enumerate}

\subsubsection{Canonical-Type of Sheaf Cohomological Third Quantisation of ADG-Gravity}

In this subsection, we recall directly from \cite{rap13, rap15} a couple of facts and results about the algebraic, canonical-type of sheaf cohomological 3rd ({\it i.e.}, external spacetime manifoldless) field quantisation of the ADG-gravitational field:\footnote{In the short list that follows, we do not give any explicit calculations, or detailed structures and technical arguments, about those facts and results. The reader is directly referred to the published papers \cite{rap13,rap15} for technical and conceptual details.}

\begin{enumerate}

\item In \cite{rap13,rap15}, the vacuum Einstein ADG-gravitational field $\mathcal{F}_{Einst}=(\modl ,\conn)$ is seen to be  {\em quantum `self-complementary'} (or quantum `self-dual'), in the following sense:

\item The local sections of the vector sheaf $\modl$ represent {\em local quantum particle position states} of the ADG-field. In turn, {\em $\modl$ is the associated (self-)representation sheaf of the principal group sheaf $\aut_{\sts}\modl$ of $\modl$'s automorphisms} \cite{mall1,mall4,vas1,vas2,vas3}, physically representing $\mathcal{F}_{Einst}$'s local dynamical internal (:gauge) self-symmetries that we saw earlier. Moreover, from a geometric pre-quantisation and 2nd quantisation perspective \cite{mall5,mall6,mall4,rap13}, the local sections of line sheaves (:vector sheaves of rank 1) represent local quantum particle states of Bosons, while the sections of vector sheaves of higher rank represent quantum particle states of Fermions.

\item The action of the $\sts$-connection field $\conn$ on $\modl$'s local sections represents {\em local (:differential) wave-momentum like changes of the said local quantum particle position states} :($\modl$'s local sections).\footnote{This is conceptually in much the same way that the momentum operator $\hat{p}=d/dx$ in the usual QT represents changes in the position operator $\hat{x}$.}

\item Certain local sheaf cohomological (matrix/operator valued) characteristic forms, that fully characterise locally $\modl$ and $\conn$, are seen to obey non-trivial canonical position-momentum type of local quantum commutation relations that close within the field itself.\footnote{In the sense that the said local quantum commutation relations close locally within $M_{n}^{\bullet}(\sts(U))$ ($U\subset X$), the noncommutative matrix group sheaf that is locally isomorphic to $\aut_{\sts}\modl(U)$---the noncommutative local principal transformation self-symmetry group sheaf of the vacuum Einstein gravitational field $\mathcal{F}_{Einst}$ and the dynamics (\ref{eq9}) that it defines, as we saw earlier.}

\item Here, it must be emphasised that, although our generalised local coordinate measurements of the vacuum Einstein ADG-gravitational field in $\sts$ are commutative,\footnote{As $\sts$ is an {\em abelian} algebra structure sheaf.} the aforesaid generalised position-momentum relations are noncommutative (:quantal). This is a generalised ADG-theoretic version of Bohr's Correspondence Principle (BCP) in the usual Copenhagen Interpretation of Quantum Theory (CIQT)---cf. especially \cite{mall12}, but also \cite{mall14,mall15,mall13,mall17}.

\item {\it A fortiori}, it must also be emphasised that all of the above is accomplished without invoking, or depending on, any external (:background) to the ADG-field itself geometrical spacetime manifold like in the usual (flat) QFTs of matter in Minkowski space or in the canonical (Hamiltonian) or the covariant path integral quantisation approaches to QGR.\footnote{This 3rd quantisation of ADG-gravity evades altogether the equal-time commutation relations imposed on spacelike hypersurfaces (:in $(3+1)$-global hyperbolically decomposed spacetime manifolds) between the gravitational field and its conjugate momentum field (in either the 2nd-order Einstein-Riemann or the 1st-order Palatini-Ashtekar formalisms) \cite{schulz,loop}.} Moreover, in the said sheaf cohomological canonical local quantum commutation relations, there does not appear any {\it ad hoc} assumed Planck constant ($\hbar$) or any other fundamental spacetime scale, as {\em there is no background spacetime geometry to begin with}. Further to this point, as there is no background spacetime geometry at all in our scheme, but only the purely algebraic (:relational), purely gauge and dynamically autonomous vacuum Einstein ADG-theoretic $\sts$-connection gauge field $\mathcal{F}_{Einst}=(\modl ,\conn)$, {\em there is no need for us to quantise spacetime geometry} like for example in the Loop QG scenario in which we find quantised expressions for spacetime area and volume. In canonical QGR, for instance, since the spacetime manifold based GR is regarded as a theory of gravity based on a curved spacetime geometry ({\it i.e.}, the curvature of spacetime, expressed via the spacetime metric $g_{\mu\nu}$, {\em represents} the gravitational field strength), it is expected that a quantisation of the gravitational field should inevitably `result' in, or be accompanied by, a concomitant quantisation of spacetime itself.

\item Due to the points above, we have coined the vacuum Einstein ADG-field $\mathcal{F}_{Einst}=(\modl ,\conn)$ {\em `inherently'  quantum} \cite{rap13,rap15,rap19}.\footnote{We have alternatively, but equivalently, called it  {\em `innately'} or {\em `intrinsically' quantum}, or even {\em self-quantised} or {\em self-quantum} (:quantum `self-dual' or `self-complementary').}

\end{enumerate}      

\subsection{Remarks on the Role of $\sts$ as the Module of Sheaf Cohomological Coefficients in ADG and ADG-gravity}

In this subsection, and in view of our remarks above about the sheaf cohomological 3rd quantisation scenario for ADG-gravity, we discuss a deep and significant relation between the usual role that the coefficient structure $\mule$ plays in the usual cohomology theories in both Algebraic and Differential Geometry,\footnote{Normally, $\mule$ is taken to be a {\em module} and referred to as the {\em coefficient module} in cohomology theory.} and our {\em structure algebra sheaf $\sts$ of generalised arithmetics or coefficients} as Mallios originally coined them in the {\it aufbau} of the general mathematical theory of ADG \cite{mall1,mall2,mall4}.\footnote{As noted at the very beginning of the present paper, in a forthcoming paper \cite{rapX} we draw close similarities and affinities between Grothendieck's work in Algebraic Geometry and Mallios's work in Differential Geometry, both resting on and issuing from the fundamental conception of {\em a structure sheaf as the cornerstone on which the notion of `Geometric Space'---algebraic or differential---is founded} \cite{groth1,mall1,mall2,mall4}, hence the title of the present paper.}

As we saw in this paper, $\sts$ is physically interpreted as the abelian algebra sheaf of theoretical local coordinate measurements or gauge localisations of the ADG-gravitational field $\mathcal{F}_{Einst}=(\modl ,\conn)$ in all applications so far of ADG to vacuum Einstein gravity, including the present paper. {\it A fortiori}, this abelian algebra sheaf $\sts$ has hitherto been seen on the one hand to be the topologico-algebraic source of the fundamental differential geometric structure of ADG  ({\it i.e.}, the source of the algebraic $\sts$-connection $\conn$), the origin of ADG-gravity's dynamical $\sts$-functoriality, $\sts$-invariance ($\AI$) and $\aut_{\sts}\modl$-covariance, and, due to its commutativity, we also saw that it represents the generalised outcomes ($c$-numbers) of our theoretical local field coordinate measurements implementing an ADG-theoretic version of Bohr's Correspondence Principle \cite{mall14,mall15,mall13,mall12,mall17}.\footnote{The original Mallios denomination of $\sts$ as {\em generalised arithmetics} is very suitable in regarding $\sts$ as a generalised version of Bohr's $c$-cumbers---the generalised numerical outcomes (:numbers) of quantum measurements (:local gauge coordinatisations) of the ADG-gravitational field.}

\vskip 0.1in

Here is a brief outline of the main roles that the module of coefficients $\mule$ plays in the usual (sheaf) cohomology theories in Algebraic and Differential Geometry:

\begin{enumerate}

\item In the usual (singular) cohomology theories with fixed or constant $\mule$ of coefficients like $\Z$ or $\Q$ and a fixed abelian cohomology group $G$, the coefficient module and the group convey important geometrical (topological) information about the underlying space;

\item Sheaf cohomology is a generalisation of the usual cohomology theories with constant coefficient module $\mule\equiv G$, whereby now $\mule$ is a {\em sheaf of coefficients} (normally, a sheaf of abelian groups, rings or algebras) usually referred to as the {\em structure sheaf} (:our $\sts$ in ADG), that keeps track of what local data are glued together to form cohomology groups.

\item By using variable (:a sheaf!), instead of constant, coefficients, sheaf cohomology uses the local data provided by the local sections of the coefficient module structure sheaf $\mule\equiv\sts$ to determine whether a problem that can be solved locally can also be solved globally. In other words, sheaf cohomology, by contrast to cohomology with constant coefficients, determines or measures the obstructions to solving a problem globally by using local information encoded in the local sections of $\mule\equiv\sts$.

\item Sheaf cohomology is richer and more flexible than constant coefficients' (:singular) cohomology as it uses the structure sheaf $\mule\equiv\sts$ and the cohomology groups to detect and characterise the topological and the geometric properties of the underlying space, such as its characteristic (topological) invariants like dimension and other important (topological) properties like connectedness, the presence of holes and other global obstructions, {\it etc}.

\item The choice of coefficient sheaf module $\mule$ (:structure sheaf $\sts$) in a cohomology theory is crucial, because it influences the calculated cohomology groups hence it determines the nature of the invariants being measured, affecting how the theory behaves with different algebraic structures and what properties of the underlying space it can capture. {\it In toto}, {\em different choices of $\mule$ result in different properties and geometrical/topological invariants of the underlying space being revealed and computed}. 

\item The remark above brings us to the most important observation about sheaf cohomology in view of our present paper, namely that: {\em different choices of coefficient module $\mule$ or structure sheaf $\sts$} lead to to different algebraic structures and thus result in different invariants being computed. Also, a specific or particular choice of coefficient module sheaf $\mule\equiv\sts$ may be more suitable or appropriate for revealing specific geometric (topological) properties of the underlying space. By using different types of algebraic coefficient structure sheaves, one can probe different aspects of the underlying space's structure. For instance, a particular choice of $\mule\equiv\sts$ may reveal topological obstructions in the base space, while another one may not. 

\end{enumerate}

\noindent In addition to the remarks above, we read from \cite{AI1}:

\bigskip\noindent (Q3)\hskip 0.9in
\begin{minipage}{11cm}
\noindent ``In essence, selecting coefficient sheaf for a cohomology theory is like choosing the measuring tool. You select a tool [a specific structure sheaf] that is appropriate for the task of revealing the specific topological or algebraic properties you are interested in.''
\end{minipage}

\vskip 0.1in 

\noindent In other words: 

\vskip 0.1in

\noindent\fbox{%
    \parbox{\textwidth}{%
\noindent\underline{\bf Apophthegm 9:} Different choices of $\mule\equiv\sts$ result in measuring, calculating or revealing different geometric aspects and properties of the underlying space. {\em Different structure sheaves suitable for different spaces and geometric purposes}.
}%
}

\vskip 0.1in

\subsubsection{Close affinity with the PARD of ADG-Gravity, but with a Difference}

Apophthegm 9 above dovetails perfectly with our Apophthegm 5 earlier, which effectively underlies the Principles of $\sts$-Relativity ($\AR$) and $\sts$-Invariance ($\AI$) of ADG-gravity, according to which one may be able to change or switch to a structure sheaf $\sts$ more suitable or appropriate to the physical problem under focus, like for example our employment of finitary spacetime sheaves of incidence algebras to formulate a locally finite version of vacuum Einstein gravity \cite{malrap1,malrap2,malrap3} and thus evade the interior Schwartzschild singularity \cite{rap5}, or even to formulate the vacuum Einstein equations on spaces that are teeming with everywhere dense singularities by employing as structure sheaf $\sts$ Rosinger's algebras of generalised functions (:non-linear distributions) \cite{malros1,malros2,malros3}, and still retain intact the entire $\sts$-functorial ADG-theoretic differential geometric mechanism. This is the essence of Mallios's quote (Q2) given earlier.

The key difference, however, is that the $\sts$-functoriality and $\sts$-invariance of the entire ADG-theoretic differential geometric mechanism guarantees that the ADG-theoretic vacuum Einstein equations will still hold intact, and will not break down in any differential geometric sense, in spite of the geometry (:topology) of any base topological space $X$ employed. As we have emphasised throughout our works over the years, {\em ADG-gravity is a genuinely and completely background independent theory}, as the base space $X$ does not contribute at all to the essentially algebraic in character differential mechanism of ADG, which derives from the stalks of the algebra ($\sts$) and the vector ($\modl$) sheaves employed, and not from their base, merely topological in character, space $X$.

The discussion above leads us to our tenth Apophthegm:   

\vskip 0.1in

\noindent\fbox{%
    \parbox{\textwidth}{%
\noindent\underline{\bf Apophthegm 10:} The choice of structure sheaf $\sts$ (or sheaf module of coefficients $\mule$) affects the [geometrical (topological) properties of the] underlying space one chooses to study. On the other hand, given that the inherently algebraic differential geometric mechanism of ADG originates from $\sts$ and it is {\it a fortiori} $\sts$-invariant (or $\sts$-functorial) according to the principles of PARD, $\AR$ and $\AI$, the said inherently algebraic differential mechanism of ADG is independent of the space (and the properties thereof) one chooses to study. ADG, unlike CDG, is not a theory of differential geometry that derives from, and essentially requires the mediation and geometrical intrpretation afforded by, a background locally Euclidean geometrical space(time) manifold. Rather, it is a purely relational (:algebraic), more Leibnizian rather than Newtonian, type of Differential Calculus that derives directly from the algebraic relations (:structure) between the `{\em geometrical objects}'---the (local sections of the) structure algebra sheaves $\sts$ and their associated differential $\sts$-module vector sheaves $\modl$---that live on a surrogate base topological space $X$, which is devoid of any physical significance as it does not partake in the $\sts$-invariant vacuum Einstein ADG-gravitational dynamics in (\ref{eq9}).
}%
}  

\vskip 0.1in     

\noindent We now turn to the second intrinsic quantum trait of the vacuum Einstein ADG-field $\mathcal{F}_{Einst}=(\modl ,\conn)$: its {\em $\sts$-closed quantum system} character.

\subsection{The $\sts$-Closed Quantum System Character of the ADG-Gravitational Field}

In this subsection we show and argue that the vacuum Einstein ADG-gravitational field $\mathcal{F}_{Einst}$ can be regarded as an `{\em observationally autonomous physical entity}', {\it alias}, an {\em intrinsically $\sts$-closed quantum physical system} with respect to our general local coordinate measurements of it in $\sts$.

To that end, below we present a horizontally flowing bottom-up diagram-{\it cum}-flowchart depicting {\em the flow of unfolding of ADG-gravity: from the structure sheaf $\sts$ of our generalised coordinates from which the $\sts$-connection $\conn$ originates, all the way to the $\sts$-invariant, $\aut_{\sts}\modl$-covariant and $\sts$-functorial vacuum Einstein ADG-gravitational equations}, together with the interplay of all the fundamental ADG-theoretic structures that we depicted in the last two architectonic {\it aufbau} towers (Diagrams I and II) in the previous section.

\subsubsection{Diagram III: Horizontal Bottom-Up Diagram-Flowchart of the Unfolding and Aufbau of Vacuum Einstein ADG-Gravity} 

Below, the reader will find a horizontal, {\em directed from left-to-right}, diagrammatic flowchart type of depiction of the structural flow of ADG-gravity: from its source in $\sts$ on the left, to the derivation of the vacuum Einstein equation on the right.

\vskip 0.5in

\begin{tikzcd}[column sep=tiny]
& (X,\sts ,\partial)\arrow[dr, dotted, "{flat}:\, d\circ\partial \equiv d^{2}=0" description]\ar[drr, "\sts -Measurement"', "\sts -Invariance/ \aut_{\sts}\modl -Covariance", bend left=20]
&
&[1.5em] \\
(X,\sts) \ar[ur, "\partial"'] \ar[dr, "\sts^{n}"]
&
& \mathcal{S}:=(\sts,\conn ,\curv(\conn)) \ar[r, dashed, "S_{\mathcal{E}\mathcal{H}}"]
& \int_{X}tr(\curv_{Ric}(\conn))dx\;\stackrel{\delta\aconn}{\mapto}\;\curv_{Einst}(\modl)=0\\
& \mathcal{F}:=(\modl ,\conn) \ar[ur, "{curved}:\, \conn\circ\conn\equiv\conn^{2}:=\curv(\conn)\not=0" description]\ar[urr, "\otimes_{\sts}-Functoriality", "\geom-Geomorphism"', bend right=20]
&
&
\end{tikzcd}

\vskip 0.1in

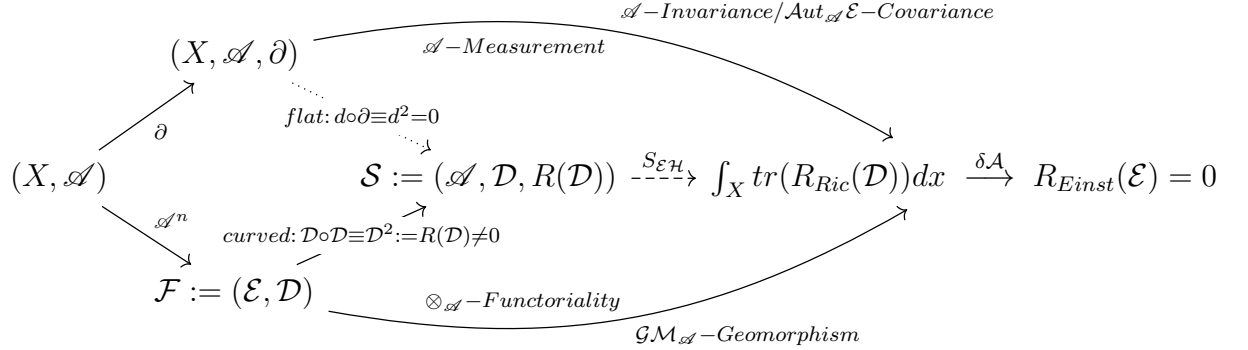
\centerline{\underline{\small\bf Diagram III: Horizontal Left-to-Right Aufbau of ADG-Gravity}}

\vskip 0.1in

\noindent {\bf Key Observation:} From Diagram III above, we especially highlight the occurrence of the {\em $\sts$-Coordinates' Measurement} process, leading to $\sts$-Invariance and $\aut_{\sts}\modl$-Covariance (top bent arrow side) of the ADG-gravitational dynamics and their equivalent functorial expressions in the guise of $\sts$-Functoriality and the $\geom:=(\otimes_{\sts},Hom_{\sts})$-geometric morphism (bottom bent arrow side) defined in (\ref{eqGM}) earlier.

\vskip 0.5in

\subsubsection{The Dynamically Autonomous, $\sts$-Circular and $\sts$-Closed Quantum System Character of the ADG-Gravitational Field $\mathcal{F}_{Einst}=(\modl ,\conn)$}

In this subsection, we want to capitalise on the following telling observation that issues from our foregoing exposition about the structure sheaf $\sts$ of generalised coordinates, and in a more ostensive way, from Diagrams I, II and especially Diagram III above:

\vskip 0.1in 

\noindent\fbox{%
    \parbox{\textwidth}{%
\noindent\underline{\bf Fundamental Observation (FO) about $\sts$:} On the one hand, as we showed and argued in the present paper, the structure sheaf $\sts$ is the origin and source of the gravitational $\sts$-connection field $\conn$, while the physical law (dynamics) that the latter defines as a differential equation (\ref{eq9}) via its curvature $\curv(\conn)$ is $\sts$-invariant. Equivalently, the dynamics is $\sts$-functorial, $\aut_{\sts}\modl$-covariant and geometric morphism $\geom$-functorial. On the other hand, as we have emphasised throughout this paper (cf. Important Note 0 earlier) and our work on applying ADG to vacuum Einstein gravity (and free Yang-Mills theories), $\sts$ physically represents ({\it i.e.}, it is physically interpreted as) the sheaf of generalised local coordinate measurements or local gauge determinations of the gravitational $\sts$-connection field in much the same way that the structure sheaf $\sts\equiv\smooth_{M}$ of smooth functions on the differential spacetime manifold $M$'s point events represents the local smooth coordinate measurement values of the ten gravitational potentials-entries of the smooth spacetime metric $g_{\mu\nu}$ relative to an atlas of local coordinate patches (:charts) covering $M$ (cf. opening quote (Q1) from \cite{fragpap}). 

\vskip 0.05in

We thus observe the following {\em fundamental $\sts$-cycle in ADG-gravity}: {\em we start from our generalised local coordinate measurements of the ADG-gravitational field in $\sts$ and we end up with an $\sts$-invariant gravitational dynamics}, a dynamics that respects and, conversely, it is respected by, our generalised local coordinate measurements of the vacuum Einstein ADG-gravitational field $\mathcal{F}_{Einst}=(\modl ,\conn)$ in $\sts$.            
}%
}

\vskip 0.1in   

\noindent We can picture the FO above by the following {\em $\sts$-cyclic `Ouroboros' diagram}:\footnote{In Ancient Greek, the word `{\em ouroboros}' means `{\em tail-biting}', a fitting name for the self-referential and self-closing $\sts$-cycle in ADG-gravity.}    

\begin{equation}\label{eq23}
\begin{tikzcd}
{\mathrm{Our\, Local\, Measurements}}\equiv\sts\!\stackrel{\bigoplus_{i=1}^{n}}{\longrightarrow}\!\modl\simeq_{\mathrm{loc.}}\!\sts^{n} \arrow[r,"\conn"] &
  (\modl ,\conn) \arrow[dr,"\conn\circ\conn"] \\
& \curv(\conn)|_{\modl}=0 \arrow[ul,dashed,bend left,"\aut_{\sts}\modl-{\mathrm{invariance}}"] & \arrow[l,"\delta S_{EH}"] \curv(\conn)
\end{tikzcd}
\end{equation}

\vskip 0.1in

\centerline{\underline{\small\bf Caption: `Ouroboros' $\sts$-Cycle Diagram 1 (OD1)}}

\vskip 0.1in

We now draw a cyclic (:circular) diagram that matches step-by-step our `Ouroboros' diagram (OD1) in (\ref{eq23}) above:

\begin{center}
 \smartdiagram[circular diagram]{1. Draw Physical Measurements~$\sts$,2. Draw Connections (Differences)~$\conn$,3. Draw Geometrical `Patterns'~$\curv(\conn)$,4. Draw Physical Laws (`Regularities')~$\curv(\conn)=0$}
 \end{center}

\vskip 0.1in

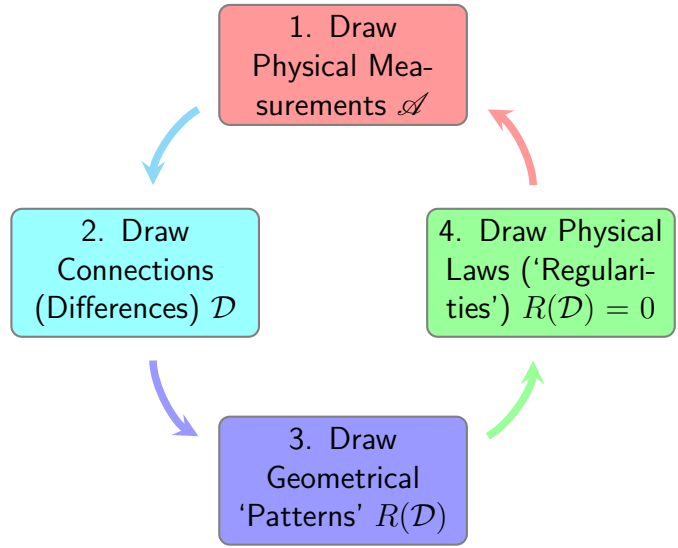
\centerline{\underline{\small\bf Caption: `Ouroboros' $\sts$-Cycle Diagram 2 (OD2)}}

\vskip 0.2in

In what follows, we explain, always from an ADG-theoretic perspective, the $\sts$-circular `self-referential' diagram (OD2) above and we juxtapose it by analogy to our first `Ouroboros' diagram (OD1) in (\ref{eq23}):

\vskip 0.2in

\begin{enumerate}

\item At the basis, we have {\em our} theory of the physical fields as partaking in the dynamical Laws of Nature that {\em we} discover by our observations of those dynamical fields and wish to mathematically describe, represent or model by (differential) geometric means---tacitly assuming, of course, that the said physical laws are local, thus that they should be somehow mathematically modelled after {\em differential} equations.

{\it En passant}, we note that in Greek, `{\em theory}' (:$\theta\epsilon\omega\rho\iota\alpha$) means `{\em generalised observation}', `{\em a general, epoptic way of looking at things}'. In our ADG-gravity, as we showed and argued earlier, {\em our entire theory of ADG-gravity is based on $\sts$}, hence our appellation of the structure sheaf $\sts$ as {\em our theoretical generalised local coordinate measurements} of the gravitational $\sts$-connection field $\conn$.

This theory, following Bohr's Correspondence Principle (BCP) dictum in the standard CIQT---which in turn fundamentally splits the World into a classical exosystem (:observer) and a quantum endosystem (:observed)---should involve commutative $c$-number like entities living in abelian algebras $\aconn$\footnote{Not to be confused with the connection's local gauge potentials that we saw in Section 1.} to represent our (quantum) field measurements. 

However, at the same time, if our theory is to be regarded as being {\em physical}, it should be {\em variable}, that is to say, {\em dynamical}. 

So, gauging or localising our generalised observations $\aconn$ into our structure sheaf of generalised arithmetics, coefficients or coordinate measurements $\sts$ (and its associated vector sheaf $\sts$-module $\modl$) relative to an open gauge $\gauge_{i}$ covering the underlying in principle arbitrary topological space $X$, entails such a dynamical variability.\footnote{By definition, a {\em sheaf} of any mathematical objects---such as sets, groups, rings, modules, {\it etc.}---is a mathematical structure of {\em localised, variable objects}---`gauged' objects that vary continuously with respect to the (topology of the) surrogate base  topological space $X$ over which the sheaves are localised \cite{bredon,macmo,mall1,mall4}. At the same time, we also abide by the general result and motto in Sheaf Theory that {\em a sheaf \underline{is} its local sections}, as the global sheaf space can be stitched up from, or result from the collation of, all the local sections' data in $\Gamma(U,\sts)$ ($U\in\gauge_{i}$ covering $X$) \cite{bredon,macmo,mall1}.} This dynamical variability is then expressed via the $\sts$-connection field $\conn$, which derives from $\sts$ as we discussed earlier in connection with (\ref{eq3}).

\vskip 0.1in

\noindent\fbox{%
    \parbox{\textwidth}{%
\underline{\bf Fundamental Apophthegm of ADG-field Theory:} Since, as we argued earlier, ADG as a theory of {\em Differential} Geometry---which is fundamentally based on the notion of Mallios's $\sts$-connection---effectively boils down to $\sts$, we coin (local sections of) the latter (relative to our choice of a system $\gauge_{i}$ of local open gauges covering $X$), {\em our theoretical generalised observations or local coordinate measurements/determinations}.
}%
}

\vskip 0.1in     

\item Thus, from our dynamical (localised or gauged) generalised measurements in $\sts$, that are {\it a fortiori} organised into the vector sheaf $\modl$, we draw/derive `{\em connections}' or `{\em differences}'\footnote{Both metaphorically and literally speaking.} and  thereby we generate $\partial$ (from $\sts$ in equation (\ref{eq1})) and $\conn$ (from $\modl$ in equation (\ref{eq3})) in terms of  which the {\em field dynamics}---Nature's Field Law of vacuum Einstein gravity---is represented {\em differential} geometrically, by {\em differential} equations proper as in (\ref{eq9}).\footnote{Thus, in the $\sts$-cyclic `Ouroboros' diagram (OD2) above, the transition arrow $1.\rightarrow 2.$ is analogous to the $\cons$-linear and Leibnizian flat and curved connection defining sheaf morphisms $\partial :\, \sts\rightarrow\Omega$  and $\conn :\, \modl\rightarrow\Omg\otimes_{\sts}\modl\equiv\Omg(\modl)$ in equations (\ref{eq1}) and (\ref{eq3}), respectively.} 

\item From the said connections we then draw/derive (geometrical) patterns/regularities, thus the said dynamics will be properly observable if it is expressed solely in terms of observable (:measurable; $\otimes_{\sts}$-tensorial or $\sts$-functorial) `quantities'.\footnote{Thus, in the $\sts$-cyclic `Ouroboros' diagram (OD2) above, the transition arrow $2.\rightarrow 3.$ is analogous to the curvature defining morphism $\curv\equiv \curv(\conn):=\conn^{1}\circ\conn$ in equation (\ref{eq18}) that we saw early in the paper.}  

\item Indeed then, the said dynamical field laws are derived from a variational Lagrangian action principle and are expressed (differential) geometrically in terms of the principal `geometrical' object/entity---the curvature $\curv(\conn)$ of the $\sts$-connection $\conn$, which is an $\sts$-functorial observable (:an $\otimes_{\sts}$-tensor).\footnote{Thus, in the $\sts$-cyclic `Ouroboros' diagram (OD2) above, the transition arrow $3.\rightarrow 4.$ is analogous to the variational action principle from which the vacuum Einstein equations (\ref{eq9}) that we saw earlier in the paper derive.} Moreover, as we emphasised throughout this paper, that field dynamics is $\sts$-functorial and $\aut_{\sts}\modl$-covariant: in other words, the dynamics respects and in turn it is respected by our generalised local coordinate measurements (:arithmetics) in $\sts$.  

\item Finally, the geometrical observable object (:curvature) via which the field laws are expressed yields, upon local observation/measurement relative to a given $U\in\gauge_{i}$, generalised coordinate $c$-numbers (:arithmetics or coefficients) in the abelian structure sheaf $\sts$,\footnote{That is, effectively, from (\ref{eq19}), the local value of $\curv(\conn)$ at some $U\in\gauge_{i}$, may be taken to be, for example: $tr(\mathcal{E}nd\modl |_{U})=tr[M_{n}(\sts(U)]$, which is effectively a local section of $\sts$ in $\sts(U)\equiv\Gamma(U,\sts)$.} thus the $\sts$-cycle closes in itself.\footnote{In the $\sts$-cyclic `Ouroboros' diagram (OD2) above, this corresponds to the transition arrow $4.\rightarrow 1.$, thus closing the $\sts$-cycle.} 

\end{enumerate}

\vskip 0.1in

\noindent $\bullet$ {\bf Note:} The reader should notice that points 1-4 in the analysis of the $\sts$-cyclic `Ouroboros' diagram (OD2) above match item-for-item the entries and arrow-flow steps of our first $\sts$-cyclic `Ouroboros' diagram (OD1) in (\ref{eq23}) earlier; more notably:

\vskip 0.1in    

\centerline{\em Both diagrams start-from-$\sts$-and-end-with-$\sts$!}

\vskip 0.1in

\noindent which leads us to the following Apophthegm, our 11th one and second fundamental principle of ADG-gravity:

\vskip 0.1in

\noindent\fbox{%
    \parbox{\textwidth}{%
\noindent\underline{\bf Apophthegm 11: The Principle of `$\sts$-Recycling' (PAR)} ADG fundamentally starts with and originates from, the structure sheaf $\sts$ and its differential $\sts$-module $\modl$, then defines an $\sts$-connection $\conn$, and then proceeds to do Geometry via $\conn$'s geometric morphism $\geom$-image, which corresponds to the curvature $\curv(\conn)$ of the connection, which features in defining the $\sts$-functorial (:Mallios's $\sts$-invariant, $\otimes_{\sts}$-functorial or $\aut_{\sts}\modl$-covariant) vacuum Einstein ADG-gravitational purely gauge field auto-dynamics in (\ref{eq9}). In effect, {\em ADG-gravity starts from $\sts$ and ends with $\sts$}.}%
}

\vskip 0.1in

\noindent The following important remark is due here in view of the PAR above:

\vskip 0.1in

\noindent\fbox{%
    \parbox{\textwidth}{%
\noindent\underline{\bf Highlight 1:} We may cumulatively refer to the ADG-field theoretic $\sts$-Connection-Dynamics-Measurement Invariance Cycle above as {\em Generalised $\sts$-Measurements Recycling} or {\em The Principle of Conservation of Generalised Measurements in $\sts$}, and it may be thought of as the ADG-theoretic generalised analogue of Noether's Theorem for the PARD and its associated $\sts$-functoriality. As the usual statement of Noether's Theorem maintains that {\em with every continuous symmetry of the Lagrangian dynamical action principle there is an associated conserved (:invariant) dynamically measurable (:observable) quantity ({\it alias}, current) in the resulting dynamical equations of motion on the spacetime continuum}, here too, as a result of the PARD and the $\sts$-functoriality (:the $\aut_{\sts}\modl$-covariance) of the ADG-gravitational field auto-dynamics, we maintain that {\em what is conserved here is the \underline{differential geometric form} of the dynamical differential equations of ADG-gravity} (here, the vacuum Einstein equations), which is expressed via the $\sts$-invariant curvature form $\curv(\conn)$ of the connection $\conn$, which in turn originated from $\sts$ in the first place.}%
}

\vskip 0.1in

\noindent As a direct corollary of the remarks in the Highlight 1 above, we come to emphasise that:

\noindent\fbox{%
    \parbox{\textwidth}{%
\noindent\underline{\bf Highlight 2:} The curvature $\curv(\conn)$ of the $\sts$-connection field $\conn$ is the abstract ADG-field theoretic analogue of a conserved current in Noether's Theorem, in the sense that it is the {\em differential geometrical form-invariant, $\otimes_{\sts}$-tensorial observable or $\sts$-invariant conserved quantity} associated with the PARD and the $\sts$-functoriality (:the $\aut_{\sts}\modl$-covariance) of the ADG-gravitational field auto-dynamics in (\ref{eq9}). We will come to discuss this more in the sequel, when we remark on {\em sheaf cohomology with coefficients in $\sts$ and characteristic form $\curv(\conn)$}.}%
}

\vskip 0.1in

\section{Epilogue: Summary of Results by `Justifying' the Title of the Paper with a Novel Mathematical and Philosophical Outlook Towards Future Quantum Gravity Research}

In this concluding section, first we summarise the basic tenets and results of this paper effectively by `justifying'  its title. Thus, there are two issues to highlight and to discuss here:

\begin{enumerate}

\item On the one hand, in the title of the present paper we contended that {\em ADG-gravity is a general theory of gravity}, {\em an abstract and generalised General Relativity}; and on the other,

\item That {\em ADG-gravity is based on, and that it derives from, the structure sheaf $\sts$ of generalised arithmetics, coefficients or coordinates.} 

\item Moreover, in connection with both features of ADG-gravity above, we also contended that ADG-gravity has certain `inherent', `intrinsic' or `innate' traits that can shed light and potentially resolve some key conceptual and structural issues in current QG research that appear to be incurably problematic and insuperable obstacles on our way to QG exactly because they are based on the background geometrical manifold dependent concepts, methods and constructions of CDG, as well as on the manifold's physical interpretation as a {\em spacetime continuum} in the CDG-based GR. We discuss the potential outlook of ADG-gravity in future QG research at the very end of the paper with various pertinent quotes and heuristic philosophical smatterings based on them.

\end{enumerate}

\subsection{ADG-Gravity as a Generalised General Theory of Relativity} 

Below, we give a shortlist of eight facts and results that corroborrate and support the first part of our claim that, as ADG is an abstract and generalised CDG as shown and explained in this paper, {\em ADG-gravity is a general theory of relativistic gravity}, {\sl an abstract and generalised GR} so to pun:

\begin{enumerate}

\item ADG-gravity, unlike GR, is not CDG-based; hence, it is manifestly {\em not} background geometrical spacetime manifold dependent. In other words, {\em ADG-gravity is a genuinely background independent theory of gravity}.

\item ADG-gravity is based on an arbitrarily chosen algebra structure sheaf $\sts$ of generalised, possibly also non-functional, coordinates,\footnote{And it is not necessarily based on the usual structure sheaf $\smooth_{M}$ of smooth coordinate functions on a differential point set manifold $M$, like the CDG-based GR is.}---the structure algebra sheaf itself being localised on a general and in principle arbitrary base topological space $X$, which serves as a surrogate space for the sheaf theoretic localisation (:gauging) of $\sts$ and $\modl$---as long as $\sts$ and $\modl$ provide us with linear and Leibnizian algebraic $\sts$-connections $\partial$ (:flat $\sts$-connection) and $\conn$ (:curved $\sts$-connection), which in turn act categorically as sheaf morphisms on (the local sections of) the relevant algebra $\sts$ and vector (:$\sts$-module) $\modl$ sheaves, respectively.

\item ADG-gravity, as a dynamical theory, is based solely on the fundamental notion of an algebraic $\sts$-connection field variable $\conn$, the cornerstone structure of ADG regarded as an abstract and general mathematical theory of {\em differential} geometry, as it was repeatedly highlighted and explained in this paper. As also emphasised in this paper, the notion of a Mallios $\sts$-connection $\conn$ is purely algebraic (:homological algebraic, {\it i.e.}, sheaf and category-theoretic) and thus ADG is a purely algebraic (:relational) theory of differential geometry that is not at all dependent on a background geometrical locally Euclidean substratum for either its formal technical (:calculational and structural) support or for its conceptual-semantic (physico-philosophical) interpretation.

\item ADG-gravity is not a theory of curved spacetime geometry---{\it i.e.}, the ADG-gravitational field is not represented by the smooth spacetime metric $g_{\mu\nu}$ as in the usual spacetime manifold based GR (:2nd order formalism), but rather, it is a pure gauge theory of a gravitational $\sts$-connection vacuum Einstein field $\mathcal{F}_{Einst}=(\modl ,\conn)$. In this respect, it is similar to the Palatini-Ashtekar formalism (:1st order formalism), but without the spacetime metric implicit in the vierbein variables there. As we explained in Section 1, in ADG-gravity, the $\sts$-metric $\rho$ is an auxiliary and `bonus' structure, while its compatibility with $\conn$ is an optional condition.\footnote{The reader should note here that, because $\conn$ is the sole dynamical variable in the theory, we do not talk about {\em the compatibility of the connection $\nabla$ with the metric $g_{\mu\nu}$} as in the usual Christoffel theory (GR). Rather, we talk about {\em the compatibility of the $\sts$-metric $\rho$ with the connection $\conn$}.}

\item The invariance group of ADG-gravity, by contrast to the CDG and spacetime manifold $M$ based GR, is not the spacetime diffeomorphism group $\mathrm{Diff}(M)$, but the principal group sheaf $\aut_{\sts}\modl$ of the local vacuum Einstein field $\mathcal{F}_{Einst}=(\modl ,\conn)$-automorphisms, the local self-symmetries of the vacuum Einstein ADG-gravitational field equations (\ref{eq9}). As we saw earlier, for a vector sheaf $\modl$ of rank $n$, the principal group sheaf $\aut_{\sts}\modl$ of automorphisms of $\modl$ is locally isomorphic to $M_{n}(\sts(U))^{\bullet}=\mathcal{G}\mathcal{L}(4,\sts(U))$, which is the ADG-theoretic matrix group generalisation of the general linear symmetry group $GL(4,\R)$ of GR.

\item From the fact above, it follows that in ADG-gravity there is no distinction between external (spacetime) and internal (gauge) symmetries like in the usual, both classical and quantum, field theories of matter. The vacuum Einstein ADG-gravitational field $\mathcal{F}_{Einst}=(\modl ,\conn)$ is a dynamically autonomous entity whose dynamical symmetries are purely internal (gauge) or intrinsic to the field itself, without recourse to or dependence on an external (to the gravitational field itself) geometrical spacetime continuum. The $\aut_{\sts}\modl$-symmetries of the ADG-gravitational vacuum Einstein field $\mathcal{F}_{Einst}=(\modl ,\conn)$ and of the dynamical equations (\ref{eq9}) that it defines via its curvature form are {\em internal (gauge) self-symmetries}.

\item For any chosen structure sheaf $\sts$ of generalised coordinates, the $\sts$-functoriality and $\sts$-invariance of the ADG theory secure and guarantee that the vacuum Einstein equations in (\ref{eq9}) hold no matter how `problematic', `pathological' or `singular' the chosen and employed structure sheaf $\sts$ \cite{malros1,malros2,malros3,malrap1,malrap2,malrap3,rap5} may seem to be from the classical background geometrical manifold based perspective of CDG, or equivalently, from the vantage of the usual Differential Calculus or Analysis \cite{clarke3,clarke4}.

\item The generalisation of the PGC of GR, which is represented by the invariance group $\mathrm{Diff}(M)$ of spacetime diffeomorphisms of the background geometrical differential spacetime manifold $M$, to the PARD, which, as we saw earlier, is represented by certain functorial, natural transformation type of morphisms between the relevant ADG-theoretic sheaf categories involved. The generalised general relativistic meaning of the PARD is that one may change (naturally transform) freely between differential structure sheaves $\sts$ of generalised coordinates, but the intrinsically relational (:algebraic) differential geometric mechanism of ADG remains invariant, so that the dynamical vacuum Einstein ADG-gravitational equations---which are functorially expressed via the curvature form $\curv(\conn)$ of the $\sts$-connection $\conn$, which is an $\otimes_{\sts}$-tensor---hold intact, do not break down in any differential geometric sense and are not assailed by singularities or unphysical infinities like the manifold and CDG-{\it cum}-Analysis based GR \cite{clarke3,clarke4}. In fact, as we highlighted earlier, the vacuum Einstein ADG-gravitational equations are seen to hold on pointless finitary-reticular \cite{malrap1,malrap2,malrap3,rap5,rap7} and densely singular \cite{malros1,malros2,malros3} spaces that seem unmanageably granular, pathological and problematic from the CDG-Analytic and manifold based perspective of GR.

\end{enumerate}

\subsection{ADG-Gravity Originates and Derives from the Structure Sheaf $\sts$}

As in the previous subsection, below we support and corroborate our second claim in the title of the present paper, namely that {\em ADG-gravity originates and derives from the structure sheaf $\sts$ of generalised arithmetics, coefficients and coordinates}. We then list seven important issues in QG research on which ADG-gravity may shed light and go quite some way to resolve. Thus,

\noindent\fbox{%
    \parbox{\textwidth}{%
\noindent\underline{\bf Apophthegm 8:} ADG-gravity originates and derives from $\sts$, because the fundamental notion and foundational structure of an $\sts$-connection (:flat $\partial$ or curved $\conn$) on which the entire ADG-theoretic edifice rests and is built upon as a mathematical theory of {\em Differential} Geometry proper, and via which the law of vacuum Einstein ADG-gravity is formulated as a {\em differential} equation proper hence also the physical theory is expressed as a pure gauge field theory as explained in the last subsection, has its source or domain of definition, and it derives from, the structure algebra sheaf $\sts$, as the defining expressions for $\partial$ in (\ref{eq1}) and for $\conn$ in (\ref{eq3}) show, respectively. This is essentially the aftermath and gist of Apophthegms 1 and 2 earlier.
}%
}

\vskip 0.1in

\subsection{The Need for New Mathematical Ideas and Physico-Philosophical Concepts: Implications for Current and Future Quantum Gravity Research} 

We commence this subsection with two quotes that generally underpin the formidable task that theoretical and mathematical physicists have undertaken in the last seven decades (at least!) of trying to arrive at a conceptually sound, internally self-consistent and calculationally finite Quantum Theory of Gravity (QG).

The first is a `pessimistic' one by Sidney Coleman in \cite{coleman}:

\bigskip\noindent (Q3)\hskip 0.9in
\begin{minipage}{11cm}
\noindent ``{\small ...Quantum gravity is notoriously a subject where problems vastly outnumber results...}''
\end{minipage}

\vskip 0.1in

The second quote, by Gerard 't Hooft in \cite{thooft}, highlights the conceptual-philosophical essence of those ``{\em vastly more numerous than} [technical] {\em results QG problems}'', rather than their technical (:mathematical or calculational) nature:

\bigskip\noindent (Q4)\hskip 0.9in
\begin{minipage}{11cm}
\noindent ``{\small ....The problems of quantum gravity are much
more than purely technical ones. They touch upon very essential
{\em philosophical} issues...}''
\end{minipage}

\vskip 0.1in

Now that we have established that ADG-gravity is an entirely algebraic, genuinely background geometrical spacetime manifold independent, pure gauge field theory of gravity, we discuss briefly seven potential applications in the form of structural-calculational (:mathematical) and conceptual (:physico-philosophical) consequences that it may have for certain central problems and issues in current and future QG research. 

\subsubsection{The background differential spacetime manifold just gets in our way towards QG}
 
If one wishes to unite General Relativity (GR), the classical relativistic field theory of gravity, with Quantum Theory (QT) thus arrive at a conceptually sound and self-consistent, as well as  calculationally finite, QG by using differential geometric concepts, structures and techniques, the background geometrical locally Euclidean spacetime continuum (:manifold) of CDG just gets in the way in the form of singularities of GR, the non-renormalisable unphysical infinities of QFT in the flat Minkowski spacetime continuum, and also in the form of the $\mathrm{Diff}(M)$-based problems of time and inner product in both covariant and canonical quantisation approaches to QG. Singularities and their associated infinities are arguably due to the fact that one can pack an uncountable infinity of events or field degrees of freedom in a finite spacetime volume. On top of those background $M$ and $\mathrm{Diff}(M)$-related problems, we have deep conceptual problems in Quantum Cosmology associated with the Measurement Problem in QT and its distiction between classical exosystems (observers) and quantum endosystems (observed), {\it contra} our regarding the Universe as being a closed, dynamically autonomous system, with no external observer measuring it `from without' as it were. Associated with the latter is the existence of  so-called fundamental spacetime scales---the Planck length  ($l_{P}=\sqrt{Gh/c^{3}}$) and the Planck time ($t_{P}=\sqrt{Gh/c^{5}}$)---below which, supposedly, the elusive QG is in force---{\it i.e.}, quantum gravitational effects become significant, but above which the usual flat QFTs of matter of the Standard Model are in force, and at even larger cosmological scales the spacetime manifold based gravity of GR is the dominant force shaping the structure and dynamics of the Universe. Furthermore, the said fundamental scales are built from the dimensional constants $G$, $c$ and $\hbar$, therefore them too are also {\em not} dimensionless (having the physical dimensions of length and time, respectively), hence they would render {\em non-renormalisable} any perturbative expansion in an attempt to finitise (:renormalise) gravity below Planck scale. Thus, such dimensional cut-off scales are impediments and fragmentations to a finitistic, coherent, unitary and unified conception of Nature in the sense that it has been moreover argued that a truly fundamental theory, such as the elusive QG that we are after, is expected to do away with such so-called fundamental spacetime scales, and even more drastically, perhaps do away with the notions of space and time altogether \cite{unzicker}.\footnote{See pertinent quotes from \cite{unzicker} and our analysis of them below.}

\vskip 0.1in

Here is an outline of the seven persistently pressing QG  research issues that we wish to briefly address and cast some different light on from the perspective of ADG-gravity:

\begin{enumerate}

\item {\em The problem of Singularities}: singularities in the classical, CDG-based field theory of gravity (GR) are viewed as {\it loci} in spacetime where the gravitational field equations of Einstein somehow break down and physically measurable geometrical quantities such as the spacetime curvature blow up without bound \cite{clarke1,clarke2,clarke3}. It is widely accepted that singularities and the associated breakdown of the physical law of the classical relativistic field theory of gravity (GR) is due to its assumption of a background smooth spacetime manifold, which can in principle pack an uncountable infinity of events in a finite spacetime region (volume). For example, it is generally supposed that only the true quantum theory of gravity will be able to reveal what happens in the interior (:past the horizon) of, say, a Schwarzschild black hole, in the vicinity of the inner point-mass singularity, where Einstein's field law supposedly breaks down as a differential equation and hence physical predictability is supposedly impossible.

\item {\em The problem of Unphysical Infinities}: the singularities of GR are formidable `obstacles' to the physical law of gravity way before quantisation of the gravitational field becomes an issue. However, the spacetime continuum is also responsible for the similarly unphysical infinities that assail the usual quantum field theories of matter, which are also based on a spacetime continuum; albeit, in the absence of gravity, on a flat one (:Minkowski space). Moreover, if one wishes to apply the usual CDG or Differential Calculus-based analytical techniques, one gets nonsensical ultraviolet divergences and infinities that come from contributions of the uncountably infinite degrees of freedom of the continuous matter and gauge fields. The heuristic process of renormalisation to remove by hand as it were the unphysical infinities, has been criticised as being theoretically {\em ad hoc}.

\item {\em The problem of Background Independence}: in a nutshell, in view of the fact that the background differential spacetime manifold $M$ of GR is a fixed sructure in the theory, the basic intuition is that  in the realm of QG that fixed ether-like background spacetime continuum will give way to a more reticular-finitistic (:discrete) and, perhaps more importantly, dynamically variable structure. Field quantities, cannot be referred to a fixed background (Minkowski) metric with respect to which, for example,m the perturbation expansion (:renormalisation) will be referred to, especially when the metric itself is supposed to be the basic dynamically variable in GR.

\item {\em The Problem of Time and the Inner Product Problem in (both canonical and covariant) QG research}: both are due to the $\mathrm{Diff}(M)$-invariance of the classical theory (GR) which, when carried over to the (both canonically or covariantly) quantised theory, gives us formidable technical and conceptual problems, like for example: (i) in defining physically meaningful states that are annihilated by the Hamiltonian operator (constraint), which is the generator of timelike diffeomorphisms (dynamical evolutions) in the theory; and, (ii) in defining a positive definite inner product between those physical states so that the theory has a cogent and consistent probabilistic physical interpretation, one that in principle does not allow for physically nonsensical negative norm states.

\item {\em The Problem of Defining Autonomous, Closed Dynamical (Field) Systems and the Measurement Problem in QC:} in Quantum Cosmology (QC) research, the fundamental assumption is that the Universe is a closed quantum system. On the other hand, the usual Copenhagen Interpretation of QT fundamentally splits the world into a classical exosystem (the observer) and a quantum endosystem (the observed) by the famous Heisenberg {\it schnitt} underlying the Quantum Measurement Problem. Thus, for QC we need to develop a theory of closed, dynamically autonomous systems, with quantum traits built in, for which there is no distinction between (classical) external observer and (quantum) observed system as there is nothing standing `outside' the quantum universe `observing' it.\footnote{In this regard, see for example Wheeler's paper in the celebrated volume \cite{wheeler}.}

\item {\em The Problem of a Fundamental Spacetime Scale:} hand in hand with defining dynamically autonomous systems, with some intrinsic quantum traits, comes the problem of having a fundamental spacetime scale---that of Planck length and Planck time, which are `concocted' from the constants of the three basic theories of Nature that we posssess: Special Relativity ($c$), General Relativity ($G$) and Quantum Theory ($\hbar$).\footnote{Recall from above that  the Planck length is expressed as $l_{P}=\sqrt{Gh/c^{3}}\approx 1.6\times 10^{-35}m$; while the Planck time as $t_{P}=\sqrt{Gh/c^{5}}\approx 5.4\times 10^{-44}s$.} This is supposed to be a fundamental Heisenberg {\it schnitt} type of cut-off scale below which the elusive QG is supposed to hold, but above which GR and the usual (flat) QFTs of matter are supposed to be in force. There have recently been aired theories maintaining that the fundamental constants may be regarded as hindrances to genuine unification, while a fundamental (field) theory, one that is unitary and holistic across all our fundamental field theories (:electromagnetism and Yang-Mills), should,  `deep down inside the quantum deep', be spacetime scale independent \cite{unzicker}. In fact, in the same line of thought, it has been argued that our usual conceptions and concepts of space and time, especially in their spacetime continuum guise, cannot and therefore should not be carried over in the QG domain \cite{unzicker}. In fact, Wheeler in the aforementioned paper \cite{wheeler} in connection with the dynamically autonomous and observationally closed system conception of the Universe, posits that  ``{\em in the realm of QG there is No Space and No Time}''.\footnote{See the Postscriptum section at the end of the paper in which we briefly discuss current ideas on {\em Spacetimelessness} in current QG research.}

\item {\it Mutatis mutandis} for Quantisation: one would expect an intrinsic and `self-referential' (:autonomous) theoretical scheme for quantisation, one that goes directly to and involves solely the force fields in-themselves, without the mediation of an external (to the fields themselves) background geometrical spacetime continuum structure, which is reminiscent of the luminipherous ether of Maxwellian electrodynamics, {\em which acts but is not acted upon} \cite{df2}. At the same time, if there is any physical spacetime geometry at all, that will be the outcome of the algebraic field dynamics---{\em a `spectral geometry' built into the algebra structure sheaf $\sts$ from which the ADG-gravitational $\sts$-connection field $\conn$ arises} \cite{malzaf1}, not a passive kinematical stage on which the fields dynamically interact and dynamically propagate, but by and in itself it has no directly observable effects. The {\it a priori} absence of a spacetime interpretation in ADG-gravity makes the quest for quantising spacetime beg the question in the first place: {\em a scheme that directly quantises the fields themselves is in no need of quantising a fiducial spacetime background that does not physically exist anyway}.

\end{enumerate}     

\subsection{Five Basic Questions about the Future of Quantum Gravity Research from an ADG-Gravitational Point of View}

In this subsection we pose five basic questions in order to prompt and stimulate lateral, innovative and `out-of-the-box' thinking\footnote{At least, thinking of gravity and a potential quantum description of it beyond the confines of the spacetime manifold and CDG-based GR.} about future QG research along the technically and conceptually novel lines of ADG-gravity:

\begin{itemize}

\item {\bf Question 1 (QU1):} What is the potential import, both technical and conceptual, as well as the physical significance, of having an entirely algebraic field theory of gravity, like ADG-gravity is, having no background geometrical spacetime manifold conceptual and technical dependences, geometrical representations and associated physical interpretations?

\item {\bf Question 2 (QU2):} Closely related to Q1 above, what is the potential import, both technical and conceptual, as well as the physical significance, of having an entirely algebraic (:sheaf and category-theoretic) theory of {\em Differential} Geometry, like ADG is, in which we can do Calculus---{\it i.e.}, construct, calculate and do field Physics---in the very presence of singularities and other differential geometric anomalies on which the usual manifold based CDG stumbles, stalls and in general comes short of delivering physically meaningful constructions, calculations and results in both GR (singularities) and QFT (non-renormalisable infinities).

\item {\bf Question 3 (QU3):} What is the potential import, both technical and conceptual, as well as the physical significance, of having a purely gauge field theory of gravity, like ADG-gravity is, having no external (:background to the gravitational $\sts$-connection field itself) differential spacetime manifold geometry and its own, also external to the gravitational $\sts$-connection field itself, $\mathrm{Diff}(M)$-symmetries?

\item {\bf Question 4 (QU4):} What is the potential import, both technical and conceptual, as well as the physical significance, of having a completely dynamically autonomous and closed system-like structure and conception of a gravitational field, with innate (:intrinsic and built-in) quantum traits from the very start, like the vacuum Einstein ADG-gravitional $\sts$-connection field $\mathcal{F}_{Einst}=(\modl ,\conn)$ is? 

\item {\bf Question 5 (QU5):} What is the potential import, both technical and conceptual, as well as the physical significance, of having a completely background geometrical spacetimeless and, {\it in extenso}, spacetime scale-less theory of gravity, like ADG-gravity is, which on the one hand has no {\it a priori} background geometrical spacetime interpretation and representation of the mathematical structures involved in ADG, and on the other, it draws no fundamental scale distinctions and therefore posits no theoretically {\it ad hoc} cut-offs in terms of fundamental constants that typically delimit the range of validity of the physical law that the gravitational $\sts$-connection field itself defines in the first place? Such supposedly fundamental cut-off scales also posit `fiducial' mathematical pseudo-problems and quandaries such as whether, in the quantum deep, spacetime is continuous or discrete in nature and, related to it, whether a Quantum Theory of Gravity, regarded as Quantum General Relativity, should be accompanied by, or even result in, some kind of quantisation of the curved spacetime continuum itself in much the same way that the supposedly in principle continuous classical radiation spectrum of the hydrogen atom, upon quantisation (solution of the Schrodinger equation of its sole valence electron), revealed the reticular (:discrete) nature of its electron orbitals.

\end{itemize}

\vskip 0.1in

\noindent In what follows, we are not going to attempt to answer directly and fully the questions (QU1-5) above. Rather, in the five subsections below, one for each question posed above, we give some telling quotes that `justify' the very posing of the five questions above in the first place, followed by related apophthegms that issue directly from results or facts we have gathered from developing ADG-gravity throughout the last quarter of a century, especially with an eye towards partly addressing and partly resolving the issues highlighted by the said five questions.\footnote{Thus, in view of ADG-gravity, the five questions above are kind of `rhetorical', in the sense that ADG-gravity has already anticipated them, highlighted them, and partly answered them.}

\subsubsection{ (QU1) Einstein: A Purely Algebraic Theory for the Description of Reality in the Quantum Deep}

We support and `justify' the posing of question (QU1) above by giving the following telling sequence of quotes by Einstein:

\bigskip\noindent (Q5)\hskip 0.9in
\begin{minipage}{11cm}
\noindent ``{\small ... Is it conceivable that a [continuous] field theory
permits one to understand the atomistic and quantum structure of
reality?...[Quantum phenomena do] {\em not seem to be
in accordance with a [spacetime] continuum theory, and must lead to an attempt
to find a purely algebraic theory for the description of
reality. But nobody knows how to obtain the basis of such a theory...}}'' (1956) \cite{einst3}
\end{minipage}

\vskip 0.1in

\noindent followed by:

\vskip 0.1in

\bigskip\noindent (Q6)\hskip 0.9in
\begin{minipage}{11cm}
\noindent ``{\small ...You have correctly grasped the drawback that
the continuum brings. If the molecular view of matter is the
correct (appropriate) one; {\it i.e.}, if a part of the universe is
to be represented by a finite number of points, then the continuum
of the present theory contains too great a manifold of
possibilities. I also believe that this `too great' is responsible
for the fact that our present means of description miscarry with
quantum theory. The problem seems to me how one can formulate
statements about a discontinuum without calling upon a continuum
space-time as an aid; the latter should be banned from theory as a
supplementary construction not justified by the essence of the
problem---{\em a construction which corresponds to nothing real.
But we still lack the mathematical structure
unfortunately}\footnote{Our emphasis.}. How much have I already
plagued myself in this way [of the spacetime manifold]...}" (1916)
\cite{stachel}
\end{minipage}

\vskip 0.1in 

\noindent also followed by:

\bigskip\noindent (Q7)\hskip 0.9in
\begin{minipage}{11cm}
\noindent ``{\small ...An algebraic theory of physics is affected
with just the inverted advantages and weaknesses, aside from the
fact that no one has been able to propose a possible logical
schema for such a theory. It would be especially difficult to
derive something like a spatio-temporal quasi-order from such a
schema. {\em I
cannot imagine how the axiomatic framework for such a physics
would appear, and I don't like it when one talks about it in dark
apostrophes.} But I hold it entirely possible that the development
will lead there; for it seems that the state of any finite
spatially limited system may be fully characterized by a finite
set of numbers. This seems to speak against a continuum with its
infinitely many degrees of freedom. The objection is not decisive
only because {\em one doesn't know, in the contemporary state of
mathematics, in what way the demand for freedom from singularity
(in the continuum theory) limits the manifold of
solutions...}''}\footnote{Again, our emphasis.} \cite{stachel}
\end{minipage}              

\vskip 0.1in

\noindent and finally, he agnostically admitted:

\bigskip\noindent (Q8)\hskip 0.9in
\begin{minipage}{11cm}
\noindent ``{\small ...Your objections regarding the existence of
singularity-free solutions which could represent the field
together with the particles I find most justified. I also share
this doubt. If it should finally turn out to be the case, then I
doubt in general the existence of a rational and physically useful
continuous field theory. But what then? Heine's classical line
comes to mind: `{\sl And a fool waits for the answer}'...}" (1954)
\cite{stachel}
\end{minipage}

\vskip 0.1in

\noindent

\noindent In `response' to the points made and issues raised by Einstein in quotes (Q5-8) above, we present the folowing apophthegm, our twelveth one:

\noindent\fbox{%
    \parbox{\textwidth}{%
\noindent\underline{\bf Apophthegm 12}: ADG and its physical application offshoot, ADG-gravity, is a competent candidate for Einstein's envisaged {\em purely algebraic theory for the description of reality} (Q5) in the quantum domain, as it is a purely homological algebraic theory. Moreover, ADG-gravity is a differential geometrically formulated field theory that is not  dependent at all on a background geometrical locally Euclidean spacetime continuum---{\em it is a field theory proper that is not at all calling upon a continuum
space-time as an aid} (Q6). In fact, ADG-field theory, and ADG-gravity in particular, passes through the horns of the continuum vs discretum dichotomy and dilemma of Einstein (Q6-7), as the character of the base topological space $X$---whether reticular or continuous---on which the algebra $\sts$ and vector sheaves $\modl$ are localised, plays no role whatsoever in the inherently algebraic (:relational) differential geometric mechanism, which derives from the stalks (:the algebraic structure) of, or the algebraic relations between, the `geometrical objects' (:the $\sts$s and the $\modl$s involved) that live on that surrogate base localisation space $X$. Furthermore, in ADG-gravity, we are indeed able to show {\em the existence of
`singularity-free' solutions} (:structure algebra $\sts$ and vector sheaf spaces $\modl$ on which the vacuum Einstein field equations hold) {\em which represent the field
together with the particles} (Q8). In fact, as we explained in this paper, the vacuum Einstein field $\mathcal{F}_{Einst}=(\modl ,\conn)$ represents exactly the gravitational connection field $\conn$ together with its representation vector sheaf $\modl$ of local quantum particle states on which the vacuum Einstein equations (\ref{eq9}) hold intact. {\it A fortiori}, we saw that the ADG-gravitational field is intrinsically quantum from our purely algebraic, sheaf cohomological 3rd quantisation perspective.}%
}

\subsubsection{(QU2) Einstein-Feynman-Isham: The No-Go and Miscarrying of CDG in the Quantum Deep and the Quest for a New Mathematical Theory}

In connection with question (QU2) above, and our discussion in the previous subsection about the fact that `{\em the background differential spacetime manifold just gets in our way towards QG}', we recall two very pertinent quotes, one by Feynman and the other by Isham, that literally state that the background geometrical manifold based CDG is an inappropriate and inadequate mathematics for representing and doing (calculating) gravitational physics in the quantum deep.

\vskip 0.1in

\noindent First comes Feynman:

\vskip 0.1in

\bigskip\noindent (Q9)\hskip 0.9in
\begin{minipage}{11cm}
\noindent ``{\small{\em ...the theory that space is continuous is
wrong, because we get...infinities} [viz. `singularities'] {\em
and other similar difficulties} ...[while] {\em the simple ideas
of [differential] geometry, extended down to infinitely small, are wrong...}}''
\cite{feyn1}
\end{minipage}

\vskip 0.1in

\noindent and then Isham:

\bigskip\noindent (Q10)\hskip 0.9in
\begin{minipage}{11cm}
\noindent ``{\small{\em ...at the Planck-length scale, classical
differential geometry is simply incompatible with quantum
theory}...[so that] {\em one will not be able to use differential
geometry in the true quantum-gravity theory...}}'' \cite{ish}
\end{minipage}

\vskip 0.1in

Here, the reader must be very careful as this point is quite subtle and sensitive from an ADG-theoretic perspective: 

\noindent\fbox{%
    \parbox{\textwidth}{%
\noindent\underline{\bf Apophthegm 13}: It is not exactly that one cannot use ideas, concepts and and constructions of Differential Geometry {\it per se} in, say, QG research. After all, how else can we formulate the local laws of Nature other than as {\em differential} equations proper? Rather, it is that the point manifold based CDG and its Analysis is inadequate and comes short in addressing QG issues, as it is marred by singularities, non-removable infinities and other differential geometric pathologies coming from our {\it a priori} assumption of a smooth background geometrical manifold supporting, both mathematically and conceptually, our fundamental field-theoretic conceptions of Nature, with a spacetime continuum physical interpretation on top.}%
}

\vskip 0.1in

In the context of classical infinitesimal (differential) locality and the (differential) manifold used on which to represent the physical laws after differential equations, here is what David Finkelstein maintained in \cite{df2} {\it vis-\`a-vis} the quantum:

\vskip 0.1in

\bigskip \noindent (Q11)\hskip 0.9in
\begin{minipage}{11cm}
\noindent ``{\small ...The locality principle seems to catch
something fundamental about nature... Having learned that the
world need not be Euclidean in the large, the next tenable
position is that it must at least be Euclidean in the small, a
manifold. The idea of infinitesimal locality presupposes that the
world is a manifold. {\small\em But the infinities of the manifold
(the number of events per unit volume, for example) give rise to
the terrible infinities of classical field theory and to the
weaker but still pestilential ones of quantum field
theory}.\footnote{Our emphasis.} The manifold postulate freezes
local topological degrees of freedom which are numerous enough to 
account for all the degrees of freedom we actually observe...}''
\end{minipage}

\vskip 0.1in

Thus, in view of the main results and didactics of ADG-gravity \cite{rap14,rap15,rap19}, we come to itemise our response to quotes (Q5-11) one by one, as follows:

\begin{itemize}

\item (Q5): We have in our hands a purely algebraic (:sheaf and categorical) theory for a field theoretic description of vacuum Einstein gravity (and free Yang-Mills theories), with quantum features built into our theory and theoretical formalism from the very start.

\item (Q6): We can indeed formulate gravity differential geometrically, with inherent or innate quantum characteristics built into the formalism, without at all calling upon a spacetime continuum as an aid, which anyway, in Einstein's words, {\em corresponds to nothing real}.

\item (Q7): In ADG and ADG-gravity, we have indeed drawn the basis of a novel {\em axiomatic mathematical framework} \cite{mall2} in which to formulate QG \cite{mall14,mall15} in the very presence of singularities and in an infinities-free fashion as {\em there are no infinities in Algebra}. Infinities creep into our calculations ({\it i.e.}, ultimately into our Calculus) via the locally Euclidean continuum that supports and mediates our standard Differential Calculus' calculations and associated Analysis. The background manifold is the carrier space of all our infinitesimal (:differential) calculations and, as a result, their physically unacceptable infinities yielding pathologies.

\item (Q8): In ADG-gravity, we indeed have a theoretical scenario according to which we can {\em represent a field together with its quantum particles and the dynamical law that guides them} in an inherently and manifestly singularities and infinities-free fashion.

\item (Q9): It is not exactly that differential geometric ideas are {\em wrong} in the quantum domain as Feynman put it. Rather, it is that when differential geometric ideas, constructions and calculations are applied in the quantum deep (:below Planck scale) via the mediation of a spacetime continuum to support and geometrically represent as well as to physically interpret our calculations ({\it i.e.}, our Calculus), we get physically non-sensical singularities, regarded as regions where the physical law breaks down, together with non-removable (:non-renormalisable)  unphysical infinities, hence CDG appears to miscarry, or even better, be out of its depth, with Quantum Theory. To use a pun, {\em CDG miscarries into the quantum domain exactly because of its carrier background locally Euclidean spacetime}.

\item (Q10): Isham's words are valid insofar as we insist on applying the background geometrical smooth manifold based CDG and Analysis to QG research. A purely algebraic, genuinely background manifold independent and infinities-free differential geometry like ADG fares differently in the quantum domain as it is able to address problems and resolve issues that CDG is simply unable to due to its {\it a priori} assumption of a base spacetime manifold.

\item (Q11): In ADG, we are still able to formulate the local laws of physics differential geometrically ({\it i.e.}, as differential equations proper), but without at all the employment of a background locally Euclidean spacetime (manifold). All our ADG sheaf-theoretic concepts, constructions and calculation tools are {\em purely algebraic} and {\em strictly local}, as ``{\em the  methods of sheaf theory are essentially algebraic and local}'' \cite{bredon}.

\end{itemize}

\noindent In view of our last remarks with respect to quote (Q11) above, we close this subsection with some `prophetic' words of Rudolph Haag from \cite{haag}, which anticipate and in retrospect ({\it i.e.}, post ADG) reinforce the use of {\em sheaf theory} in axiomatic (algebraic and local) QFT:\footnote{In the usual axiomatic formulation of flat QFT using local operator $C^{*}$-algebras and Von Neumann algebras over Minkowski spacetime ({\it i.e.}, in the absence of gravity).}

\bigskip\noindent (Q12)\hskip 0.9in
\begin{minipage}{11cm}
\noindent ``{\small {\bf Germs.} {\small\em We may take it as the
central message of Quantum Field Theory that all information
characterizing the theory is strictly local i.e. expressed in the
structure of the theory in an arbitrarily small neighborhood of a
point}.\footnote{Our emphasis.} For instance in the traditional
approach the theory is characterized by a Lagrangean density.
{\small\em Since the quantities associated with a point are very
singular objects, it is advisable to consider neighborhoods. This
means that instead of a fiber bundle one has to work with a sheaf.
The needed information consists then of two parts: first the
description of the germs, secondly the rules for joining the germs
to obtain the theory in a finite region}\footnote{Again, emphasis
is ours.}...}''
\end{minipage}

\noindent This is precisely how we think about stitching and collating all our generalised local coordinates' measurement data in $\sts(U)$,\footnote{Strictly speaking, the {\em germs} of the local continuous sections of $\sts$ in $\sts(U)\equiv\Gamma(U,\sts)$, which inhabit the {\em stalks} of the structure sheaf $\sts$ \cite{bredon,mall1,mall4}.} as the open $U$ ranges through the system $\gauge_{i}$ of local open gauges covering the base topological space $X$.

\subsubsection{(QU3) In the spirit of Feynman: Gravity as a Pure Gauge Theory}

In connection with question (QU3) above, the idea here is that instead of thinking geometrically about gravity in the original GR fashion due to Einstein, as if the spacetime metric field $g_{\mu\nu}(x)$ (and its derivatives) is partaking in the curvature of a background smooth spacetime manifold $M$ ($x\in M$),\footnote{In what amounts to the theoretical {\it clich\'e}, that: {\em GR taught us that gravity is not a `proper force', but it is represented by the curvature of the spacetime continuum, which is the `fabric' of the Universe}.} one could alternatively {\em view the gravitational field as some kind of gauge field} (like, for instance, the Yang-Mills ones).\footnote{{\it Prima facie}, the theoretical advantage of adopting a gauge-theoretic ({\it i.e.}, connection, not metric, based) viewpoint on gravity is that it would appear more suitable towards a `unified field theoretic' conception of the fundamental forces of Nature, as the other three fundamental forces of matter---the GUTs of electroweak and strong interactions---are known to be gauge (:Yang-Mills) field forces.} That is to say, regard the gravitational field as a connection $\conn$---classically, {\it \`a-la} CDG, a connection on a smooth vector bundle over a smooth base spacetime manifold $M$. This is how Feynman intuited it below, thus he avoided having to learn upfront some ``{\em fancy schmanzy differential geometry}'' as he coined it, that, anyway, {\em gets in the way of Physics} as he was quoted to maintain in quote (Q7) above:

\bigskip\noindent (Q13)\hskip 0.9in
\begin{minipage}{11cm}
\noindent ``...Thus it is no surprise that Feynman would
recreate general relativity from a non-geometrical viewpoint. The
practical side of this approach is that one does not have to learn
some `{\em fancy-schmanzy}' (as he liked to call it) differential
geometry in order to study gravitational physics. (Instead, one
would just have to learn some quantum field theory.) However, when
the ultimate goal is to quantize gravity, Feynman felt that the
geometrical interpretation just stood in the way. From the field
theoretic viewpoint, one could avoid actually defining upfront the physical meaning of quantum geometry, fluctuating
topology, space-time foam, {\it etc.}, and instead look for the
geometrical meaning after quantization... Feynman certainly felt
that the geometrical interpretation is marvellous, `{\em but the
fact that a massless spin-$2$ field can be interpreted as a metric
was simply a coincidence that might be understood as representing
some kind of gauge invariance'}\footnote{Our emphasis of Feynman's
words as quoted by Bryan Hatfield in \cite{feyn2}.}...''
\end{minipage}

\vskip 0.1in

\noindent In view of (Q13) above, we distill our ADG-theoretic stance towards {\em gravity as a gauge theory} to the following Apophthegm:

\noindent\fbox{%
    \parbox{\textwidth}{%
\noindent\underline{\bf Apophthegm 14}: From an ADG-theoretic point of view, gravity is a background spacetime manifold independent, purely gauge theory that is formulated solely in terms of an algebraic $\sts$-connection field $\conn$ acting categorically as a sheaf morphism on (the local sections of) its associated representation vector sheaf $\modl$ of local quantum gravitational  particle (`graviton') states. The combination of the two into the `unitary' pair $\mathcal{F}_{Einst}=(\modl ,\conn)$ represents, always from an ADG-theoretic standpoint, the {\em dynamically autonomous, intrinsically quantum and external spacetimeless vacuum Einstein gravitational gauge field}, as we have maintained and shown throughout this paper.}%
}

\vskip 0.1in

\noindent This `unified', or better, {\em `unitary'}, purely algebraic, purely gauge, background spacetime manifoldless, dynamically autonomous and intrinsically quantal conception of the ADG-gravitational field had been originally intuited very early on in the trilogy \cite{malrap1,malrap2,malrap3}, further elaborated subsequently in \cite{rap11,rap7,rap13,rap14} and further refined and distilled recently in \cite{rap15,rap19}.

\subsubsection{(QU4-5) Cutting the Gordian Knot: No Background Geometrical Spacetime Manifold, No Inner Product Problem, No Problem of Time, No Need to Quantise Spacetime}

The title of this subsection strikes at the very core of what ADG-gravity can accomplish in addressing arguably the three most important issues that Quantum General Relativity (QGR) encounters.\footnote{By QGR one normally understands the attempt at quantising GR along canonical lines, with Loop Quantum Gravity (LQG), based on Ashtekar's new spin-connection variables in the first order Palatini formulation of GR, being the main approach to QGR \cite{loop,rocci,rovelli,schulz}.} 

Thus, very briefly:

\begin{itemize}

\item Since ADG-gravity does not involve at all a background geometrical spacetime manifold and its $\mathrm{Diff}(M)$ symmetry group, there is no {\em Problem of Time}. That is, there is no Hamiltonian operator, which normally acts as the generator of dynamical time evolution (:temporal diffeomorphisms), that has to annihilate, as a primary constraint, the physical states in the theory,\footnote{This is the content of the Wheeler-DeWitt equation: $H\Psi_{phys.} =0$.} since there is no {\it a priori} background spacetime continuum to begin with.

\item Similarly, because ADG-gravity is background spacetime manifoldless, there is no need to find a $\mathrm{Diff}(M)$-invariant inner product to calculate the physical amplitudes for the dynamical transitions between the aforesaid physical states. In ADG-gravity, physical states are the local sections of $\modl$ on which the connection sheaf morphism $\conn$ acts and defines its own $\aut_{\sts}\modl$-covariant dynamical vacuum Einstein equations (\ref{eq19}) via its own $\aut_{\sts}\modl$-invariant curvature form.

\item {\it Mutatis mutandis}, since ADG-gravity is not based on a background spacetime continuum, it is not impeded, let alone breaks down, by singularities and their associated analytical infinities \cite{clarke3}.

\item Moreover, as we also contended earlier, since no spacetime continuum is involved at all in ADG-gravity, there is no {\it a priori} need to quantise it; furthermore, there is no cut-off spacetime scales (like Planck's) to invoke in order to regularise a spacetime continuum based field theory. The ADG-gravitational field is inherently quantum---{\it i.e.}, it is intrinsically quantised, {\it alias}, sheaf cohomologically 3rd quantised.

\end{itemize}

\noindent The last point above brings us to the next subsection.

\subsubsection{(QU5) No Fundamental Planck Length and Time Scales or Fundamental Constants: A New Mathematical Theory is Needed for a Genuinely Background Spacetimeless Gauge Field Theory}

In \cite{rap14,rap15,rap19} we highlighted and emphasised the need to develop new mathematics (:concepts and structures) more `suitable' for QG research. With the abandonment of any background geometrical spacetime manifold structure in the purely algebraic and purely gauge (:$\sts$-connection based) field theoretic ADG-gravity, we also abandon the idea of the existence of the fundamental spacetime scales of Planck that are built from the three fundamental constants $G$, $c$ and $\hbar$ mentioned earlier.\footnote{Recall again from earlier that the Planck length is given by the expression $l_{P}=\sqrt{Gh/c^{3}}\approx 1.6\times 10^{-35}m$; while the Planck time is $t_{P}=\sqrt{Gh/c^{5}}\approx 5.4\times 10^{-44}s$.} In this respect, we quote Unzicker from \cite{unzicker}: 

\bigskip\noindent (Q14)\hskip 0.9in
\begin{minipage}{11cm}
\noindent ``...Despite all progress, however, physics still needs some fundamental constants to describe nature's behaviour. And it is precisely here that our knowledge, which is certainly far advanced, reaches its limits. Unlike most physicists, I am convinced that these constants of nature do not represent an absolute limit to our knowledge, but mark our currently still limited understanding. Ultimately, these constants of nature are arbitrary, unexplained numbers that have allowed academics to find peace of mind by declaring the unexplained to be unexplainable. However, a thorough historical and methodological reflection forces us to consider an alternative: The alleged existence of fundamental constants simply means that we have not yet understood the laws of nature down to their origin. There are no constants of nature, just as there are no gods...'' 
\end{minipage}

\noindent and further down in the book \cite{unzicker}, Unzicker highlights the importance of `{\em Spacetimelessness}' and, like Einstein urged us above, of the need to {\em develop a new mathematical theory} in order to address fields and their particle quanta together in a unified fashion:

\bigskip\noindent (Q15)\hskip 0.9in
\begin{minipage}{11cm}
\noindent  ``{\em ...A thorough analysis of the history of physics leads to the conclusion that there is a serious problem with what have been considered the basis of reality for centuries: Space and Time. These may be the most accessible concepts for human perception, but are probably unsuitable for a basic understanding of nature...}

\vskip 0.07in

 ...Yet a new perspective unfolds that clarifies which problems of fundamental physics can and must be solved in order to achieve a satisfactory understanding of reality. {\em Ultimately, we search for mathematical objects whose properties describe the various physical phenomena in purely mathematical terms...}

\vskip 0.07in

...[Mathematicians, physicists and even non-specialists,] once they become familiar with the historical-methodological approach outlined in the following chapters, will easily understand that {\em physics needs a new paradigm that goes beyond the concepts of space and time}...''

\end{minipage}

 \vskip 0.1in

\noindent We hereby contend that ADG and ADG-gravity provide us with such a new paradigm of theory construction that ``{\em goes beyond the concepts of space and time}'' and focuses directly on the dynamical algebraic relations between the physical fields themselves, without recourse or reference to an external to them and fixed, ether-like spacetime continuum, with all its theoretically arbitrary fixed parameters (constants) and fundamental cut-off scales based on them, as well as its inherent singularities and unphysical field infinities. 

For, already back in the early 20th century, Albert Einstein came to `warn' us in \cite{einst2} about the potential `unphysicality' of the traditional concepts of {\em Space} and {\em Time}, as follows:\footnote{This Einstein quote (Q13) can be found on page 10 of \cite{rap19}.}

\bigskip\noindent (Q16) \hskip 0.9in
\begin{minipage}{11cm}
\noindent ``{\small Time and space are modes by which {\em
we}\footnote{Our emphasis.} think, not conditions in which we
live.}''
\end{minipage}

\vskip 0.1in

\noindent At the same time, ADG is a type of essentially algebraic, {\em Relational Mathematics}, that seems to be tailor-cut for the mathematics needed in QG research as Mallios contends in \cite{mall15}. To further reinforce this point, we recall back in 1999, the year after Mallios's first 2-volume monograph {\em Geometry of Vector Sheaves} (:ADG) was published \cite{mall1}, that there was a Russian referee's/reviewer's report about (the first Russian translation of) the monograph-book by saying [quoting his remarks almost verbatim from memory below]:\footnote{If this author's memory serves him well, the aforementioned Russian reviewer/referee was the late Professor Alexander Khelemskii, Department of Mechanics and Mathematics, Lomonosov Moscow State University (Russia).}

\bigskip\noindent (Q17)\hskip 0.9in
\begin{minipage}{11cm}
\noindent ``...This book is a more than welcome addition to new theoretical developments in Differential Geometry, especially nowadays that the need has arisen to move away from a smooth background [differential manifold] `space' and focus directly on the [algebraic] relations between the `geometrical objects' that live on that `space'...''
\end{minipage}

\vskip 0.1in

\noindent Ultimately, it may well turn out to be that concepts like {\em Space}, {\em Time} and their fusion into the {\em Spacetime Continuum} of classical relativistic (GR) and quantum relativistic field theory (QFT) and their mathematical modelling by CDG-theoretic means are inappropriate for addressing QG issues and they hinder our progress on that research front. Yet again, Einstein comes to warn us about our almost
religious abiding by old, tried-and-tested concepts \cite{einst7}:

\bigskip\noindent (Q18)\hskip 0.9in
\begin{minipage}{11cm}
\noindent ``{\small ...Concepts which have proved useful for
ordering things easily assume so great an authority over us, that
we forget their terrestrial origin and accept them as unalterable
facts. They then become labelled as `conceptual necessities', `a
priori situations', etc.\footnote{Think for instance of the
apparently fundamental notion of the `{\em spacetime continuum}'.} The
road of scientific progress is frequently blocked for long periods
by such errors. It is therefore not just an idle game to exercise
our ability to analyze familiar concepts, and to demonstrate the
conditions on which their justification and usefulness depend, and
the way in which these developed, little by little...}"
\end{minipage}

\vskip 0.1in

In closing, we quote Ernst Straus' reminiscing about Einstein's, widely regarded as unfinished and incomplete, `{\em unified field theory}' program, or perhaps better, {\em vision}, in \cite{straus}:

\bigskip\noindent (Q19)\hskip 0.9in
\begin{minipage}{11cm}
\noindent ``{\small ...Einstein's quest for the ultimate correct [unified or unitary] field theory is generally considered to
have failed. I think that this did not really surprise Einstein,
because {\em he often entertained the idea that vastly new mathematical
models would be needed, that possibly the field-theoretical
approach through the kind of mathematics that he knew and in which
he could do research would not, could not, lead to the ultimate
answer, that the ultimate answer would require a
kind of mathematics that probably does not yet exist and may not
exist for a long time}.\footnote{Our emphasis.} However, he did not have the slightest
doubt that an ultimate theory does exist and can be discovered.}''
\end{minipage}

\noindent The need of developing {\em new mathematics for QG research}, always in connection with and under the prism of ADG-gravity, has been amply emphasised in previous works of ours with a philosophical slant \cite{rap14,rap15,rap19}. Especially for the purpose of our discussion here, we borrow and emphasise from \cite{rap14} {\em the need to develop new, abstract and axiomatic mathematics which will serve as the very foundational substrate on which to build a conceptually sound and calculationally efficacious QG}. To this end, we quote we would like to borrow from Ludwig Faddeev's paper \cite{faddeev} some
telling remarks made by Paul Dirac in
\cite{dirac3}:\footnote{The quotation below is split into two
paragraphs (I and II), on which we comment separately following it.}

\vskip 0.1in

\bigskip\noindent (Q20)\hskip 0.9in
\begin{minipage}{11cm}
\noindent ``{\small ...The steady progress of physics requires for
its theoretical foundation a mathematics that gets continually
more advanced. This is only natural and to be expected. What,
however, was not expected by the scientific workers of the last
century was the particular form that the line of advancement of
the mathematics would take, namely, it was expected that the
mathematics would get more complicated, but would rest on a
permanent basis of axioms and definitions, {\em while actually the
modern physical developments have required a mathematics that
continually shifts its foundation and gets more abstract...It
seems likely that this process of increasing abstraction will
continue in the future and that advance in physics is to be
associated with a continual modification and generalization of the
axioms at the base of mathematics rather than with logical
development of any one mathematical scheme on a fixed
foundation.}\footnote{Our emphasis.} {\bf (I)}

There are at present fundamental problems in theoretical physics
awaiting solution [...]\footnote{Dirac here mentions a couple of
outstanding mathematical physics problems of his times. We have
omitted them.} the solution of which problems will presumably
require a more drastic revision of our fundamental concepts than
any that have gone before. Quite likely these changes will be so
great that it will be beyond the power of human intelligence to
get the necessary new ideas by direct attempt to formulate the
experimental data in mathematical terms. The theoretical worker in
the future will therefore have to proceed in a more indirect way.
{\em The most powerful method of advance that can be suggested at
present is to employ all the resources of pure mathematics in
attempts to perfect and generalise the mathematical formalism that
forms the existing basis of theoretical physics, and {\sl
after}\footnote{Dirac's own emphasis.} each success in this
direction, to try to interpret the new mathematical features in
terms of physical entities}\footnote{Again, our emphasis
throughout.}...} {\bf (II)}''
\end{minipage}

\vskip 0.1in

\begin{itemize}

\item {\bf (I)} The words from this paragraph to be highlighted
with ADG-gravity in mind are: `{\em a mathematics that gets more
abstract}' and `{\em advance in physics is to be associated with a
continual process of abstraction {\rm [leading to a]} modification
and generalization of the axioms at the base of mathematics}'.

Indeed, our {\em axiomatic Abstract Differential Geometry} essentially involves an abstraction of
the fundamental notions of modern differential geometry ({\it e.g.},
connection), resulting in an entirely algebraic (:sheaf-theoretic)
modification and generalisation of the latter's basic axioms
\cite{mall1,mall2,mall4}. And it is precisely this abstract and
generalised character of ADG and its offshoot application, ADG-gravity, that makes us hope that their further development and 
application could advance significantly (theoretical) physics, and
in particular, QG research. 

\item {\bf (II)} In this paragraph, what should be highlighted is on
the one hand Dirac's prompting us `{\em to generalize the
mathematical formalism that forms the existing basis of
theoretical physics}', and on the other, `{\em to try to interpret
the new mathematical features in terms of physical entities}'.
Again, ADG comes to fulfill Dirac's vision, since {\em the} (or at
least the bigger part of the) mathematics that lies at the heart
of current theoretical physics---namely, (the formalism of) {\em the smooth manifold based
Classical Differential Geometry} (CDG)---is
abstracted and generalised, while {\em after} this abstraction and generalisation
has been achieved, the physical application and interpretation (of
ADG's novel concepts and features) has been carried out,
especially in the theoretical physics' field of quantum gauge
theories and gravity research. 

\vskip 0.1in

Following Dirac's words above, we believe that {\em ADG is a powerful theory and method for advancing QG reasearch} indeed.
\end{itemize}

\vskip 0.1in

\section{A Brief Postscriptum on the Fundamental Background Spacetime-lessness of ADG-Gravity vis-\`{a}-vis the Metaphysics of Quantum Gravity Research} 

This brief postscript is an after-the-maths after-math and a $\mu\epsilon\tau\alpha^{'}$-the-physics meta-physical discussion,\footnote{$\mu\epsilon\tau\alpha^{'}$ in Greek means {\em after}. A double pun is intended here.} about the {\em Fundamental Ontological Background Spacetime-lessness} (FOBS) posited and advocated in ADG-gravity.

The metaphysics of QG, as it has been recently expounded in detail in \cite{lebihan}, centers around the philosophical contention that the notions of {\em Space}, {\em Time}, but perhaps more importantly, their classical relativistic merging into a {\em Spacetime Continuum} fundamentally assumed in both SR and GR, as well as in the flat QFTs of matter, are {\em not} fundamental concepts---{\it i.e.}, ones having no basic {\it a priori} ontological status in current QG research.\footnote{See the analytical discussion and the extensive references in \cite{lebihan} for everything in connection with that paper that we mention in this section about the non-fundamental character of the notion of {\em spacetime} in QG research.} 

At the very beginning of \cite{lebihan}, the authors make it clear that:

\bigskip\noindent (Q21)\hskip 0.9in
\begin{minipage}{11cm}
\noindent ``{\small ...Approaches to quantum gravity are not yet fully worked-out theories. Nevertheless,
they already provide a certain partial understanding of physical reality in different ways.

{\em Remarkably, they do so with a striking similarity: they virtually all deny the existence of
some features usually regarded as essential to the existence of spacetime}\footnote{Our emphasis.} (or space and/or
time) such as its four-dimensionality, the existence of distances and durations between
events, or even the very partial ordering of events.

This observation is particularly noteworthy, considering the pervasive influence of
spatial and temporal organisation on the human mind across various facets of daily life
and theoretical thinking, ranging from most ancient religions to contemporary scientific
worldviews. 

{\em The metaphysics of quantum gravity takes the puzzling observation that
physics could teach us that space and time are not fundamental as its starting point.}\footnote{Our emphasis.} It
draws on resources from traditional metaphysics to tackle a set of issues related to the
possible non-fundamentality of spacetime, and investigates its potential implications for
venerable traditional issues in metaphysics.

{\em The metaphysics of quantum gravity is a relatively small and new research field,
and thus as of now, its focus has been on explaining how spacetime could emerge from
a more fundamental and non-spatiotemporal ontology}\footnote{Again, our emphasis.} ...}"
\end{minipage}

\vskip 0.1in

\noindent Then, the authors of \cite{lebihan} give general theoretical non-approach-specific, as well as particular approach-specific, arguments about the non-fundamental ontological character of spacetime in QG research. In the approach-specific arguments, they draw ideas from arguably the three currently most popular and so far perhaps most fruitful approaches to and research programs for QG, namely, {\em string theory}, {\em loop quantum gravity}\footnote{Canonical or covariant.} and {\em causal set theory}.\footnote{See the references in \cite{lebihan} and in the present paper for those three approaches to QG.}

We are not going to delve deep into the arguments for and against the fundamental ontological character of {\em spacetime} in QG as expounded in \cite{lebihan}. Instead, by contrast to the three research programs mentioned above, we are going to state in a clear-cut way: 

\begin{itemize}

\item (i) In what sense is ADG-gravity {\em fundamentally and ontologically background spacetimeless} (FOBS); and, 

\item (ii) What is ADG-gravity's sole ontological entity and the theory's commitment to it---{\it i.e.}, the theory's only axiomatically assumed fundamental {\it ur}-structure and its {\em aufbau} based on it.\footnote{The epithet {\it ur} in German means {\em primitive, original, elementary, basic, atomic, irreducible and/or fundamental}. The German word {\it aufbau}, as we saw earlier, means {\em progressive building or construction} (from the bottom up).}

\end{itemize}

\subsection{ADG-gravity is Fundamentally and Ontologically Spacetimeless (FOBS)}

Here is a shortlist of the features of ADG-gravity that qualify the theory as being FOBS:

\begin{enumerate}

\item No background (base) locally Euclidean differential spacetime manifold is employed in the theory.

\item No background (smooth) spacetime metric is employed in the theory.

\item As a result, the theory, which is of a purely algebraic (:relational) character, does not have an upfront, {\it a priori} geometrical and physical interpretation in terms of spacetime structures and concepts, hence the notion of spacetime is not of a fundamental ontological character in the theory.

\item {\it Mutatis mutandis} then for four issues that are `of basic concern' in the three aforementioned approaches to QG, but {\em not} in ADG-gravity as we discussed earlier, namely: 

\begin{itemize}

\item (i) the issue whether spacetime is discrete or continuous; 

\item (ii) the issue of a possible quantisation of spacetime itself; 

\item (iii) the issue of fundamental constants and their featuring into the fundamental spacetime scales of Planck that are regarded as some kind of basic regularisation `cut-offs' of the base spacetime continuum, which turn it to a `reticular' substratum so as to `regularise' the continuous (ultra-violet, for example) field infinities based on it;\footnote{Throughout our works on formulating ADG-theoretically a finitary, causal and quantal version of vacuum Einstein-Lorentzian gravity \cite{malrap1,malrap2,malrap3,rap1,rap2,rap5,rap7,rap11,rap13,rap15,rap19,rapzap1,rapzap2}, we have used the eipithets {\em `discrete'/`reticular'} and the verbs {\em `to discretise'/`to reticularise'} interchangeably.} and, 

\item (iv) closely related to all three issues (i-iii) above is the issue of {\em the emergence of the classical spacetime continuum} (:the curved base manifold of GR and the flat background Minkowski space of the QFTs of matter) at scales greater than Planck's as some kind of (formal) classical/Bohr continuum/correspondence limit/principle.\footnote{In connection with our ADG-theoretic formulation of a locally finite, causal and quantal version of vacuum Einstein-Lorentzian gravity, see especially \cite{malrap3,rap5,rap7,rapzap1,rapzap2}.}

\end{itemize}

\end{enumerate}

\subsection{The Basic Ontological Entity in ADG-gravity}

In keeping with the {\em abstract}, {\em general} and essentially {\em axiomatic} character of ADG and ADG-gravity \cite{mall2,mall5,mall14}, below we state the last Apophthegm of the paper regarding {\em the basic ontological entity/entities in ADG-gravity}. In doing so, we tacitly abide by the following broad `definition' of {\em a fundamental ontological entity in a theory} as found in \cite{AI2}: 

\bigskip\noindent (Q22)\hskip 0.9in
\begin{minipage}{11cm}
\noindent ``{\small An {\em ontological entity}\footnote{Our emphasis.} in a theory is any`thing'---object, property, process, or concept---that a theory asserts or assumes to exist. {\em It defines the fundamental building blocks of that theory's reality, acting as the inventory of what must exist for the theory to be true},\footnote{Our emphasis.} such as electrons in physics, or social classes in sociology...{\em A theory is `ontologically committed' to the entities that must exist for its statements to be true}.\footnote{Again, our emphasis.} For example, a theory of gravity is committed to the existence of gravitational fields or the atomic theory of matter to the existence of atoms. The types of Entities can be physical (chairs, atoms), abstract (numbers, sets), or conceptual (social structures, minds)...''}
\end{minipage}

\vskip 0.1in

\noindent Here is our concluding apothegm about the basic ontological entity in ADG-gravity and how the theory is committed to it:

\noindent\fbox{%
    \parbox{\textwidth}{%
\noindent\underline{\bf Concluding Apophthegm (15)}: The sole basic ontological entity in ADG-gravity is the following triplet of closely entwined structures:

\vskip 0.05in

$\bullet$ (i) The {\em structure algebra sheaf} $\sts$ of abstract arithmetics, sheaf cohomological coefficients and generalised coordinates; 

\vskip 0.05in

$\bullet$ (ii) The {\em vector sheaves} (:differential $\sts$-modules) $\modl$ that are essentially based on and derive from $\sts$; but most importantly, the key ontological entity in ADG and ADG-gravity is, 

\vskip 0.05in

$\bullet$ (iii) The notion of an algebraic $\sts$-connection field $\conn$ on the relevant differential $\sts$-modules $\modl$, which: 

---(a) derives from $\sts$ as we argued in this paper; 

---(b) defines an ADG-theoretic field as the pair $\mathcal{F}=(\modl ,\conn)$ as shown in (\ref{eq6}); 

---(c) generates the vacuum Einstein dynamical field law of gravity via its curvature form as shown in (\ref{eq9}); and, 

---(d) partakes in the purely algebraic canonical type of sheaf cohomological 3rd quantisation of gravity \cite{rap13,rap15}.

\vskip 0.05in

$\bullet$ We can merge all three structures above to the folowing {\em unitary ontological triplet of ADG-gravity}: $\mathfrak{U}:=(\sts ,\modl, \conn)$; which in turn can be reduced to the pair $\mathfrak{U}:=(\sts ,\mathcal{F})$, since an ADG-field is defined as the pair $\mathcal{F}:=(\modl ,\conn)$ as we saw earlier in (\ref{eq6}).}%
}

\subsection{The Ontological Commitment and Vital Dependence of ADG-gravity on $\mathfrak{U}:=(\sts ,\modl, \conn)$}

The ontological commitment of ADG-gravity to $\mathfrak{U}:=(\sts ,\modl, \conn)$ can be reduced to the following aphorism, our third and last one:

\vskip 0.1in

\noindent\fbox{%
    \parbox{\textwidth}{%
\noindent\underline{\bf Aphorism 3:} ADG-gravity is fundamentally based on, derives from, and can be reduced down to, the basic unitary ontological triplet $\mathfrak{U}:=(\sts ,\modl, \conn)$.
}%
}

\vskip 0.1in

\noindent which reflects the very title of the present paper. 

\subsubsection{A brief concluding note on the Principle of Pure Field Realism}

Elsewhere, during past philosophical musings on ADG-gravity \cite{malrap3,rap11,rap13,rap14} and more recently in \cite{rap15,rap19}, we maintained that, because the vacuum Einstein ADG-gravitational field $\mathcal{F}_{Einst}:=(\modl ,\conn)$ is, as we showed and argued in detail in this paper, the following four-fold entity:

\begin{enumerate}

\item An external (background) spacetime manifoldless entity (:structure);

\item A relational, dynamically autonomous and self-symmetric structure;

\item An intrinsically quantum and closed gauge field system;

\item An $\sts$-invariant (:$\sts$-functorial) structure with respect to our generalised coordinate measurements in $\sts$;

\end{enumerate}

\vskip 0.1in

\noindent\fbox{%
    \parbox{\textwidth}{%
\noindent ADG-gravity is a theory of gravity whose underlying philosophy may be coined {\em Background Spacetimeless and Fundamental Spacetime Scale-less Pure Gauge Field Realism} \cite{malrap3,rap5,rap7,rap11,rap13,rap14,rap15,rap19}.
}%
}

\vskip 0.1in

\noindent 

\section*{Acknowledgments}

The present paper is lovingly dedicated to my teachers, mentors, colleagues and travelling companions in the quest, Professors {\em David Finkelstein, Jim Lambek, Tasos Mallios} and {\em Steve Selesnick}, who, over the span of three and a half decades, showed me and taught me the following fundamentally {\sl Quadriunal Character of Mathematical Physics Research}: {\em to question everything, to doubt everything, to appreciate my ignorance,} but perhaps most importantly, {\em to humour my knowledge}.

\hskip 0.02in

\centerline{$<.><.><.><.><.><.><.><.><.>$}

\hskip0.02in

\end{document}